\documentclass[tradiabstract]{aa}
\usepackage{hyperref}
\usepackage{txfonts}
\usepackage{graphicx,subcaption}
\usepackage{float}
\usepackage{natbib}
\usepackage[english]{babel}
\usepackage{color}
\usepackage{longtable}
\usepackage{rotating}
\usepackage{amsmath}

\begin{document}

\title{CONCERTO: High-fidelity simulation of millimeter line emissions of galaxies and [CII] intensity mapping}

\author{M.~B\'ethermin\inst{1,2} \and 
A.~Gkogkou\inst{1} \and
M.~Van Cuyck\inst{1} \and
G.~Lagache\inst{1} \and
A.~Beelen\inst{1} \and
M.~Aravena\inst{3} \and
A.~Benoit\inst{4,5} \and
J.~Bounmy\inst{6,5} \and
M.~Calvo\inst{4,5} \and
A.~Catalano\inst{6,5} \and
B.~de Batz de Trenquelleon\inst{1} \and
C.~De Breuck\inst{7} \and
A.~Fasano\inst{1} \and 
A.~Ferrara\inst{9} \and
J.~Goupy\inst{4,5} \and 
C.~Hoarau\inst{6,5} \and
C.~Horellou\inst{10} \and
W.~Hu\inst{1} \and
A.~Julia\inst{1} \and
K.~Knudsen\inst{10} \and
J.-C.~Lambert\inst{1} \and
J.~Macias-Perez\inst{6,5} \and
J.~Marpaud\inst{6,5} \and
A.~Monfardini\inst{4,5} \and
A.~Pallottini\inst{9} \and
N.~Ponthieu\inst{8,5} \and
Y.~Roehlly\inst{1} \and
L.~Vallini\inst{9} \and
F.~Walter\inst{11} \and 
A.~Weiss\inst{12}}


\institute{
Aix Marseille Univ, CNRS, CNES, LAM, Marseille, France, \email{matthieu.bethermin@lam.fr} \and
Universit\'e de Strasbourg, CNRS, Observatoire astronomique de Strasbourg, UMR 7550, 67000 Strasbourg, France \and 
N\'ucleo de Astronom\'ia, Facultad de Ingenier\'ia y Ciencias, Universidad Diego Portales, Av.  Ej\'ercito 441, Santiago, Chile
\and
Univ. Grenoble Alpes, CNRS, Grenoble INP, Institut N\'eel, 38000 Grenoble, France
\and 
Groupement d'Interet Scientifique KID, 38000 Grenoble and 38400 Saint Martin d'H\'eres, France
\and 
Univ. Grenoble Alpes, CNRS, LPSC/IN2P3, 38000 Grenoble, France
\and 
European Southern Observatory, Karl Schwarzschild Straße 2, 85748 Garching, Germany
\and
Univ. Grenoble Alpes, CNRS, IPAG, 38400 Saint Martin d'H\'eres, France
\and
Scuola Normale Superiore, Piazza dei Cavalieri 7, 56126 Pisa, Italy
\and
Chalmers University of Technology, Department of Space, Earth and Environment, Onsala Space Observatory, SE-43992 Onsala, Sweden
\and
Max-Planck-Institut für Astronomie, Königstuhl 17, D-69117 Heidelberg, Germany
\and
Max-Planck-Institut für Radioastronomie, Auf dem Hügel 69, D-53121 Bonn, Germany
}

\date{Received ??? / Accepted ???}

\abstract{The intensity mapping of the [CII] 158\,$\mu$m line redshifted to the submillimeter window is a promising probe of the z>4 star formation and its spatial distribution into the large-scale structure. To prepare the first-generation experiments (e.g., CONCERTO), we need realistic simulations of the submillimeter extragalactic sky in spectroscopy. We present a new version of the simulated infrared dusty extragalactic sky (SIDES) including the main submillimeter lines around 1\,mm (CO, [CII], [CI]). This approach successfully reproduces the observed line luminosity functions. We then use our simulation to generate CONCERTO-like cubes (125--305\,GHz) and forecast the power spectra of the fluctuations caused by the various astrophysical components at those frequencies. Depending on our assumptions on the relation between star formation rate and [CII] luminosity, and the star formation history, our predictions of the z$\sim$6 [CII] power spectrum vary by two orders of magnitude. This highlights how uncertain the predictions are and how important future measurements will be to improve our understanding of this early epoch. SIDES can reproduce the CO shot noise recently measured at $\sim$100 GHz by the mmIME experiment. Finally, we compare the contribution of the different astrophysical components at various redshift to the power spectra. The continuum is by far the brightest, by a factor of 3 to 100 depending on the frequency. At 300\,GHz, the CO foreground power spectrum is higher than the [CII] one for our base scenario. At lower frequency, the contrast between [CII] and extragalactic foregrounds is even worse. Masking the known galaxies from deep surveys should allow to reduce the foregrounds to 20\,\% of the [CII] power spectrum up to z$\sim$6.5. However, this masking method will not be sufficient at higher redshifts. The code and the products of our simulation are released publicly and can be used for both intensity mapping experiments and submillimeter continuum and line surveys.}


\titlerunning{SIDES: line emission and [CII] intensity mapping forecasts}

\authorrunning{B\'ethermin et al.}

\maketitle


\section{Introduction}

Our understanding of the star formation history in the Universe has dramatically evolved during the last two decades (see \citealt{Madau2014} for a review). We have now access to the rest-frame UV light from young massive stars escaping the galaxies up to z$\sim$10 \citep[e.g.,][]{Schenker2013,Bouwens2015,Ishigaki2018} and to the UV energy absorbed by dust and reprocessed into the far-infrared up to z$\sim$7 \citep[e.g.,][]{Gruppioni2020,Khusanova2021,Fudamoto2021,Zavala2021,Wang2021}, even if there are still significant discrepancies between authors in the far-infrared (see Sect.\,\ref{sect:SFRD}). The combination of these two pieces of information allows us to reconstruct the full evolution of the star formation rate density (SFRD), i.e. the total mass of stars formed per time unit in a comoving volume of Universe over about 13\,Gyr.

In contrast, we know much less about the spatial distribution of the star formation in the large-scale structures at z$\gtrsim$2. Most of the studies use the stellar mass function and abundance matching (sometimes combined with clustering measurements) to measure the relation between stellar mass versus halo mass \citep[e.g.,][]{Behroozi2013,Moster2013,McCracken2015,Cowley2018,Legrand2019,Moster2018,Behroozi2019}. These studies showed that star formation is most efficient in dark-matter halos of $\sim 10^{12}$\,M$_\odot$. However, this only provides information about the integrated star formation history of the halo. The link between star formation rate (SFR) and halo mass is less analyzed and the works are mostly focused on z$\sim$2 \citep[e.g.,][]{Magliocchetti2011,Lin2012,Bethermin2012a,Wang2013,Bethermin2014,Ishikawa2016}. These studies are usually based on the angular clustering of galaxies selected in the optical and/or the far-infrared regimes. At higher redshift, the studies are limited to the clustering of Lyman break galaxies \citep[e.g.,][]{Kashikawa2006,Lee2009,Ishikawa2017}, which is biased towards unobscured star formation. Measurements of dusty galaxies selected in the submillimeter domain remains difficult because of the small samples and the large beam ($>$10") of single-dish instruments \citep{Cowley2017,Wilkinson2017}.

The anisotropies of the cosmic infrared background (CIB), i.e. the integrated emission of dust in galaxies across cosmic times, provide an alternative probe of the spatial distribution of the SFR in the high-redshift Universe \citep[e.g.,][]{Lagache2007,Viero2009,Planck_CIB2011,Amblard2011,Planck_CIB2013}. The modeling of these anisotropies at various frequencies can thus provide constraints on both the SFRD evolution and the host halos of the dust-obscured star formation \citep{Penin2012a,Shang2012,Viero2013,Bethermin2013,Maniyar2018,Maniyar2021}. However, the CIB comes from a large redshift range and significant degeneracies between redshift slices exist. Depending on the emission redshift, the observed peak of the dust emission is located at different frequencies. Consequently, lower frequencies have a higher contribution from higher redshift. The degeneracies between redshift slices can thus be broken by modeling several frequencies simultaneoulsy. However, this becomes extremely difficult at z$>$4.

These degeneracies no longer exist if we use a line and spectroscopy instead of the continuum and broad-band photometry. Indeed, spectroscopy isolates naturally a thin redshift slice, while the photometric signal is the sum of galaxy emissions at all redshifts. Intensity mapping experiments aim to detect the large-scale collective line emission from galaxies using wide-field spectral mapping. The [CII] line at 158\,$\mu$m is one of the main cooling lines of the interstellar medium \citep[e.g.,][]{Tielens1985,Wolfire2022}. Observational studies at low redshift \citep[e.g.,][]{De_Looze2014, Herrera-Camus2015} and high redshift \citep[e.g.,][]{Capak2015,Schaerer2020,Carniani2020} found a non-evolving and nearly-linear empirical correlation between [CII] luminosity and SFR. Both numerical simulations \citep{Vallini2015,Olsen2017,Lupi2020,Pallottini2022} and semi-analytical models \citep{Lagache2018,Popping2019,Yang2021sam} predict a weakly evolving relation with redshift. The [CII] line is thus a suitable tracer of the high-z star formation at large scales. Several intensity mapping experiments (\citealt{Kovetz2017} for a review) aim to probe the [CII] emission from z$>$4 galaxies redshifted in the submillimeter domain: the Carbon [CII] line in post-reionization
and reionization epoch project \citep[CONCERTO,][]{CONCERTO2020}, the instrumentation for the tomographic ionized-carbon intensity mapping experiment \citep[TIME,][]{Crites2014}, and the Fred Young submillimeter telescope \citep[FYST, formerly CCAT-prime,][]{Stacey2018}. Since they aim to target large scales, these experiments are all installed on single-dish telescope with a limited angular resolution ($\gtrsim$20") but a large field of view ($\gtrsim$10') and have an intermediate spectral resolution (R$\gtrsim$100).

The interpretation of this new type of data opens new challenges, such as characterizing the transfer function from the astrophysical signal to the data cubes through the instrument and the analysis pipeline, and separating the [CII] line from other astrophysical components as the dust continuum or lower-z CO lines \citep[e.g.,][]{Sun2018,Cheng2020}. To prepare our analysis pipelines, we need realistic simulations of the submillimeter extragalactic sky in spectroscopy. Predictions from hydrodynamical simulations are limited to small volumes \citep{Pallottini2015,Hernandez-Monteagudo2017}, the current forecasts are thus based on either analytical approaches based on halo occupation distribution models \citep{Gong2012,Yue2019,Yang2021} or empirical recipes to predict the [CII] emission of galaxies hosted by dark-matter halo simulations \citep{Silva2015,Yue2015,Chung2020}. There are huge differences between these models. For instance, there are more than two orders of magnitude between the amplitudes of the [CII] power spectrum at z$\sim$6 predicted by the various models (see Sect.\,\ref{sect:compare}). It is crucial to use the latest measurements of these empirical relations to have reliable forecasts. 

In this paper, we present a new simulation dedicated to submillimeter intensity mapping, which was calibrated and tested using the latest observational data from (sub-)millimeter observatories as the Atacama large millimeter/submillimeter array (ALMA) and the northern extended millimeter array (NOEMA). This simulation is an extension of the simulated infrared dusty extragalactic sky (SIDES), which starts from dark-matter halo lightcones and uses an empirical prescription to reproduce accurately a large set of mid-infrared to millimeter statistical properties of galaxies in continuum as number counts, redshift distribution, pixel histograms, or power spectra \citep{Bethermin2017}. The [CII] line and its two main lower-z contaminants (CO and [CI]) are included using new empirically-based recipes. The codes and products of this new simulation are publicly available (\url{https://cesamsi.lam.fr/instance/sides/home}) and can be used to prepare or interpret intensity mapping experiments, deep spectral scans with interferometers, or photometric surveys. 

In Sect.\,\ref{sect:sides}, we describe the new version of the SIDES simulation and in particular the implementation of the emission lines in the model. We then compare our results with the latest constraints from deep (sub)millimeter spectroscopic surveys (Sect.\,\ref{sect:LFs}). In Sect.\,\ref{sect:compare}, we compare the intensity-mapping forecasts of our simulations with other models. Finally, we discuss the contribution of the various astrophysical components ([CII], dust, continuum, CO, [CI]) to the intensity mapping signal as a function of frequency and the effect of the masking of known galaxies in Sect.\,\ref{sect:power_spectra}.

We assume a \citet{Planck2015_cosmo} cosmology and a \citet{Chabrier2003} initial mass function (IMF).


\section{Modeling of the lines in SIDES}

\label{sect:sides}
 
The intensity mapping simulations presented in this paper are based on the simulated infrared dusty extragalactic sky (SIDES) presented in \citet[][see Sect.\,\ref{sect:sides2017} for a short description]{Bethermin2017}. This new version of the simulation now includes the main high-redshift lines observed in the millimeter domain: CO, [CII], and [CI] (see Sect.\,\ref{sect:recipe_CII}, \ref{sect:recipe_CO}, and \ref{sect:recipe_CI}, respectively). In addition, the new version of the code contains a spectral cube generator described in Sect.\,\ref{sect:data_cubes}. Contrary to the 2017 version, which was coded in IDL, the new pySIDES code is entirely written in Python and is publicly available at \url{https://gitlab.lam.fr/mbethermin/sides-public-release} (see appendix\,\ref{app:release}).

\subsection{The SIDES simulation}

\label{sect:sides2017}

The SIDES simulation is based on the Bolshoi-Planck simulation \citep{Rodriguez-Puebla2016} from which a lightcone of 1.4$\times$1.4 deg$^2$ was produced. For each halo and sub-halo, we generated a stellar mass using an abundance matching technique tuned to reproduce observed stellar mass functions \citet[e.g.,][]{Ilbert2013,Davidzon2017,Grazian2015}. We then randomly split this sample into star-forming and passive galaxies with a probability depending on their redshift and stellar mass. 

For the star-forming population, we generated star formation rates by distributing galaxies on the so-called main sequence of star-forming galaxies following observational constraints of \citet{Schreiber2015}. A small fraction of this population (3\,\% at $z>1$) is located on the starburst sequence and exhibits a SFR excess. We then derived dust continuum properties using the evolving main-sequence and starburst observed spectral energy distribution (SED) up to z=4 from \citet{Bethermin2017}. As shown in \citet{Bethermin2020}, they are also compatible with the most recent measurements between z=4 and z=6.

All the details are provided in \citet{Bethermin2017}. This model is compatible with observed continuum number counts from the mid-infrared to the millimeter domain after taking into account the angular resolution effects, the source redshift distribution, and the pixel histograms and power spectra of the \textit{Herschel} maps. The results presented in this paper are based on a new Python implementation called pySIDES using a \texttt{pandas} data structure \citep{pandas} for the catalogs. We carefully tested that the results are fully compatible with the \citet{Bethermin2017} version based on an IDL code by comparing the stellar masses, SFR, and continuum fluxes distributions in a large set of redshift slices.

\subsection{Philosophy of this extension of SIDES}

\label{sect:philo}

To prepare the data analysis and the interpretation of future submillimeter intensity mapping experiments, we need simulations with galaxy properties as close as possible from the reality. Semi-analytical models \citep[e.g.,][]{Lagache2018,Popping2019,Yang2021sam} can produce physically-motivated forecasts, but they are usually limited in volume and can be expensive in computing time. More simple approaches use relations between the halo masses and the line intensity. The intensity mapping signal is then derived by either applying such relations to dark-matter simulations \citep[e.g.][]{Yue2015,Chung2020} or deriving the expected signal analytically using halo occupation distribution models \citep[e.g.,][]{Yue2019,Yang2021}. These approaches allow to process large-volume dark-matter simulation rapidly, but the galaxy properties are usually simplified, e.g., same continuum spectral energy distributions or fixed CO line ratios for all galaxies, or no scatter in the relations used to generate the galaxy properties.

Our new SIDES simulation aims to propose an intermediate solution between these two approaches, as it is producing realistic galaxy properties and can be very efficiently applied to large dark-matter lightcones. It will also model both the [CII] emission and its foregrounds (continuum, CO, [CI]) in a consistent manner at all redshifts. At its core the model is semi-empirical, i.e., it is more physically motivated than a completely empirical model and it describes accurately the various galaxy properties relevant for intensity mapping. The previous version of the model produces continuum properties with a scatter in dust temperature and on the relation between stellar mass and SFR, together with different SEDs depending on the type of galaxy. All these recipes evolve with redshift.

In this new version of SIDES, we implemented the lines following the same philosophy. We focused on the three main lines relevant for submillimeter intensity mapping: [CII], CO, and [CI]. For [CII], we use two different empirical relations to test how our results depend on it. For CO, we use spectral line energy distributions templates, which are linked to the intensity of the UV radiation field that is used to derive the dust continuum. This allows to have diverse CO line ratios and an overall evolution with redshift. This feature is particularly important to test component separation methods \citep[e.g.,][]{Cheng2016,Sun2018,Cheng2020}, whose formalism assumes implicitly fixed line ratios. For [CI], we have less constraints and we propose new empirical recipes based on a recent observational compilation.

The new SIDES code is modular and can be easily modified to include new observational results. The version presented in this paper is a compromise between simplicity and being realistic. It was not fine tuned, since the observational constraints were overall well reproduced at the first try. Since the code is public, documented, and modular, different methods can be easily implemented by other users based on their preferences and access to new observational data. The code has been optimized to be able to produce a catalog or data cube on a laptop in a few tens of minutes, which is ideal to perform tests and explore the impact of the various parameters.

\subsection{[CII] emission}

\label{sect:recipe_CII}

The [CII] line at 158\,$\mu$m (1900.54\,GHz) is one of the brightest lines emitted by galaxies and is shifted to the millimeter domain for high-z galaxies. CONCERTO will be sensitive to [CII] at z$>$5.2. Before ALMA, there were very few constraints on [CII] at these early times, but important results were obtained in the recent years. Initially, the [CII] detection rate of high-redshift targets were low \citep[e.g.,][]{Ouchi2013,Ota2014,Maiolino2015,Wilott2015} and theoretical explanations were proposed to explain this [CII] deficit. For instance,  \citep{Vallini2015} suggested that it could be due to extremely low metallicities, while \citet{Katz2017} pointed out the high ionized gas filling factors as a possible explanation. However, \citet{Capak2015} showed using a small sample that 5$<$z$<$6 galaxies still follow the L$_{\rm [CII]}$-SFR relation observed in the local Universe \citep[][HII/starburst relation, hereafter DL14]{De_Looze2014}:
\begin{equation}
\label{eq:DL14}
\log_{10} \left ( \frac{L_{\rm [CII]}^{DL14}}{L_\odot} \right ) = 7.06 + \log_{10} \left ( \frac{\rm SFR}{\rm M_\odot \, yr^{-1}} \right )
\end{equation}
This result was confirmed by the ALPINE large program \citep{Le_Fevre2020,Bethermin2020,Faisst2020} targeting 118 normal galaxies at 4.4$<$z$<$5.9 and \citet{Schaerer2020} confirmed that the DL14 relation remains valid at high-redshift. \citet{Carniani2020} found a similar result after re-analyzing archival data and correcting for the flux loss caused by the too extended configuration used by the early observations. We thus used the DL14 relation (Eq.\,\ref{eq:DL14}) to derive [CII] luminosities from the SFR in the SIDES simulation.

As pointed by \citet{Ferrara2019}, massive high-redshift galaxies correspond much high [CII] surface brightness regime than the initial low-redshift \citet{De_Looze2014} observations and it is almost surprising that this relation remains valid under these physical conditions. While numerical simulations \citep{Arata2020,Pallottini2019,Pallottini2022} predict that the local relation is respected also for $M_\star < 10^9$\,M$_\odot$ galaxies, we currently have no strong observational support for this and we cannot guarantee that a [CII]-deficit does not appear at low-mass (M$_\star<10^{9.5}$\,M$_\odot$) or higher redshift (z$>$6), where the metallicity is expected to be much lower. To test this scenario, we also implemented the \citet[][hereafter L18]{Lagache2018} relation predicted by a semi-analytical model, which produce a lower [CII] luminosity at higher redshifts and lower SFRs:
\begin{equation}
\log_{10} \left ( \frac{L_{\rm [CII]}^{L18}}{L_\odot} \right ) = (1.4 - 0.07 z) \log_{10} \left ( \frac{\rm SFR}{\rm M_\odot \, yr^{-1}} \right ) + 7.1 - 0.07 z.
\end{equation}
This relation predicts a lower [CII] luminosity at higher redshift caused by the high intensity of the radiation field and the cosmic microwave background (CMB) effect \citep[e.g.,][]{Da_Cunha2013}.

We do not expect all the galaxies to follow exactly these relations. In the local Universe, \citet{De_Looze2014} (hereafter DL14) measured a scatter of 0.27\,dex, while \citet{Schaerer2020} found 0.28\,dex after correcting SFR estimates from dust attenuation using \citet{Fudamoto2020} IRX-$\beta$ corrections. However, the intrinsic scatter on the observed L$_{\rm [CII]}$-SFR relation is difficult to estimate observationally because of the uncertainties on SFR estimators. The SFR uncertainties are not well know at high redshift, but they are usually estimated to be around 0.2\,dex. If we combine quadratically this 0.2\,dex scatter with another 0.2\,dex scatter, we obtain 0.28\,dex, which is compatible with the observed scatter. We thus assume an intrinsic scatter of 0.2\,dex to generate [CII] luminosities from the SFR in our simulation.

Finally, the line flux in Jy\,km/s ($I_{\rm [CII]}$) is then derived from the $L_{\rm [CII]}$ using the standard formula provided in \citet{Carilli2013} -- see also \citet{Solomon1997} and \citet{Solomon2005}:
\begin{equation}
\label{eq:conv_lum2flux}
I_{\rm [CII]} = \mu \, \frac{1}{1.04 \times 10^{-3} \, D_L^2 \, \nu_{\rm obs}} \, L_{\rm [CII]},
\end{equation}
where $\mu$ is the lensing magnification provided by SIDES, $D_L$ is the luminosity distance in Mpc, and $\nu_{\rm obs}$ is the observed frequency in GHz. The same formula is also used to convert $L_{\rm [CI]}$ into $I_{\rm [CI]}$ in Sect.\,\ref{sect:recipe_CI}.

\subsection{CO emission}

\label{sect:recipe_CO}

The CO rotational transitions result in emissions at frequencies that are multiples of 115.27\,GHz, and dominate the rest-frame millimeter spectra of star-forming galaxies. This molecule is the most popular tracer of the molecular gas \citep[e.g.,][]{Solomon2005,Carilli2013}. The evolution of galaxy CO emissions from the local to the high-redshift Universe has been widely explored \citep[e.g.,][]{Magdis2012b,Tacconi2013,Saintonge2013,Daddi2015,Dessauges2015,Genzel2015,Aravena2016,Freundlich2019,Tacconi2020}. These studies showed that, at fixed stellar masses, galaxies at higher redshifts have higher CO luminosities and the gas fraction is higher in high-redshift galaxies. 

Main-sequence galaxies at various redshifts follow the same correlation between the luminosity of the ground transition of CO ($L'_{\rm CO(1-0)}$) and the bolometric infrared luminosity (L$_{\rm IR}$, integrated between 8 and 1000\,$\mu$m), which is directly proportional to SFR in the context of our model. Here, we use the L' pseudo luminosities expressed in K\,km/s\,pc$^2$. We chose to derive $L'_{\rm CO(1-0)}$ directly from L$_{\rm IR}$ for simplicity.  In our model, we use the relation calibrated by \citet{Sargent2014} for main-sequence galaxies:
\begin{equation}
\label{eq:Lco}
\log_{10} \left ( \frac{L'_{\rm CO(1-0)}}{\rm K \, km/s \, pc^2} \right ) = 0.81 \log_{10} \left ( \frac{L_{\rm IR}}{L_\odot} \right ) + 0.54.
\end{equation}
Similarly to [CII], we assume an intrinsic scatter of 0.2\,dex, since \citet{Greve2014} measured a scatter of 0.26\,dex on the $L'_{\rm CO(1-0)}$-L$_{\rm FIR}$. The line flux in Jy\,km/s ($I_{\rm CO}$) is then computed from $L'_{\rm CO}$ in K\,km/s\,pc$^2$ using a formula similar to Eq.\,\ref{eq:conv_lum2flux} \citep{Carilli2013}:
\begin{equation}
I_{\rm CO} =  \mu \, \frac{(1+z)^3 \, \nu_{\rm obs}^2}{3.25 \times 10^{7} \, D_L^2} \, L'_{\rm CO}.
\end{equation}

Starbursting systems\footnote{As described in \citet{Bethermin2017}, starbursts in SIDES are treated as a separate population. Their SFRs are drawn from a relation offset by a factor of $\sim$6 above the main sequence. Because of the scatter around both the main and the starburst sequences, there is no clear border between the two populations in the SFR-M$_\star$ plane.} do not follow this correlation. This may be caused by two phenomena: starbursts have a higher SFR for a similar gas mass \citep[e.g.,][]{Genzel2010,Daddi2010b} and the conversion factor from $L'_{\rm CO(1-0)}$ to gas mass is different in starbursts \citep{Downes1998}. Because of these effects, \citet{Sargent2014} found that their $L'_{\rm CO(1-0)}$-L$_{\rm IR}$ correlation lies below the main-sequence one with an offset of -0.46\,dex. We apply the same offset to galaxies labelled as starburst in the SIDES simulation.

To produce the flux of the other transitions, we assume a spectral line energy distribution (SLED) for each object. Both low-redshift \citep[e.g.,][]{Rosenberg2015,Kamenetzky2016} and high-redshift objects \citep[e.g.,][]{Yang2017,Canameras2018,Valentino2020b,Boogaard2020} exhibit a large variety of SLED . Our simulation does not aim to encompass all the complexity of the gas physics in high-redshift galaxies. For main sequence galaxies, we thus use an empirical approach based on \citet{Daddi2015}, who found a correlation between the CO(5-4)/CO(2-1) flux ratio and the mean intensity of the UV radiation field $\langle U \rangle$ \footnote{We use the same definition of the mean UV radiation field as \citet{Magdis2012b} based itself on \citet{Draine2007}. The unity corresponds to the solar neighborhood.}:
\begin{equation}
R_{2-1}^{5-4} = \log_{10} \left ( \frac{I_{\rm CO(5-4)}}{I_{\rm CO(2-1)}} \right ) = 0.60 \log_{10} \left ( \langle U \rangle \right ) - 0.38.
\end{equation}
We note that \citet{Rosenberg2015} also found a similar relation between the CO excitation and the 60\,$\mu$m versus 100\,$\mu$m color. In our simulation, the dust continuum SEDs are already parametrized using this $\langle U \rangle$ parameter, which is also linked to the dust temperature and increases with increasing redshift for main-sequence galaxies (see details in \citealt{Bethermin2017}). Our simulated galaxies will thus have a higher CO excitation at higher redshift. In addition, a scatter on $\langle U \rangle$ of 0.2 \,dex is already included and there will thus be naturally diverse SLEDs for a given galaxy type and redshift.

To generate all the transitions, we need to produce the full SLED. We use a linear combination of the clump and the diffuse templates from \citet{Bournaud2015}, noted $T^{clump}$ and $T^{diff}$ respectively and normalized it to unity for the 1-0 transition. These templates are computed only up to the 8-7 transition. However, \citet{Decarli2020} showed that higher transitions have a negligible contribution at CONCERTO frequencies. The flux of the CO transition between the $J_{\rm up}$ and the $J_{\rm up} - 1$ levels, $I_{CO(J_{\rm up},J_{\rm up} - 1)}$, is computed using:
\begin{equation}
I_{CO(J_{\rm up},J_{\rm up} - 1)} = I_{\rm CO(1-0)} \left ( f_{\rm clump} T_{(J_{\rm up},J_{\rm up} - 1)}^{\rm clump} + (1 - f_{\rm clump}) T_{J_{\rm up}}^{\rm diff} \right ),
\end{equation}
where $f_{\rm clump}$ the contribution of the clump component to the 1--0 transition. We choose $f_{\rm clump}$ to obtain a CO(5-4)/CO(2-1) ratio corresponding to the \citet{Daddi2015} relation. This constraint implies that:
\begin{equation}
 f_{\rm clump} = \frac{ T_{5-4}^{\rm diff} - R_{2-1}^{5-4}  T_{2-1}^{\rm diff} } { R_{2-1}^{5-4} (T_{2-1}^{\rm clump}-T_{2-1}^{\rm diff}) - (T_{5-4}^{\rm clump}-T_{5-4}^{\rm diff})}.
\end{equation}
Because of the scatter on $\langle U \rangle$, some objects have extreme values of this parameter and the $I_{\rm CO(5-4)}/I_{\rm CO(2-1)}$ ratio cannot be reproduced using a combination of these two templates with $0<f_{\rm clump}<1$. For these objects, we use a pure diffuse (low $\langle U \rangle$) or pure clump template (high $\langle U \rangle$). Because of the increasing $\langle U \rangle$ with increasing redshift, the main-sequence galaxies at higher redshifts are more excited, which agrees with the conclusion of, e.g., \citet{Daddi2015} and \citet{Boogaard2020}. A z$>4$, their excitation is close from the average SLED of submillimeter galaxies (SMGs) reported in \citet{Carilli2013} or \citet{Birkin2021}.

The method described in the two previous paragraphs is mainly based on observations of main-sequence galaxies. For the starburst galaxies, we produced an alternative version of the simulation in which we use the same mean SMG ($\gtrsim$3\,mJy at 870\,$\mu$m) SLED template of \citet{Birkin2021} for all galaxies labeled as starburst. The 8-7 transition is not provided and we assume that it has the same flux as the 7-6 transition. We compared the two approaches and found no significant difference because of the small relative contribution of starbursts to the observables related to intensity mapping. In the rest of the paper, we use the version of the model using the \citet{Birkin2021} SLED.

\begin{figure}
\centering
\includegraphics[width=9cm]{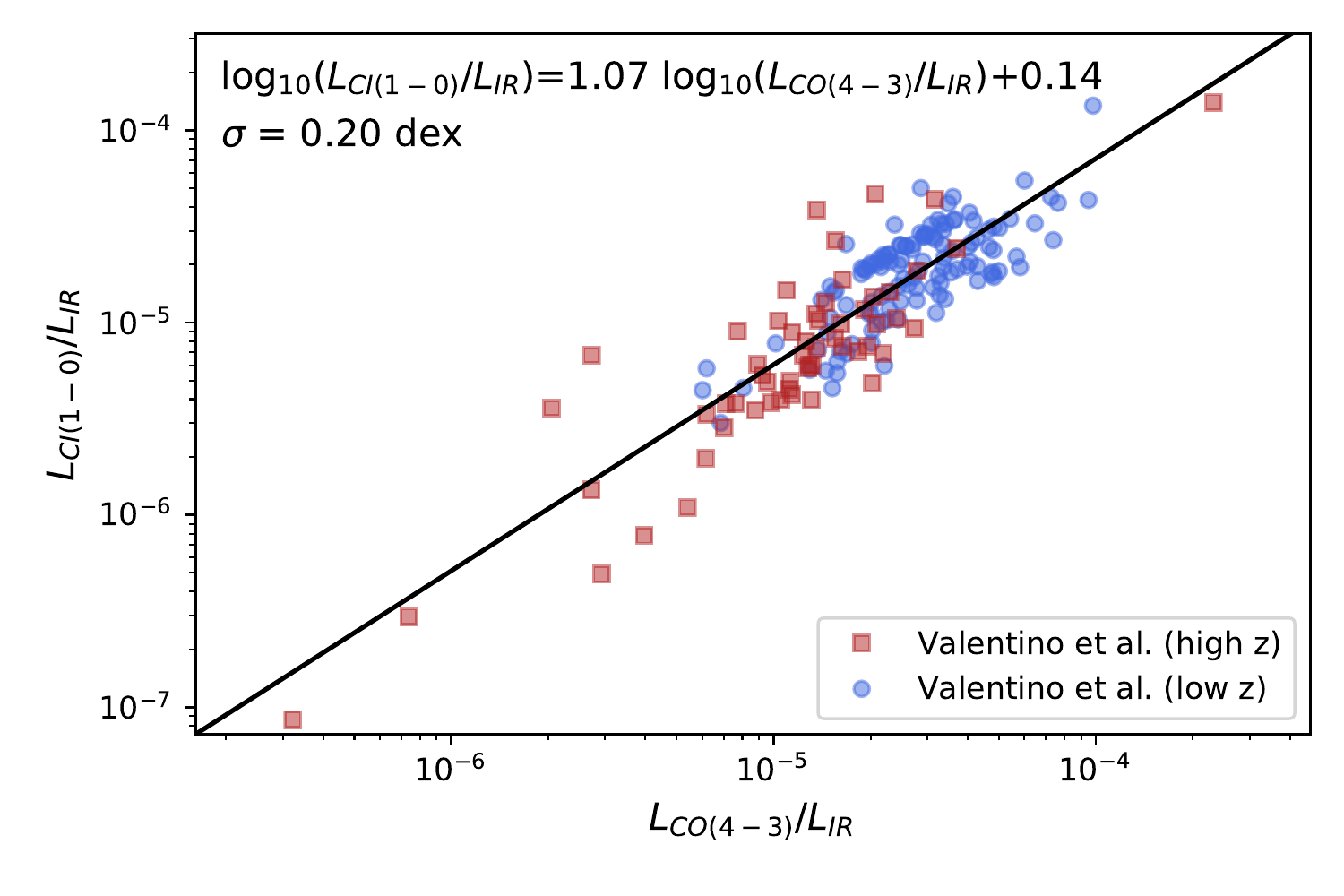}
\includegraphics[width=9cm]{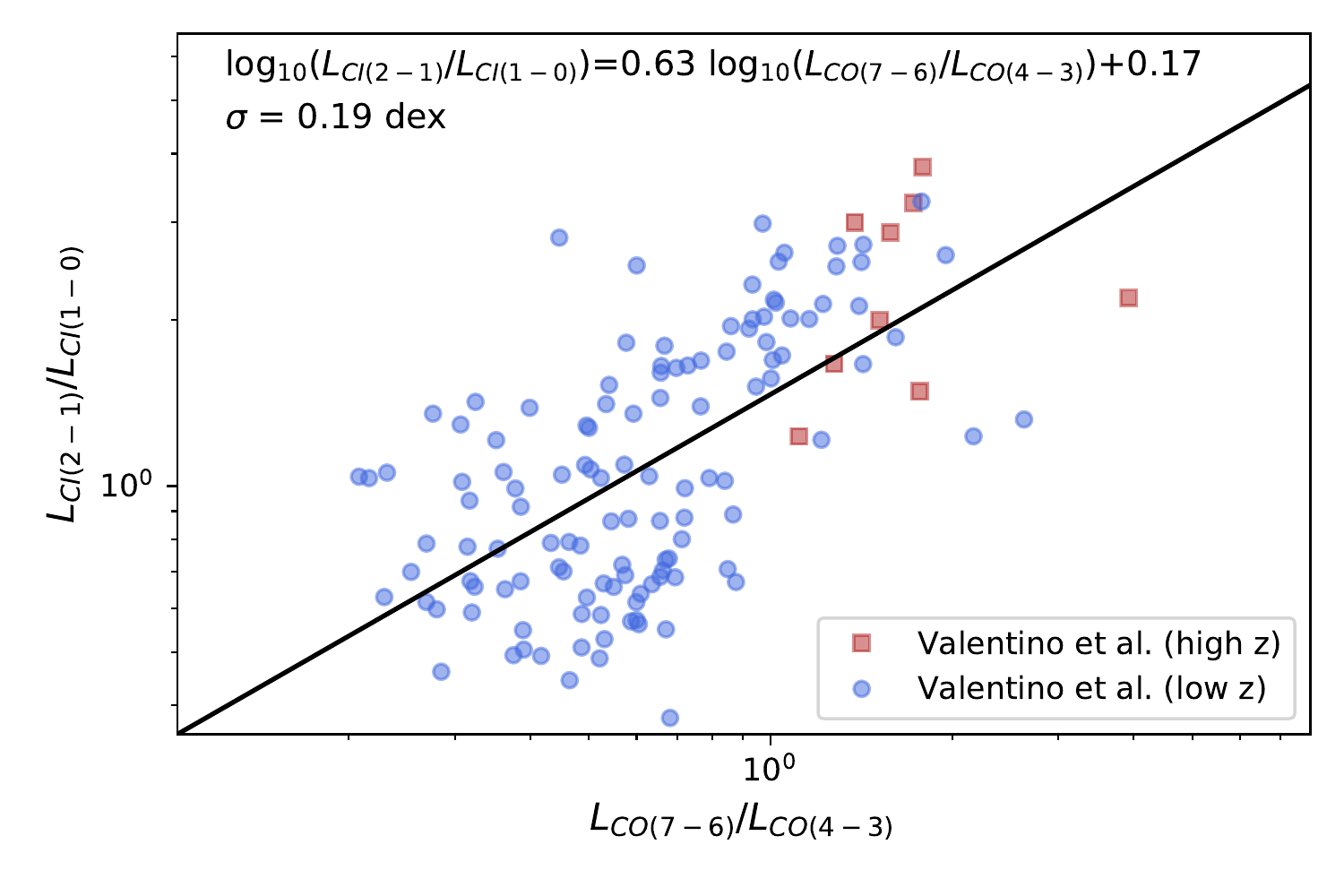}
\caption{\label{ref:CI_recipes} Upper panel: relation between the L$_{\rm [CI](1-0)}$/L$_{\rm IR}$ and L$_{\rm CO(4-3)}$/L$_{\rm IR}$ ratios. The filled blue circles and the red filled squares are respectively the low-z and high-z galaxies compiled by \citet{Valentino2020}. The solid line is our best-fit of this correlation used to generate [CI](1-0)fluxes in our simulation. Lower panel: same figure for the relation between the L$_{\rm [CI](2-1)}$/L$_{\rm [CI](1-0)}$ and L$_{\rm CO(7-6)}$/L$_{\rm CO(4-3)}$ ratio.}
\end{figure}

\subsection{[CI] emission}

\label{sect:recipe_CI}

The two [CI]($^3P_1$-$^3P_0$) transition at 492.16\,GHz rest-frame  and [CI]($^3P_2$-$^3P_1$) transition 809.344\,GHz rest-frame (hereafter abridged [CI](1-0) and [CI](2-1) respectively for simplicty) are also non-negligible contributors to the millimeter sky. The [CI](2-1) is often as bright as its CO(7-6) neighbor at 806.65\,GHz rest-frame \citep{Valentino2020} and can be much brighter in some objects \citep{Gullberg2016}. It can potentially contaminate the CO(7-6) signal and be a problem for CO-decontamination methods assuming a constant CO SLED. The [CI](1-0) is usually a factor $\sim$2 fainter than the CO(4-3) line \citep{Alaghband_Zadeh2013,Bothwell2016,Valentino2020,Bisbas2021}.

For the SIDES simulation, we calibrated empirical relations using the compilation of [CI] data from \citet{Valentino2020}. After exploring various correlations, we found two reasonably tight relations ($<$0.2\,dex of scatter) presented in Fig.\,\ref{ref:CI_recipes}. In this section and contrary to Eq.\,\ref{eq:Lco}, all luminosities are expressed in normal power units, i.e. in multiples of watts such as solar luminosities\footnote{We use only ratios here. Consequently, watts or solar luminosities can be used indistinctively.}.

We found that the ratio between the [CI](1-0) line luminosity (L$_{\rm [CI](1-0)}$) and the total infrared luminosity (L$_{\rm IR}$) and the ratio between L$_{\rm CO(4-3)}$ and L$_{\rm IR}$ are correlated with dispersion of 0.2\,dex (see Fig.\,\ref{ref:CI_recipes} upper panel):
\begin{equation}
\log_{10} \left (  \frac{L_{\rm [CI](1-0)}}{L_{\rm IR}} \right) = 1.07 \, \log_{10} \left (  \frac{L_{\rm CO(4-3)}}{L_{\rm IR}} \right) + 0.14.
\end{equation}
We used the CO(4-3) transition instead of a lower-energy one because of the larger high-z observational dataset available to calibrate our relation. For the typical range of values of $L_{\rm CO(4-3)}/L_{\rm IR}$ in the \citet{Valentino2020} study (10$^{-6}$-10$^{-4}$), the mean $L_{\rm [CI](1-0)}/L_{\rm CO(4-3)}$ will be in the range between 0.5 and 0.7. The slightly superlinear slope suggests that [CI](1-0) is fainter than CO(4-3) in objects with a low line-to-continuum ratio as starbursts or main-sequence galaxies on the higher end in terms of star-formation efficiency. This is expected, since [CI](1-0) is tracing more diffuse gas than CO(4-3) and thus tends to be relatively brighter in less extreme objects. We thus generated [CI](1-0) luminosities and fluxes in our simulation using this relation and the CO(4-3) derived using the method described in Sect.\,\ref{sect:recipe_CO}. We added a 0.2\,dex scatter following the observational constraints.

The ratio between the [CI](2-1) and the [CI](1-0) luminosity is directly linked to the kinematic temperature of the gas \citep{Papadopoulos2004}. This parameter is not included in our simulation. We thus use the CO excitation as a proxy for it. Using the \citet{Valentino2020} compilation, we found the following correlation (see Fig.\,\ref{ref:CI_recipes} lower panel):
\begin{equation}
\log_{10} \left (  \frac{L_{\rm [CI](2-1)}}{L_{\rm [CI](1-0)}} \right) = 0.63 \, \log_{10} \left (  \frac{L_{\rm CO(7-6)}}{L_{\rm CO(4-3)}} \right) + 0.17.
\end{equation}
This relation has a scatter of 0.19\,dex. In the simulation, we generated the [CI](2-1) luminosities from the [CI](1-0) ones using this relation and its scatter.

The two [CI] line fluxes in Jy\,km/s ($I_{\rm [CI]}$) are finally computed from the [CI] luminosities using Eq.\,\ref{eq:conv_lum2flux}.

\begin{figure*}
\centering

\begin{subfigure}{0.24\textwidth}
\includegraphics[width=\textwidth]{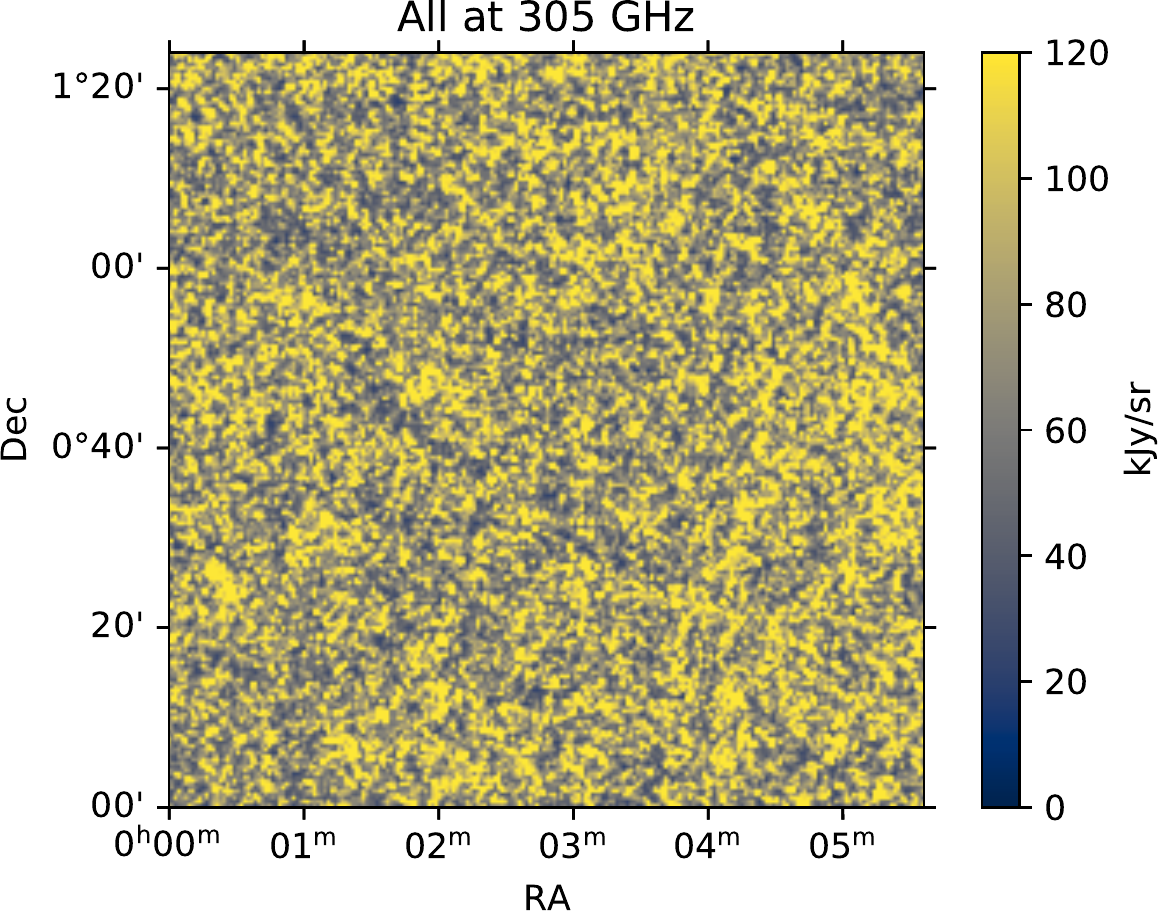}
\end{subfigure}
\hfill
\begin{subfigure}{0.24\textwidth}
\includegraphics[width=\textwidth]{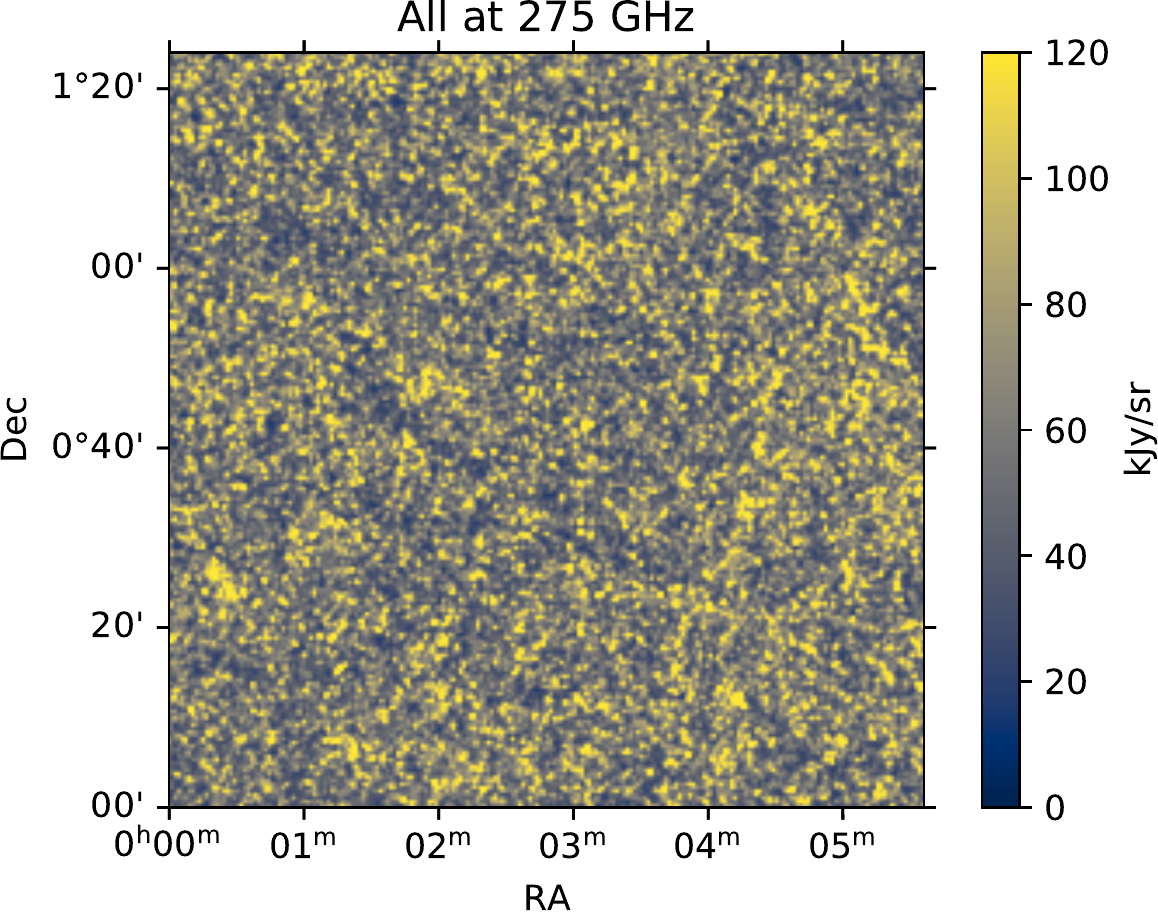}
\end{subfigure}
\hfill
\begin{subfigure}{0.24\textwidth}
\includegraphics[width=\textwidth]{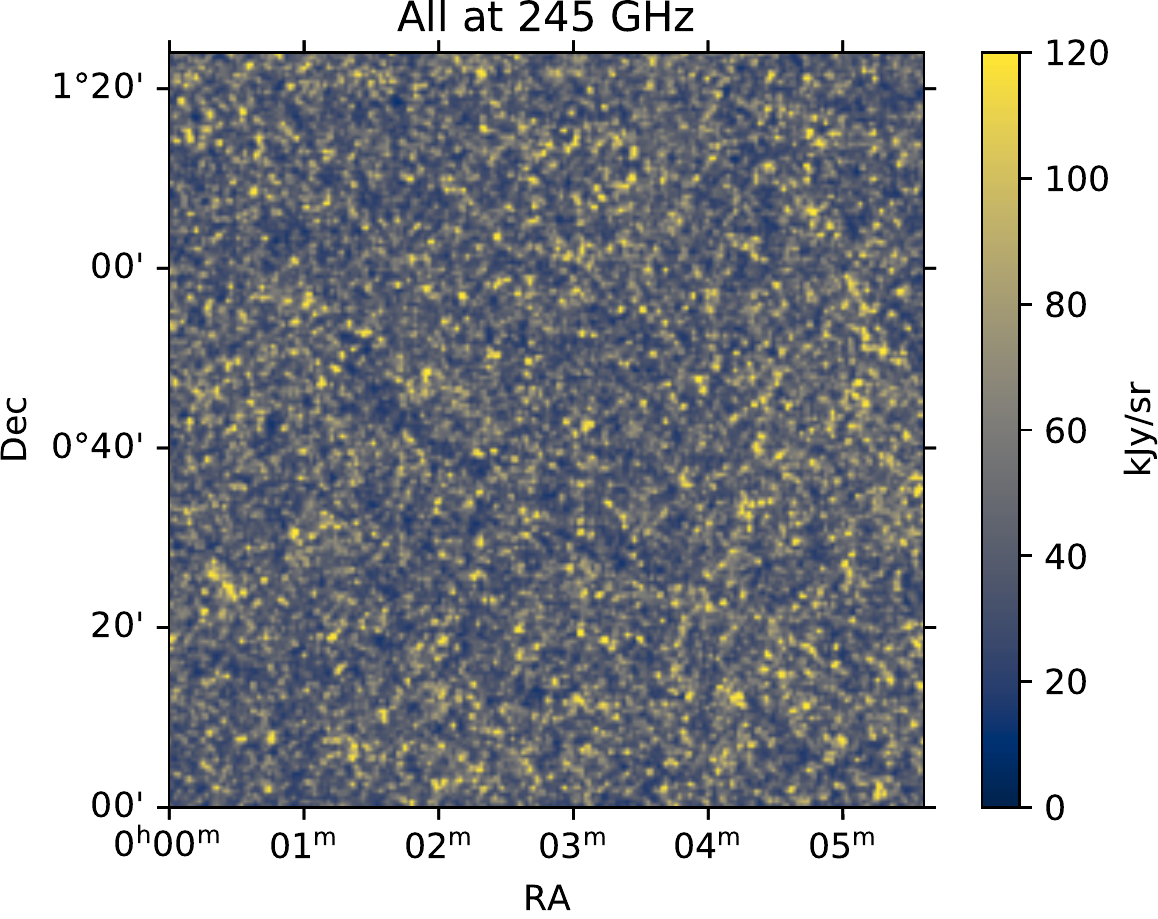} 
\end{subfigure}
\hfill
\begin{subfigure}{0.24\textwidth}
\includegraphics[width=\textwidth]{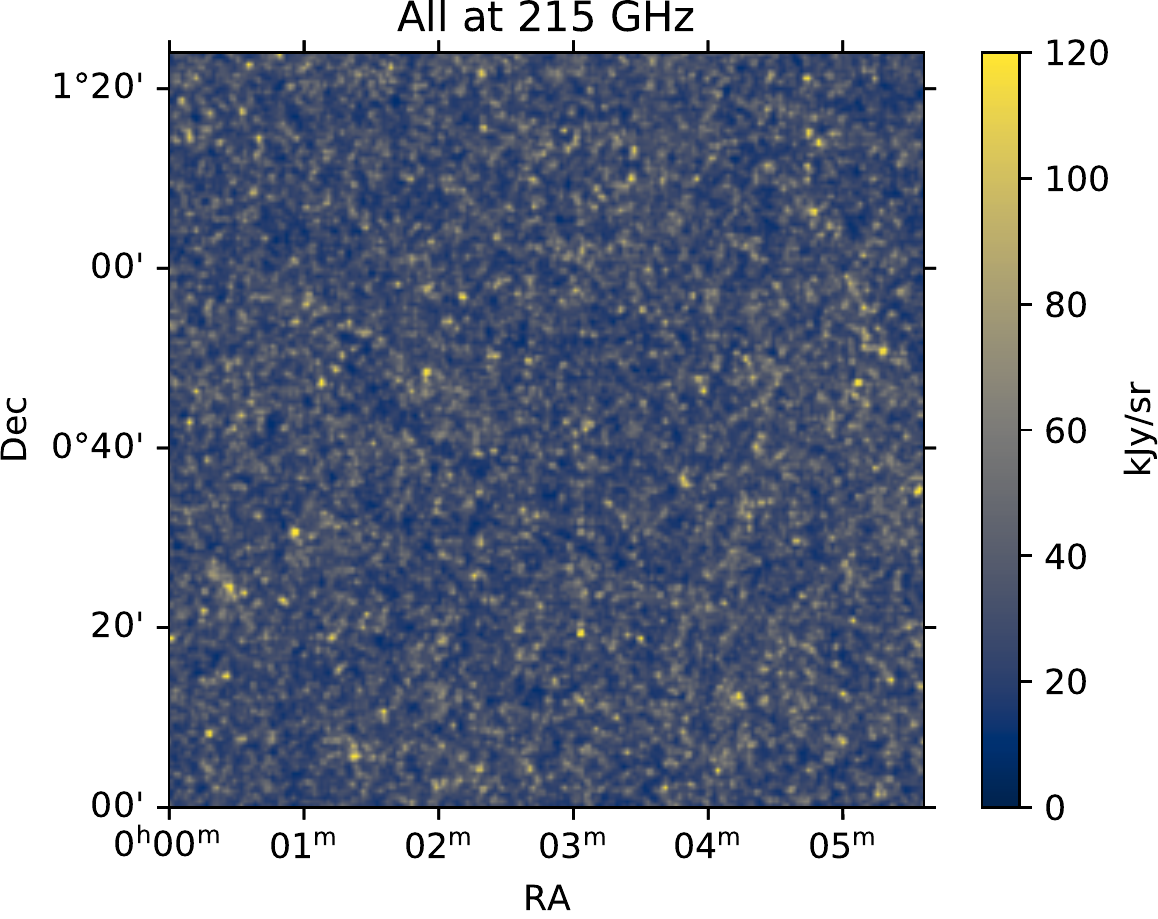}
\end{subfigure}

\vfill

\begin{subfigure}{0.24\textwidth}
\includegraphics[width=\textwidth]{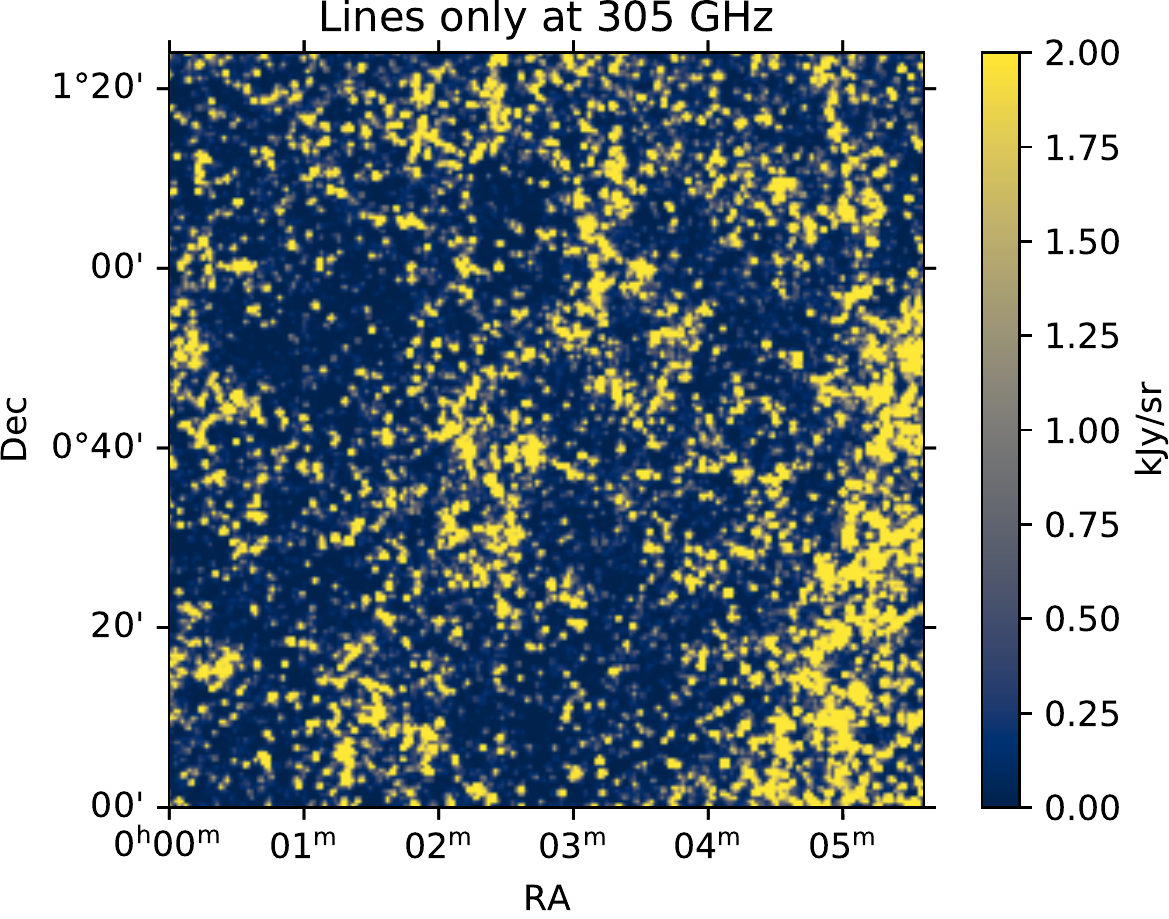}
\end{subfigure}
\hfill
\begin{subfigure}{0.24\textwidth}
\includegraphics[width=\textwidth]{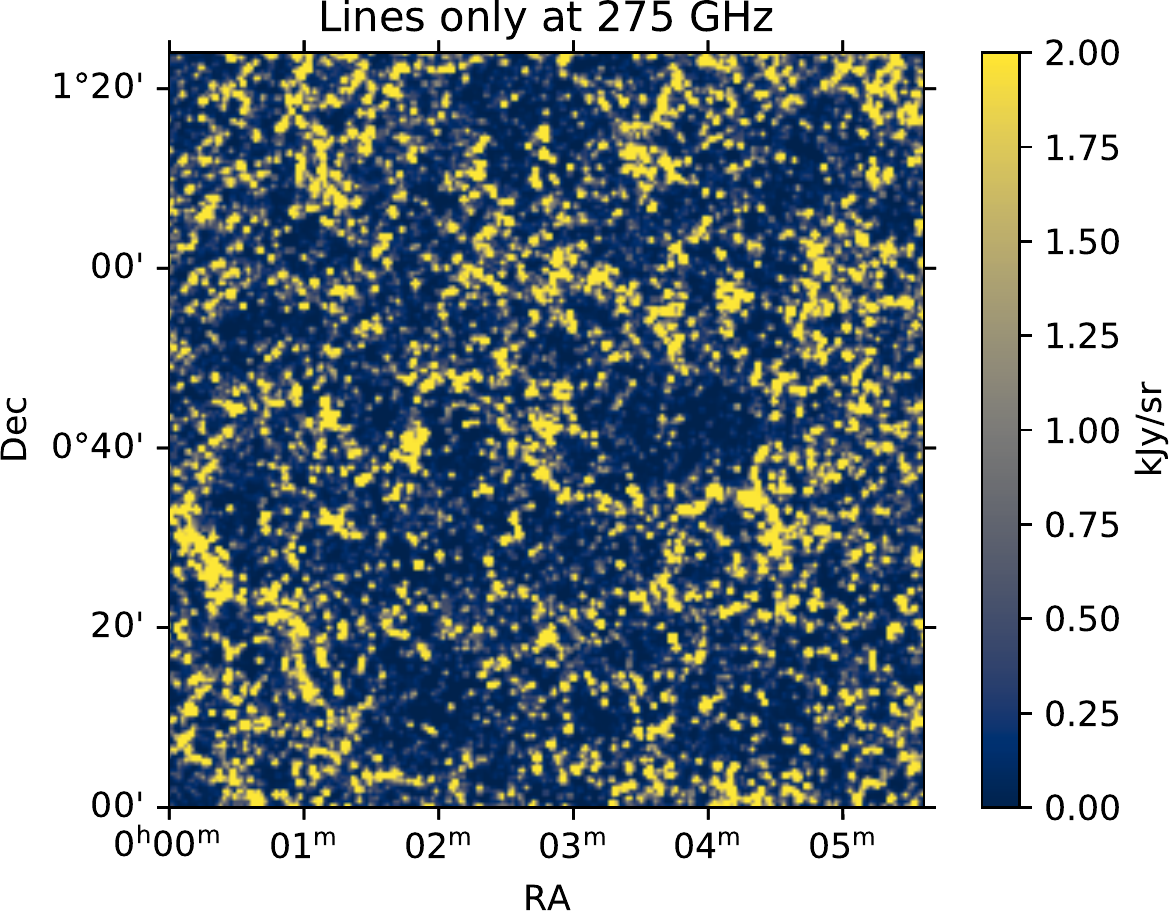}
\end{subfigure}
\hfill
\begin{subfigure}{0.24\textwidth}
\includegraphics[width=\textwidth]{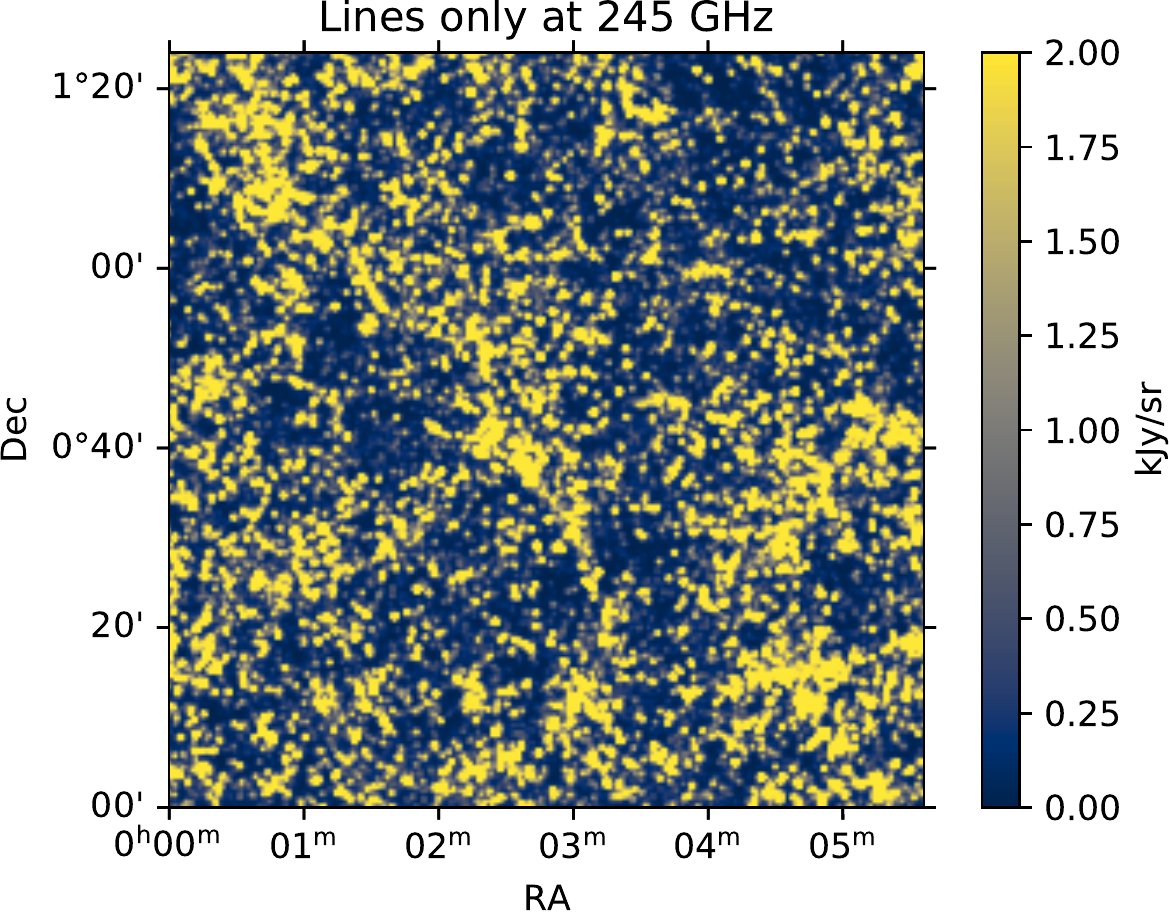} 
\end{subfigure}
\hfill
\begin{subfigure}{0.24\textwidth}
\includegraphics[width=\textwidth]{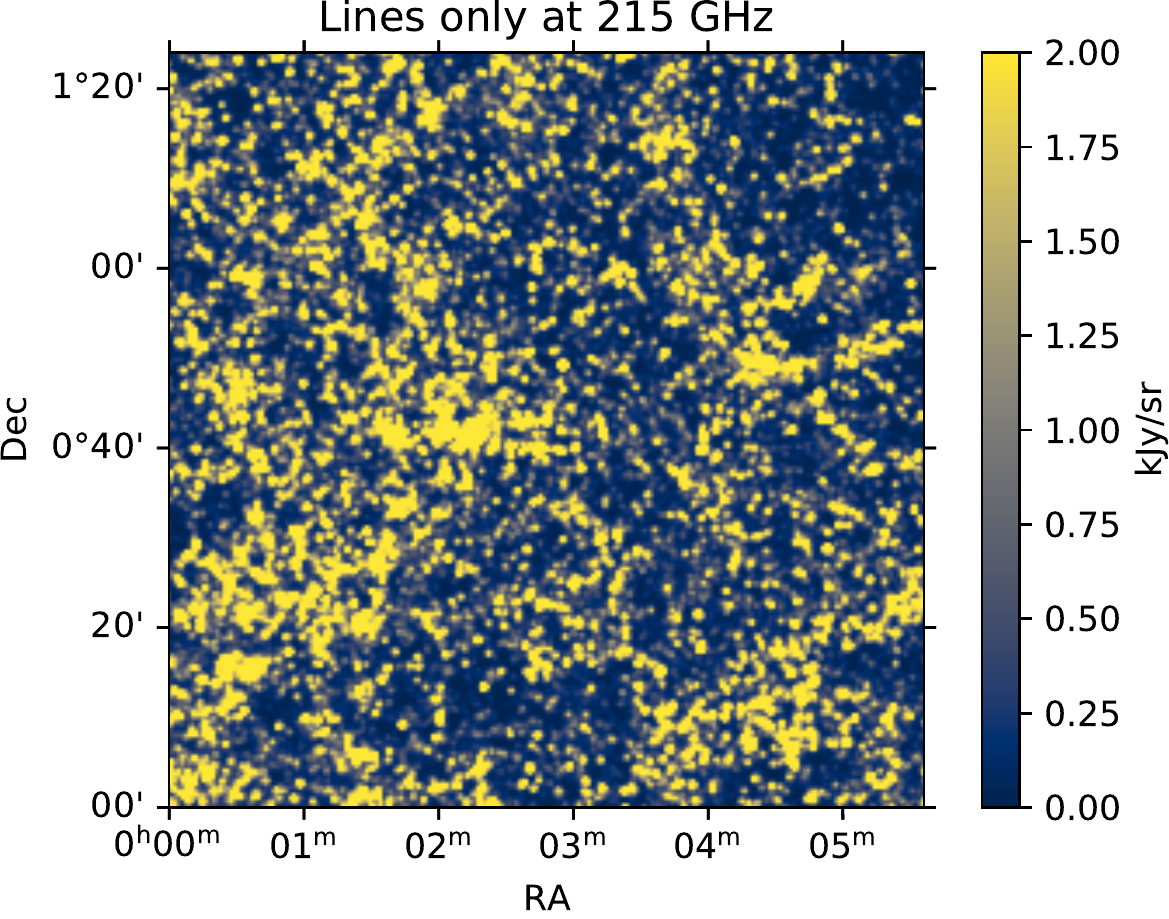}
\end{subfigure}

\vfill

\begin{subfigure}{0.24\textwidth}
\includegraphics[width=\textwidth]{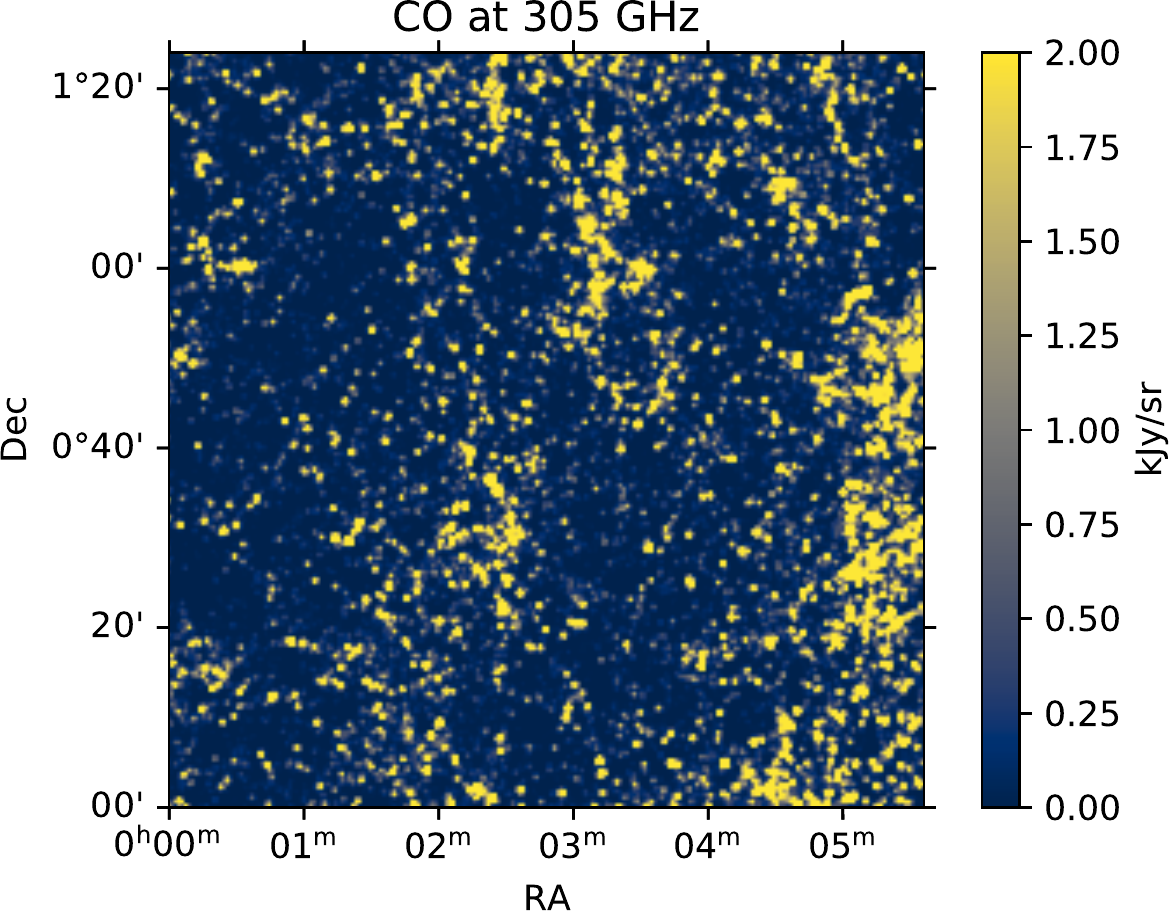}
\end{subfigure}
\hfill
\begin{subfigure}{0.24\textwidth}
\includegraphics[width=\textwidth]{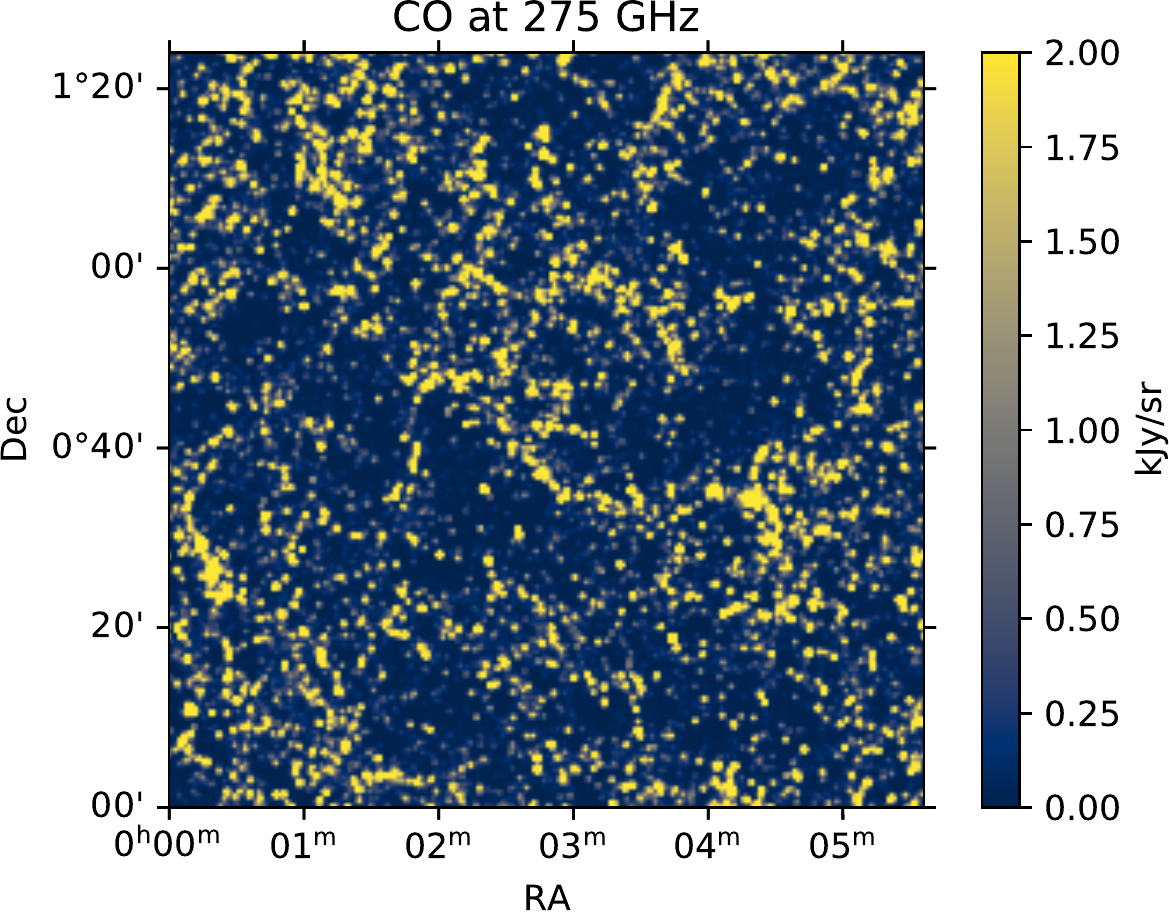}
\end{subfigure}
\hfill
\begin{subfigure}{0.24\textwidth}
\includegraphics[width=\textwidth]{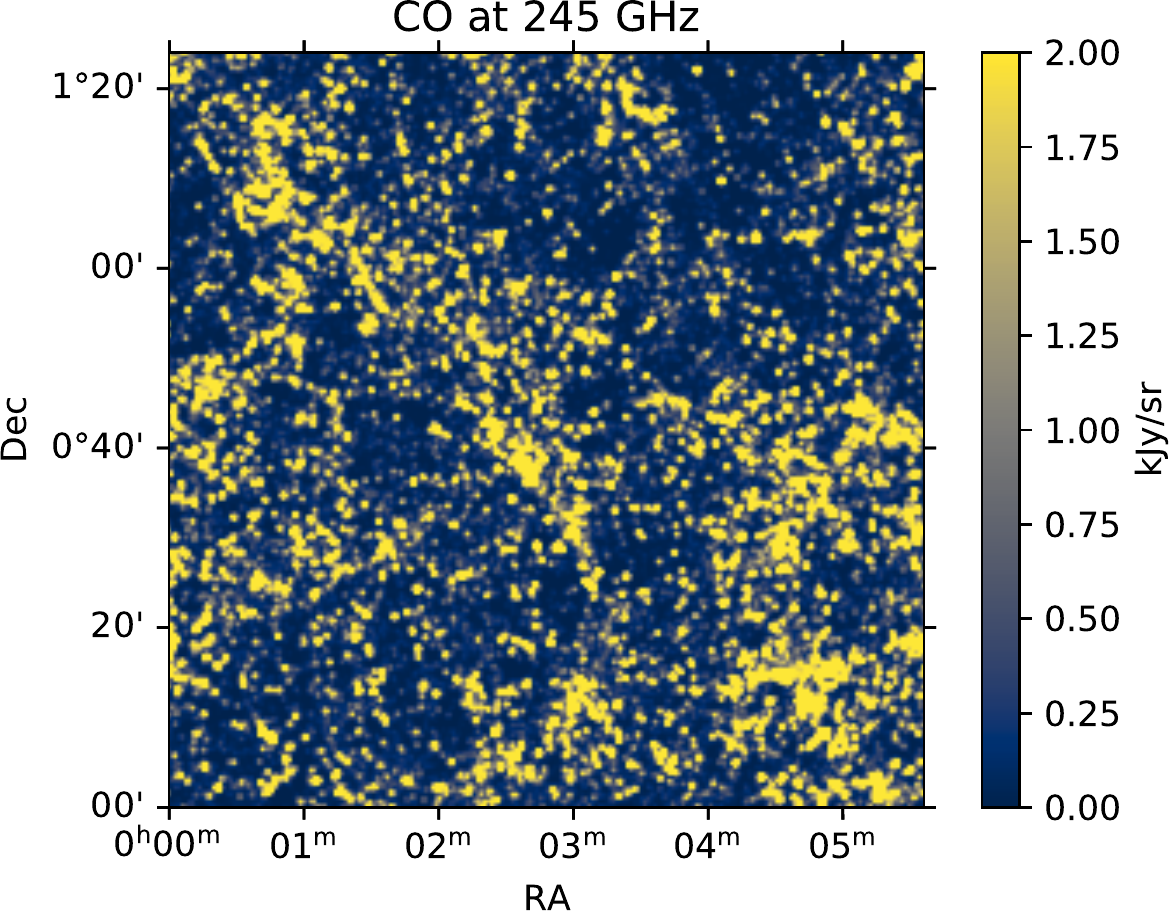} 
\end{subfigure}
\hfill
\begin{subfigure}{0.24\textwidth}
\includegraphics[width=\textwidth]{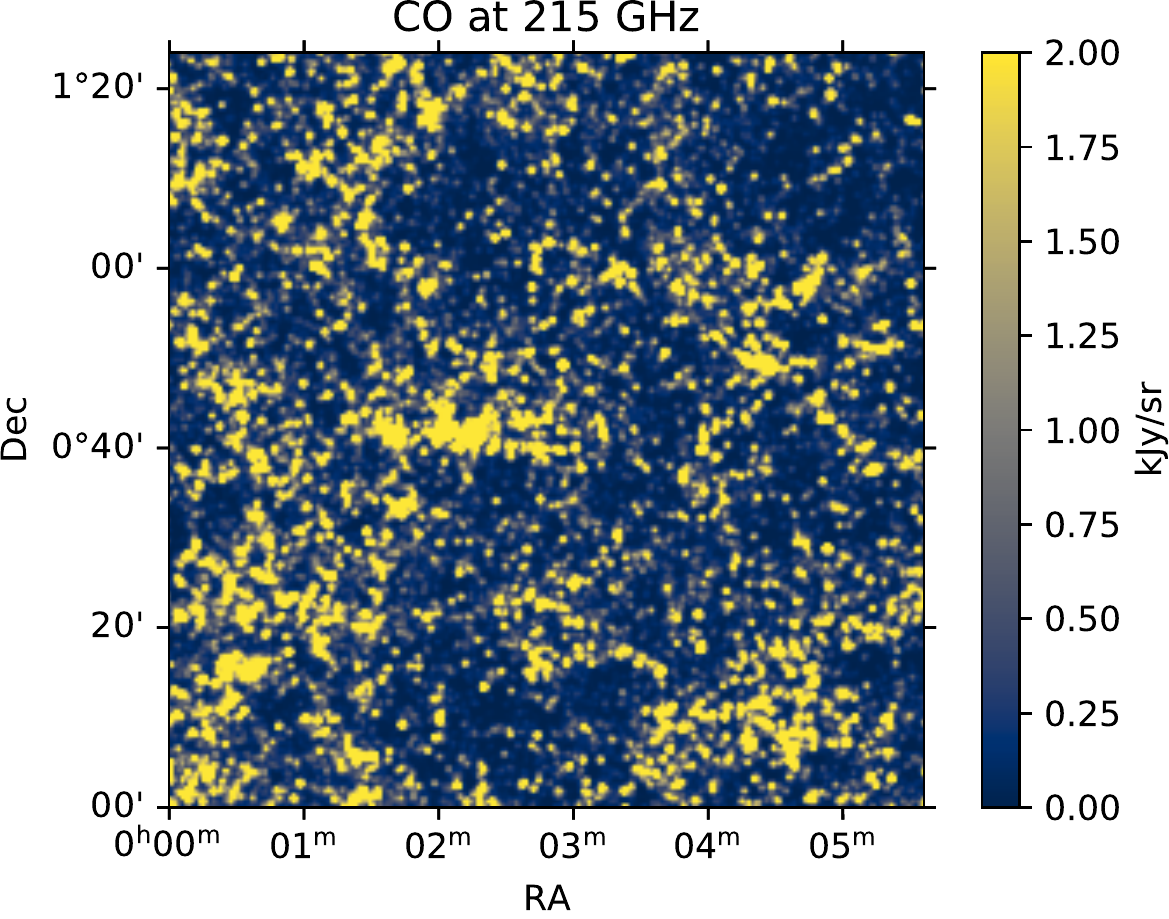}
\end{subfigure}

\vfill

\begin{subfigure}{0.24\textwidth}
\includegraphics[width=\textwidth]{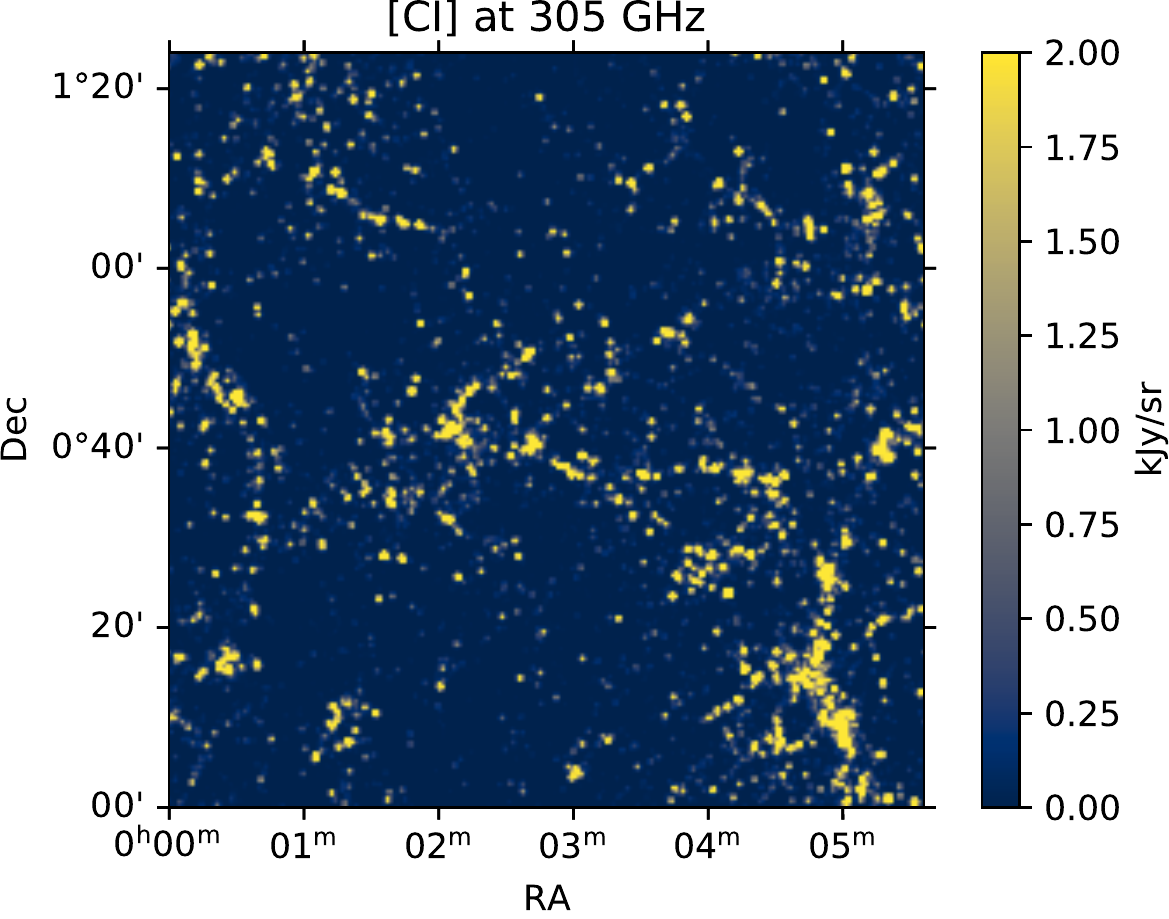}
\end{subfigure}
\hfill
\begin{subfigure}{0.24\textwidth}
\includegraphics[width=\textwidth]{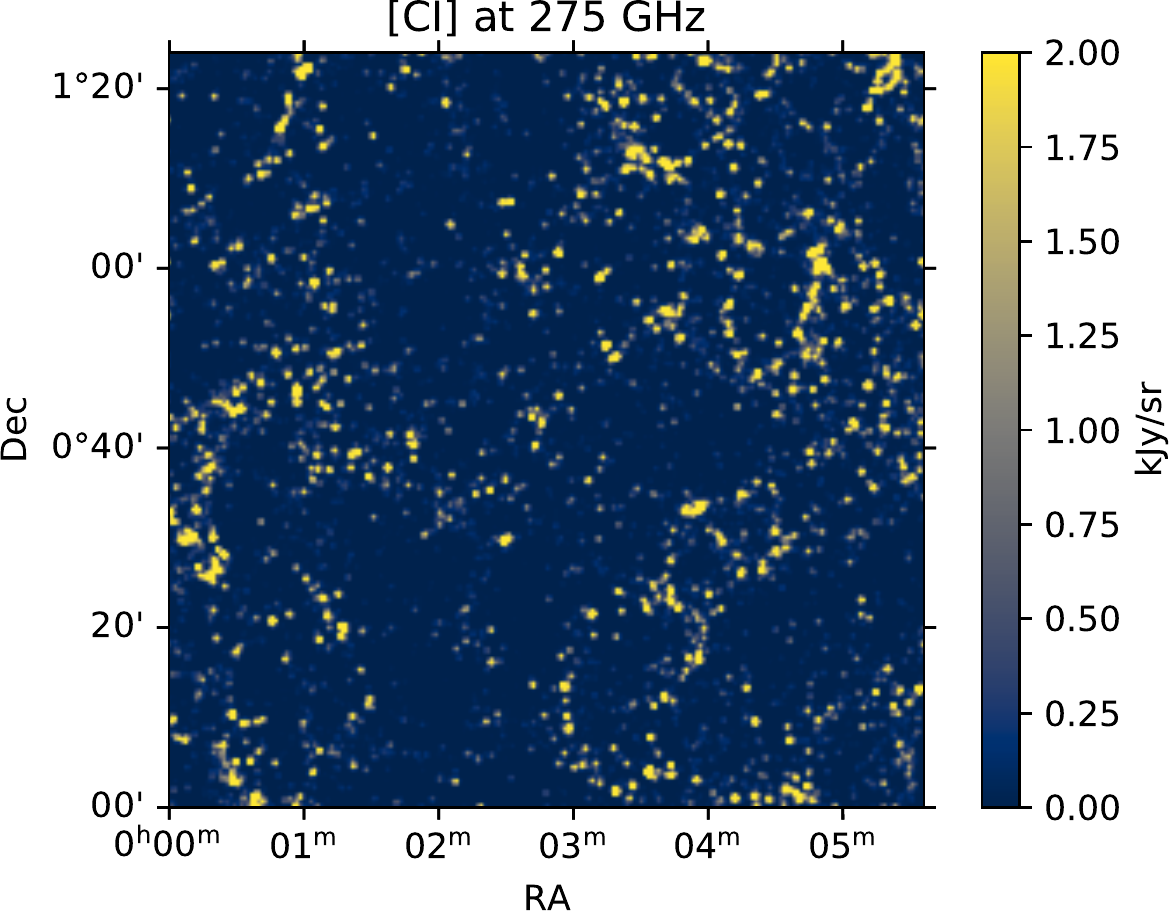}
\end{subfigure}
\hfill
\begin{subfigure}{0.24\textwidth}
\includegraphics[width=\textwidth]{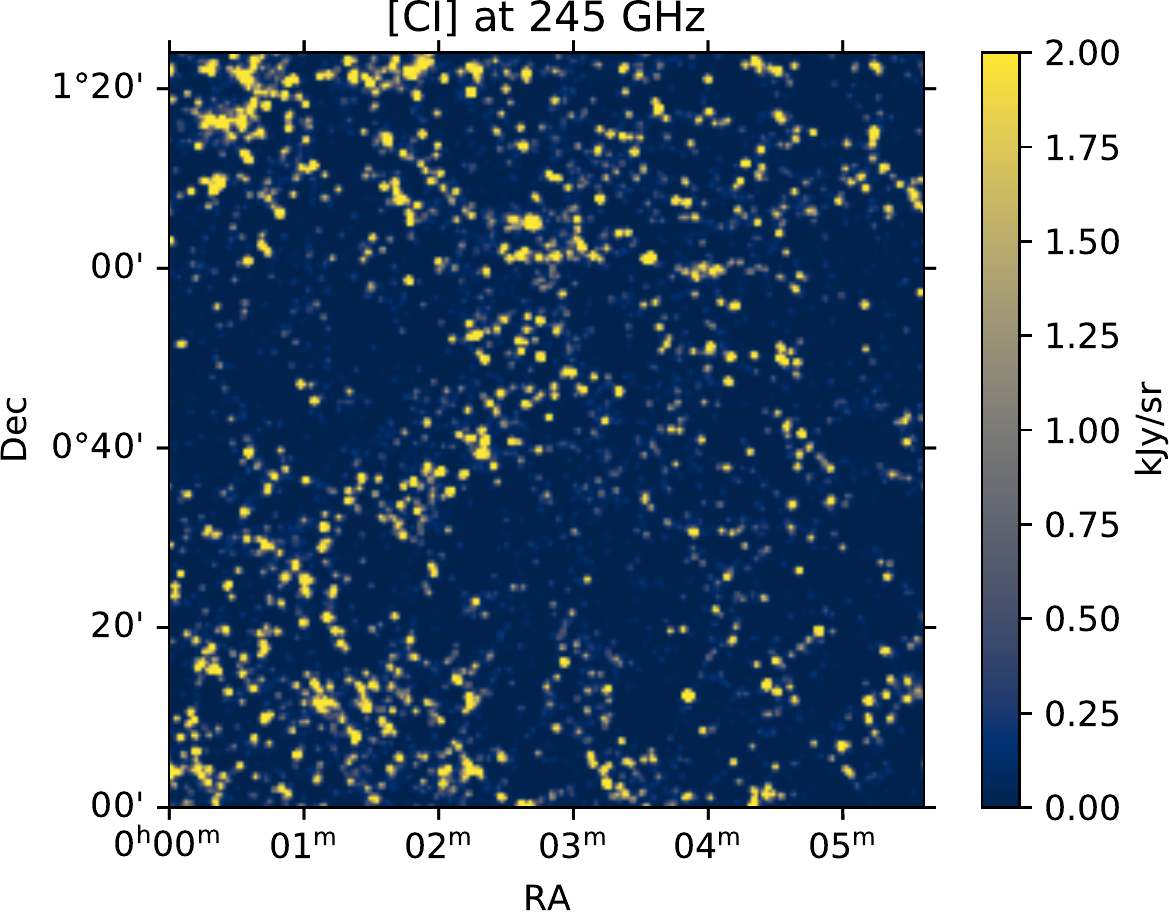} 
\end{subfigure}
\hfill
\begin{subfigure}{0.24\textwidth}
\includegraphics[width=\textwidth]{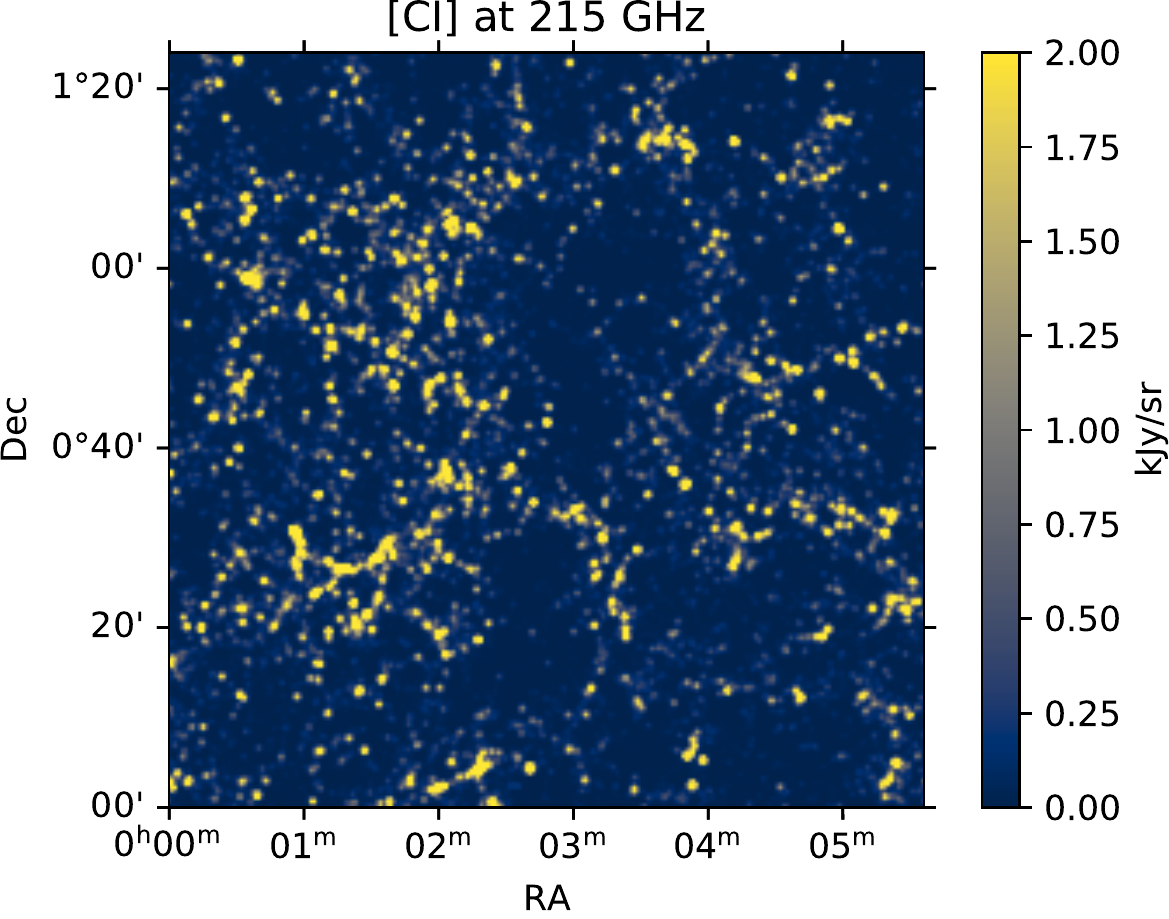}
\end{subfigure}

\vfill

\begin{subfigure}{0.24\textwidth}
\includegraphics[width=\textwidth]{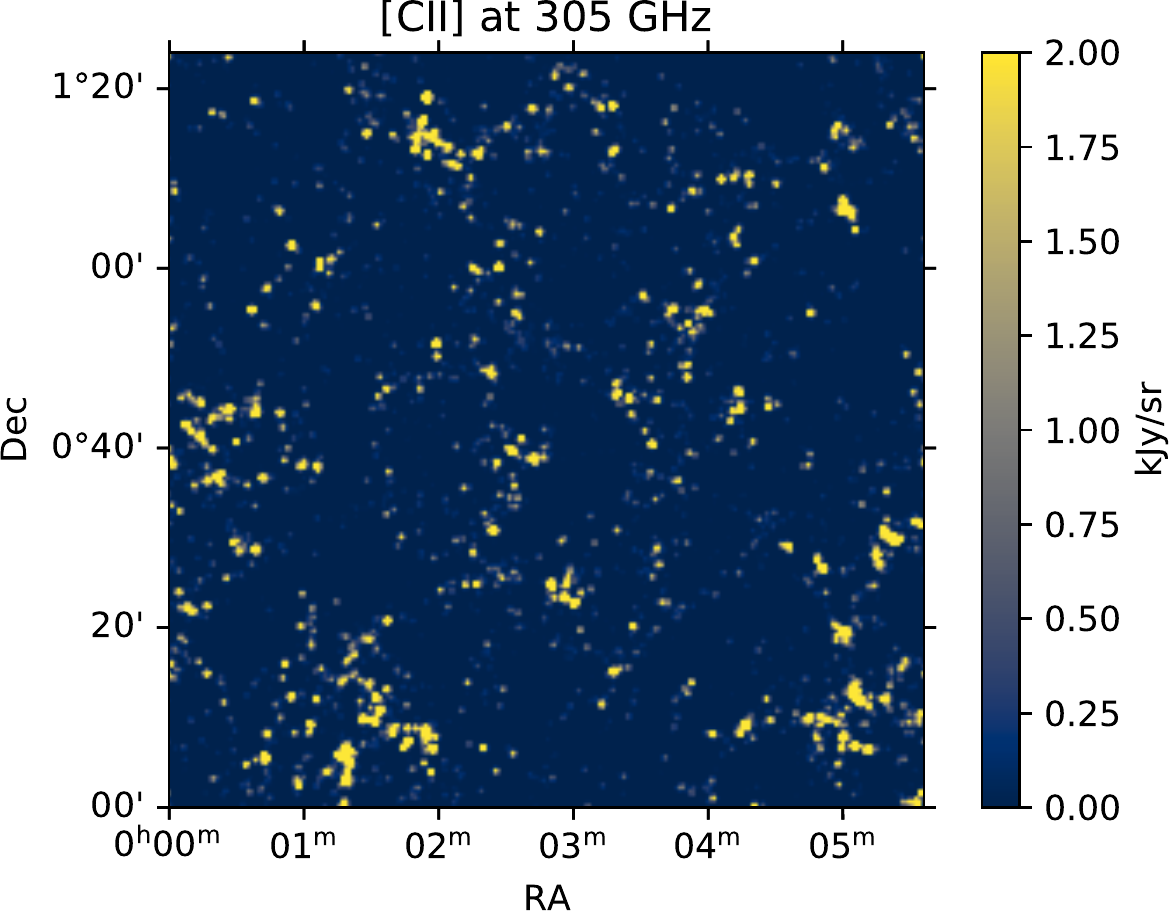}
\end{subfigure}
\hfill
\begin{subfigure}{0.24\textwidth}
\includegraphics[width=\textwidth]{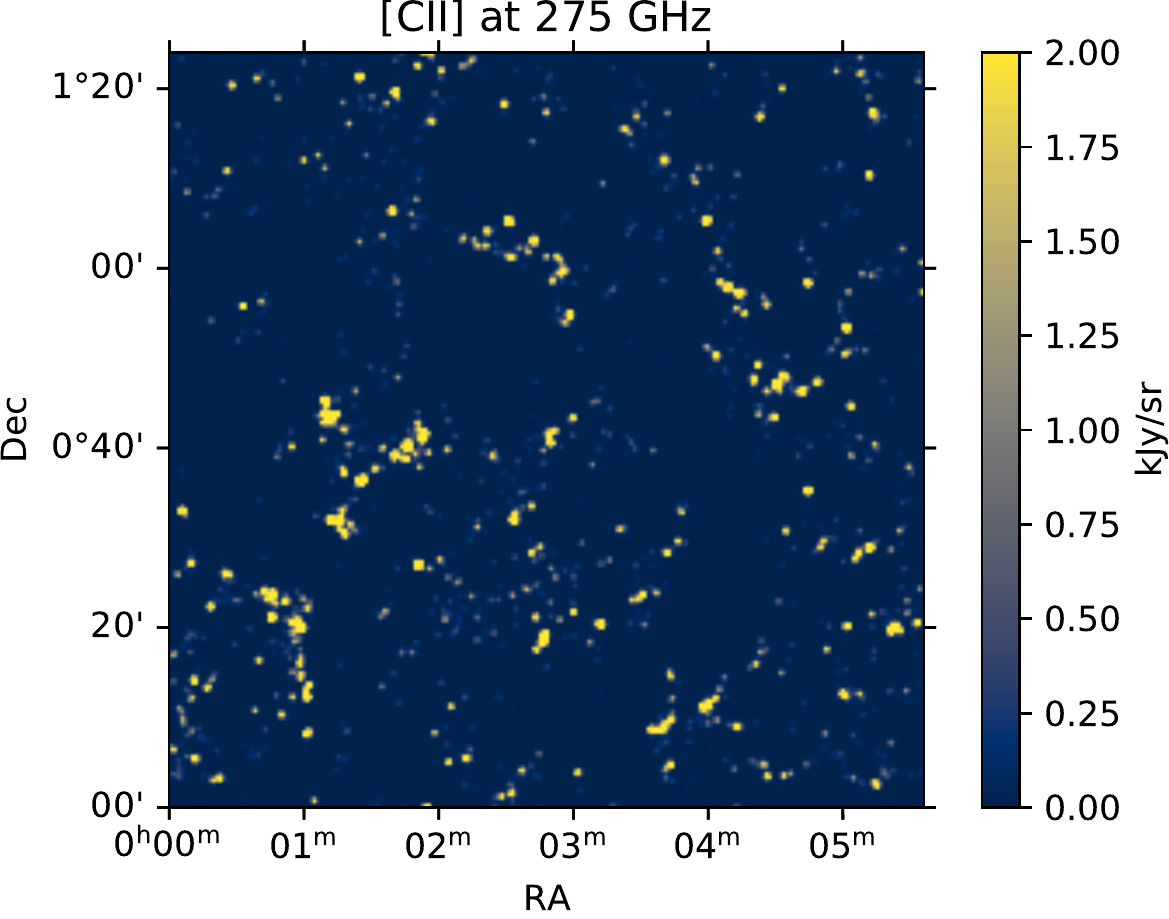}
\end{subfigure}
\hfill
\begin{subfigure}{0.24\textwidth}
\includegraphics[width=\textwidth]{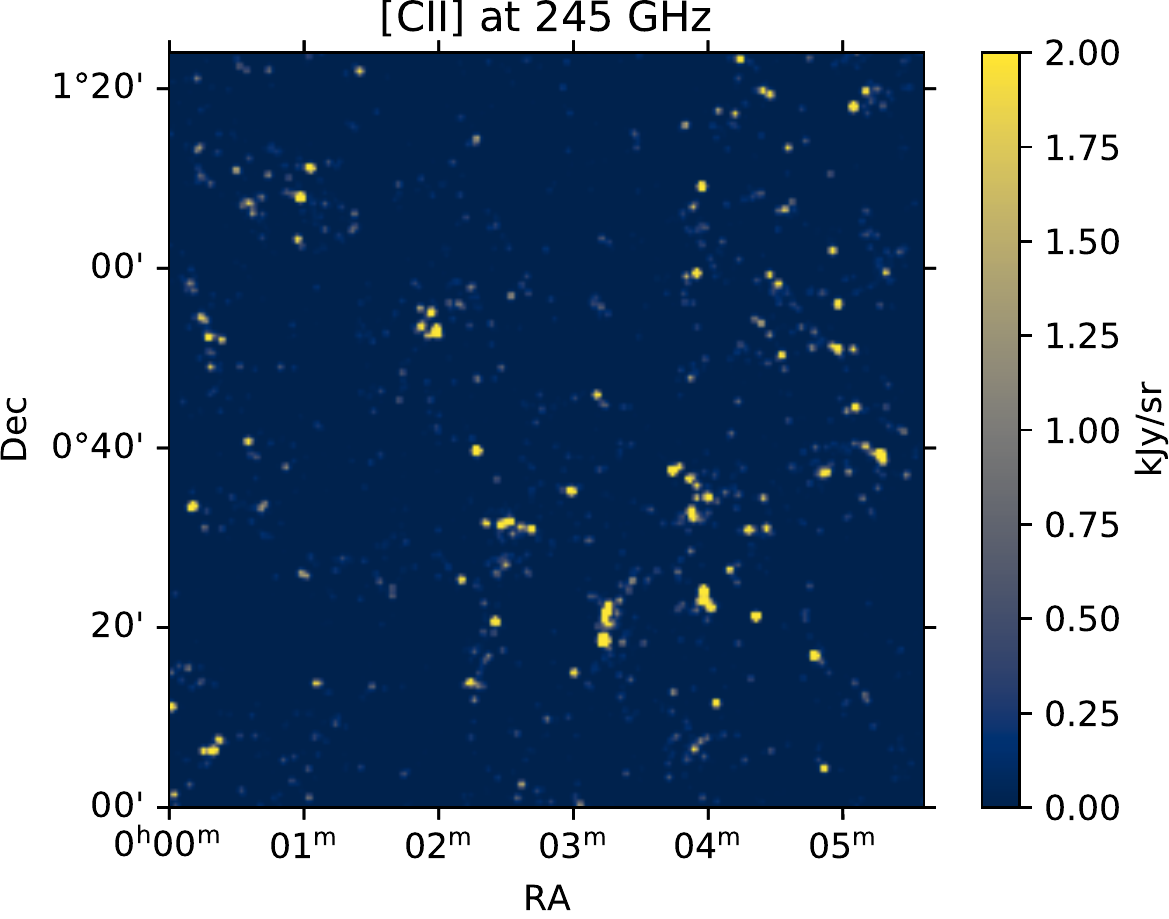} 
\end{subfigure}
\hfill
\begin{subfigure}{0.24\textwidth}
\includegraphics[width=\textwidth]{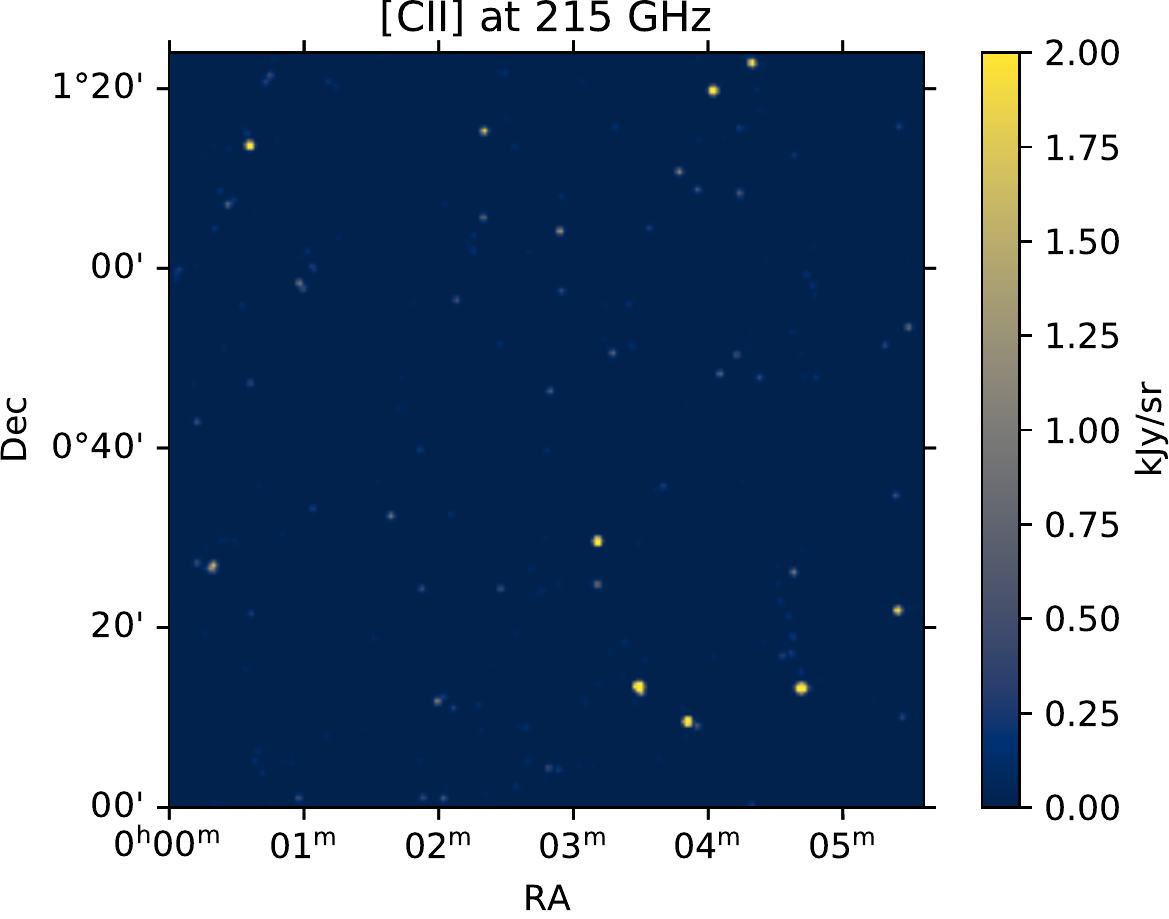}
\end{subfigure}

\caption{\label{fig:slices} Slices of simulated CONCERTO data cubes. These cubes are produced using a Gaussian beam and have no instrumental noise. Each slice has a 1\,GHz width. The four columns from the left to the right correspond to 305\,GHz ($z_{\rm [CII]}$ = 5.23, $\Delta z_{\rm [CII]}$ = 0.020), 275\,GHz ($z_{\rm [CII]}$ = 5.91, $\Delta z_{\rm [CII]}$ = 0.025), 245\,GHz ($z_{\rm [CII]}$ = 6.75, $\Delta z_{\rm [CII]}$ = 0.032), and 215\,GHz ($z_{\rm [CII]}$ = 7.83, $\Delta z_{\rm [CII]}$ = 0.041). The five rows from top to bottom show the cubes containing all the components including the continuum, all the lines, the CO lines only,  the [CI] lines, and [CII] line. To make comparisons easier, the color scale of the four panels with continuum (top row) is the same. All the other panels share another common color scale. We assume the DL14 [CII]-SFR relation in this figure.}
\end{figure*}

\subsection{Data cubes}

\label{sect:data_cubes}
From the line and continuum properties produced in the simulated catalog, we built simulated CONCERTO cubes. The simulated cubes covers the frequency range between 125\,GHz and 305\,GHz. The width of the spectral elements is fixed to 1\,GHz over the entire bandpass. We set the cube pixel size to 5\,arcsec to properly sample the beam. The pySIDES cube generator can be easily adapted to produce simulated observations for other instruments from the a few tens of GHz to the THz.

We first produced the cubes associated to each line. These first cubes are not smoothed by the instrumental beam. They are used in this paper to derive the intrinsic power spectra of the simulation (see Sect.\,\ref{sect:power_spectra}). We first compute the flux density associated to all the sources in a given voxel. We neglect the width of the line and place the entire flux of a line in the spectral element, where its central observed frequency $\nu_{\rm rest}^{\rm line}/(1+z)$ is located. The surface brightness density of a voxel B$_\nu^{\rm voxel}$ expressed in Jy/sr is then:
\begin{equation}
\label{eq:cubes}
B_\nu^{\rm voxel} = \frac{1}{\Delta \upsilon_{\rm element} \, \Omega_{\rm pixel}} \sum_{k=1}^{N_{\rm sources}} I_k^{\rm line} = \frac{\nu_{\rm element}}{c \, \Delta \nu_{\rm element} \, \Omega_{\rm pixel}} \sum_{k=1}^{N_{\rm sources}} I_k^{\rm line},
\end{equation} 
where $c$ is the speed of light and $\Omega_{\rm pixel}$ is the solid angle of a pixel in steradians. $\Delta \upsilon_{\rm element}$, $\nu_{\rm element}$, and $\Delta \nu_{\rm element}$ are the velocity width, the central frequency, and the frequency width of the voxel, respectively\footnote{The conversion between the velocity width, the frequency width, and the redshift width (used hereafter) are obtained in the following way: $\delta \upsilon = c \frac{\delta \nu_{\rm obs}}{\nu_{\rm obs}} = c \frac{\nu_{\rm rest} \delta z}{(1+z)^2 \nu_{\rm obs}} = c \frac{\delta z}{(1+z)}$.}. $I_k^{\rm line}$ is the flux in Jy\,km/s of the k-th source in the voxel. We remark that the $1/\Delta \nu_{\rm element}$ will be compensated by the number of sources in the voxel being proportional to $\Delta \nu_{\rm element}$. In practice, we first convert the spatial and spectral sky coordinates of the lines into cube coordinates using the astropy \citep{astropy:2013, astropy:2018} world coordinate system (WCS) package. We then use the \texttt{histogramdd} 3-dimensional histogram routine of the numpy package \citep{numpy} using the line fluxes as weights to generate the cube and normalize each of its frequency slices using Eq.\,\ref{eq:cubes}. This operation is performed for each transition of each line.

To produce the continuum cube, we computed the flux density of each galaxy at the central frequency of each spectral element using a parallelized code. For each frequency, we then produce a map using the \texttt{histogram2d} histogram routine applying the flux density at this frequency as weight. The maps corresponding to all the frequency slices are then stacked to produce the cube. The final cube is created by summing the line cubes and the continuum cube.

We then produced cubes including the instrumental beam by convolving them by the beam. Each frequency slice is treated separately, since the beam size varies with frequency. For simplicity, we assumed Gaussian beams with a full width at half maximum (FWHM) of 1.22\,$\lambda/D$, where $\lambda$ the wavelength and the $D$ the diameter of the telescope. Since CONCERTO is installed on APEX, we use $D$=12\,m in this paper.

In Fig.\,\ref{fig:slices}, we show various frequency slices of the cubes. In the total cube including both the lines and the continuum (top panels), the large-scale structures are not obviously visible even if a trained eye can see that the source distribution is not Poissonian. We also remark that the slices at various frequencies are remarkably similar. It is not surprising, since these maps are dominated by the continuum and for most of the sources at these frequencies are in the Rayleigh-Jean regime. 

In the cube containing only the lines (second row), we start to see the filamentary structures, while they appear more clearly in the cubes containing a single species (third to fifth rows). The CO is widely distributed over the map and the [CII] emission is located in a couple of dense regions. It is expected, since star formation at high-redshift where [CII] is emitted is more clustered than at lower redshift where the CO comes from \citep[e.g.,][]{Bethermin2013,Maniyar2018}. We can also note that there is much less [CII] emission at lower frequency, while CO is stronger. The power spectrum from each component and the implication for CONCERTO will be discussed in details in Sect.\,\ref{sect:power_spectra}. 

\begin{figure}
\centering
\includegraphics[width=9cm]{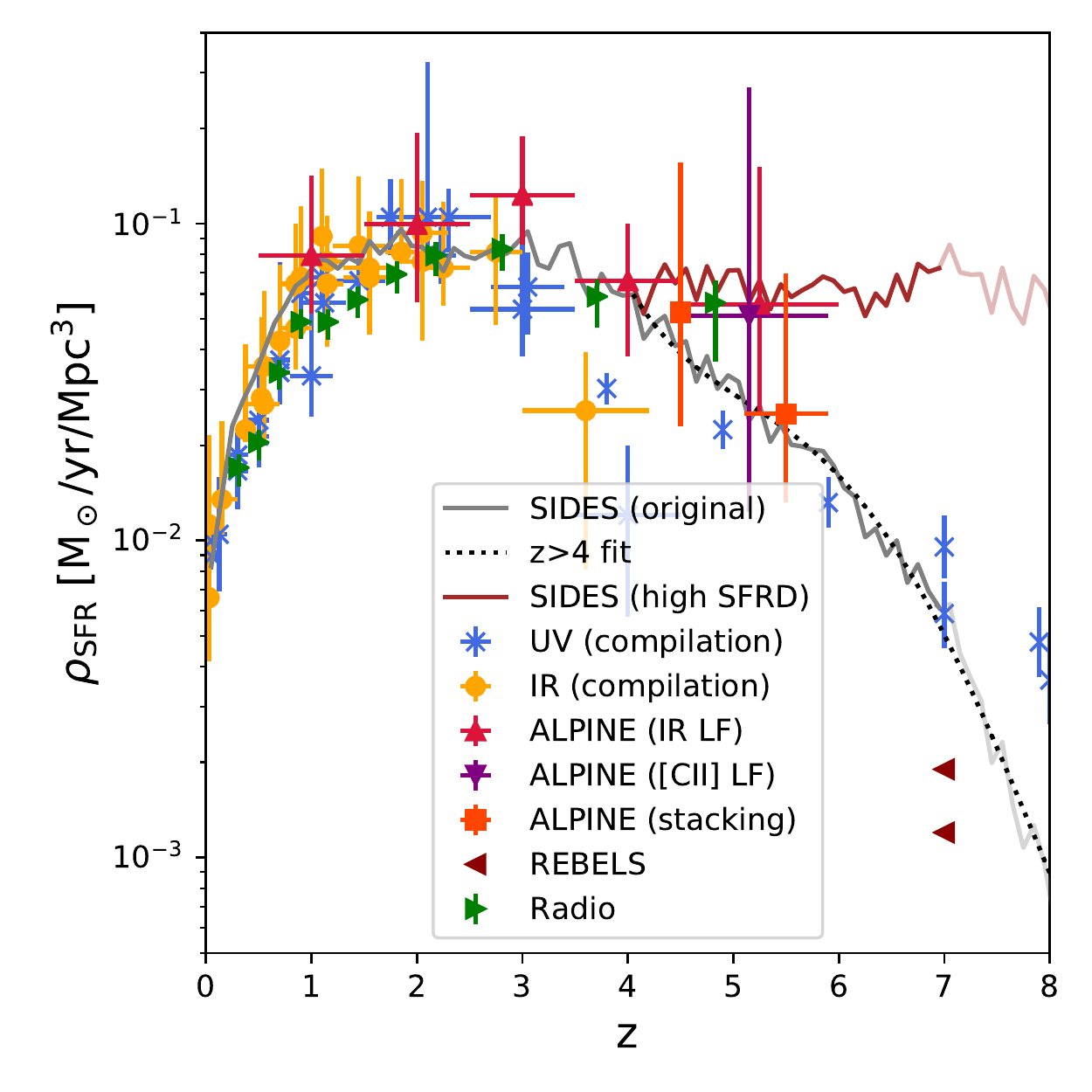}
\caption{\label{fig:SFRD} Evolution of the star formation rate density. The grey solid line is derived from the SIDES simulation and the brown solid line is an alternative model at z$>4$ for which we boosted the SFRs to obtain a flat SFRD at high-redshift ("high SFRD" model, see Sect.\,\ref{sect:SFRD}). The SFR boosting factor to obtain a flat SFRD is computed from the fit of the z$>$4 original model (dotted line). The compilations of UV-derived and infrared-derived measurements by \citet{Madau2014} are represented by blue crosses and orange filled circles, respectively. We also show the measurement from the ALPINE survey based on the luminosity function of serendipitous continuum detections (\citealt{Gruppioni2020}, red upwards triangles), the [CII] luminosity function of serendipitous detection (\citealt{Loiacono2021}, purple downwards triangles), and an indirect method using the stacking of the target sample (\citealt{Khusanova2021}, dark orange squares). The dark red left-facing triangles are the estimates of the obscured SFRD from REBELS by \citet{Fudamoto2021} using the continuum (low value) and the [CII] (high value). The constraints from the radio from \citet{Novak2017} are shown using green right-facing triangles. As discussed in Sect.\,\ref{sect:discussion_models}, the SFRD at z$>$7 could be underestimated by our simulation because of its halo mass limit and a paler color is thus used.}
\end{figure}

\subsection{Alternative model with a higher star formation rate density at z$>$4}

\label{sect:SFRD}

\textit{Hubble} space telescope (HST) deep surveys provided constraints on the z$>$4 SFR density (SFRD) \citep[e.g.,][]{Bouwens2007, Bouwens2012,Schenker2013} using dust-corrected UV data. They found that the SFRD decreases with increasing redshift at z$>$3. This is compatible with the predictions of the latest semi-analytical models \citep[e.g.,][]{Lagos2020} and hydrodynamical simulations \citep[e.g.,][]{Pillepich2018}. However, the discovery of dusty galaxy population without counterparts seen by the \textit{Hubble} space telescope \citep[e.g.,][]{WangT2019,Talia2021,Fudamoto2021} showed how important long-wavelength data are to obtain the full picture of the star formation. Based on a combination of aggressively deblended \textit{Herschel} data and modeling, \citet{Rowan-Robinson2016} claimed that SFRD is flat even above z$=$3 (see also the discussion in \citealt{Casey2018}), but another analysis using a similar approach found results more compatible with the UV-corrected estimates \citep{WangL2019}.

In Fig.\,\ref{fig:SFRD}, we compare the SFRD from our simulation (grey solid line) with the various observations. Below z$=$4, our  simulation is inside the cloud of observational points. Up to z$\sim$7, our model is compatible with the dust-corrected UV measurements. In contrast, the SFRD from SIDES is systematically lower than the recent constraints from ALPINE \citet{Gruppioni2020,Loiacono2021,Khusanova2021} and the radio \citep{Novak2017}, but it remains compatible at the 1.5\,$\sigma$ level. At z$\sim$7, the SFRD in SIDES is a factor of $\sim$2 above the estimate of the obscured SFRD from REBELS based on their two serendipitous detections \citep{Fudamoto2021}. Considering that they applied a factor of $\sim$4 correction for the clustering and that there are only two objects, this factor of 2 difference cannot be considered as significant.

To study the impact of an extreme scenario with a flat SFRD at z$>$4, we produced a version of the SIDES model with a flat SFRD by multiplying all the SFRs in the SIDES simulation by a factor $C_{\rm SFR}$ varying with redshift. To compute this factor, we fitted the decimal logarithm of the SIDES SFRD versus z by a fourth order polynomial considering only the z$>$4 points (dotted line in Fig.\,\ref{fig:SFRD}) and obtained the following correction:
\begin{equation}
C_{\rm SFR}(z) =  \begin{cases} 1 & \textrm{if z $\leq$ 4,}\\  10^{-0.013206 z^4 + 0.34472 z^3 - 3.2180 z^2 + 13.149 z - 19.79} & \textrm{else.} \end{cases}
\end{equation}
This version of the simulation is compatible with the ALPINE and radio data, but is one order of magnitude higher than the UV constraints at z$=$7. Using this model with the DL14 relation, we can thus derive an optimistic upper limit on our intensity mapping predictions called hereafter "high SFRD" model.

\begin{figure*}
\centering
\includegraphics[width=15cm]{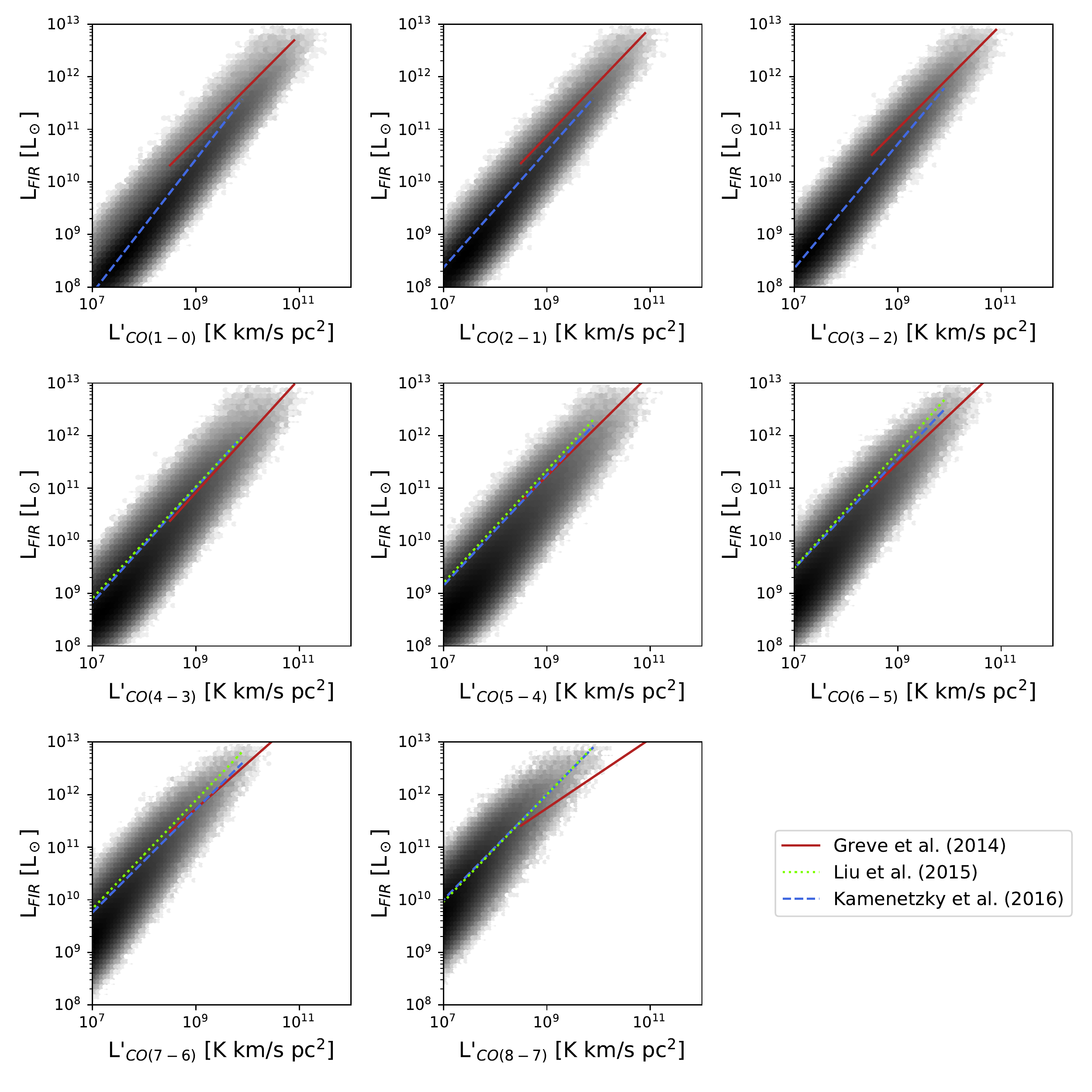}
\caption{\label{fig:CO_scale} Relation between the CO and the far-infrared luminosity integrated between 40 and 400\,$\mu$m. The number of density of SIDES galaxies is coded using a grey scale.  The best-fit observed relations of \citet[][local and high-z (U)LIRGs]{Greve2014}, \citet[][low-z sample]{Liu2015}, and \citet[][low-z sample]{Kamenetzky2016} are represented by a red solid, a green dotted, and a blue dashed line, respectively. These relations are plotted only in the luminosity range in which they were measured. Note that the \citet{Liu2015} study does not include the three lowest rotationnal transitions of CO.}
\end{figure*}

\begin{figure}
\centering
\includegraphics[width=9cm]{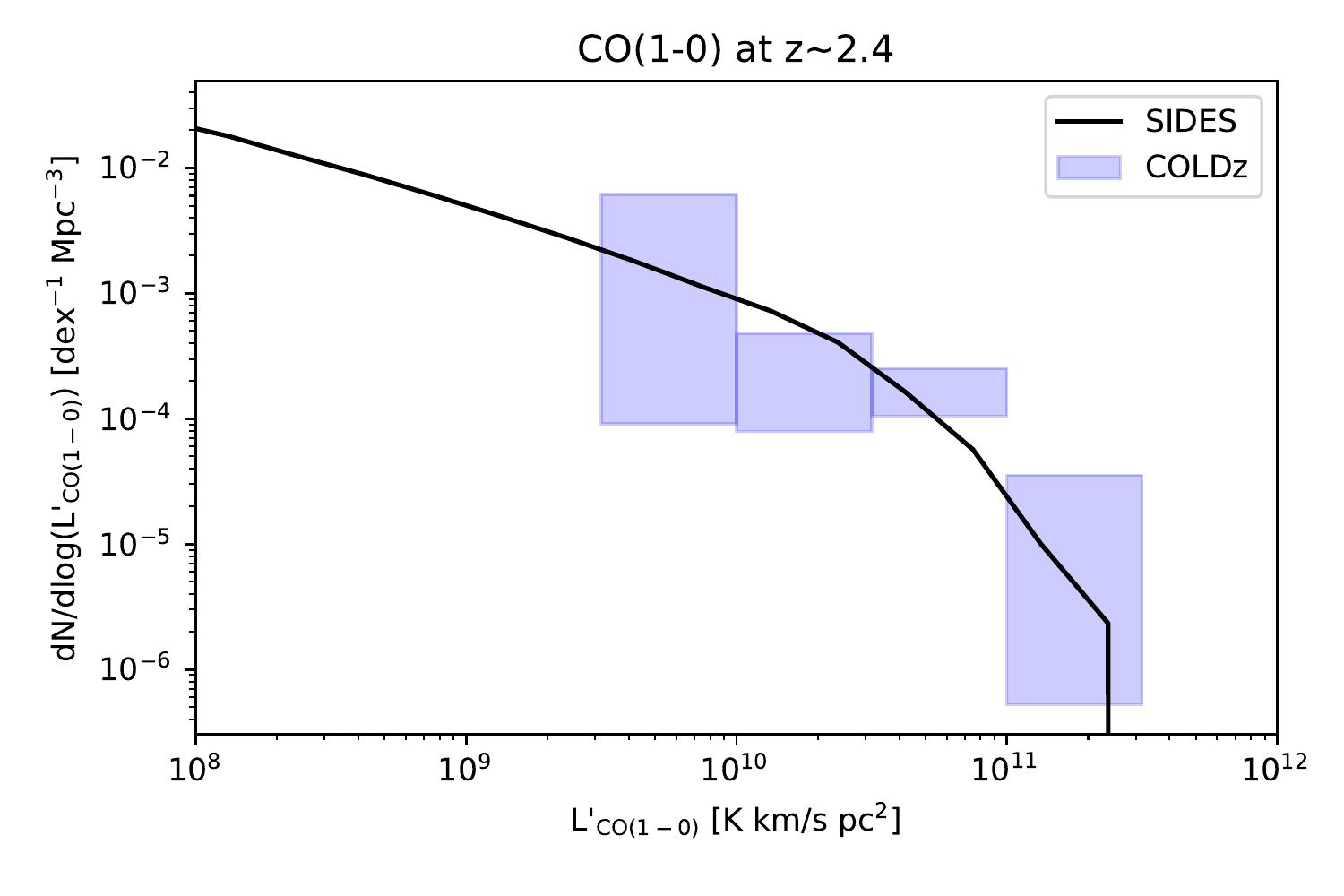}
\caption{\label{fig:CO10_LF} CO(1-0) luminosity function at z$\sim$2.4. The black solid line is computed using SIDES and the blue filled rectangles are the measurements of \citet[][51+9\,arcmin$^2$]{Riechers2019} using the JVLA COLDz survey. The width of the rectangles indicates the bin size and the height shows the 1-$\sigma$ confidence region.}
\end{figure}

\begin{figure*}
\centering
\includegraphics[width=18cm]{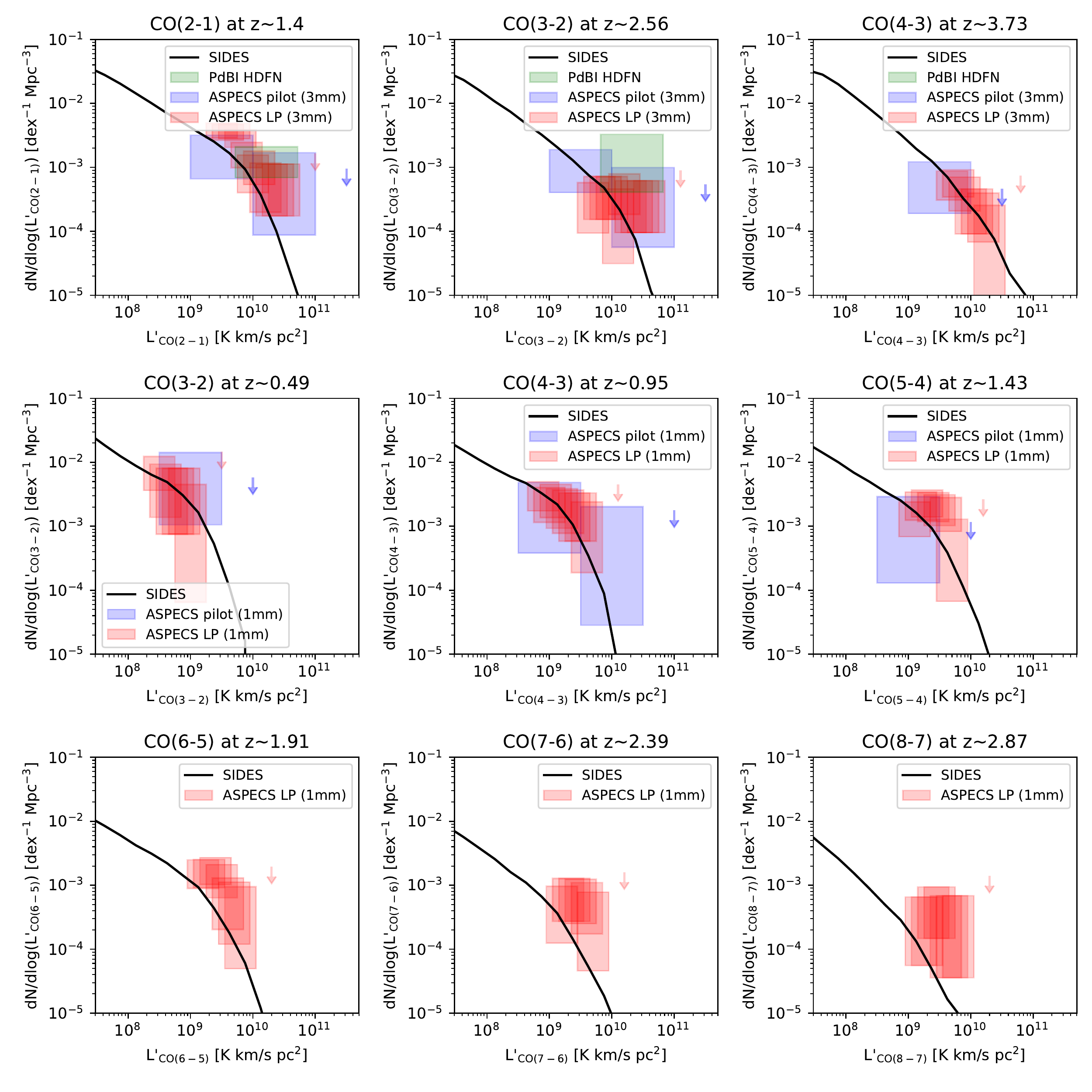}
\caption{\label{fig:CO_LF} Comparison between CO luminosity from the SIDES simulations (black solid lines) and the observations for various transitions and redshifts (see details above each panel). Observational data are represented by filled rectangles, whose width indicates the bin size and the height shows the 1-$\sigma$ confidence region. The PdBI HDFM program of \citet[1\,arcmin$^2$]{Walter2016} is in green. The ALMA/ASPECS pilot program is in blue \citep[1\,arcmin$^2$]{Decarli2016} and the large program in red \citep[4.6\,arcmin$^2$]{Decarli2019,Decarli2020}.}
\end{figure*}


\section{Comparison with observations from deep surveys}

\label{sect:LFs}

In order to validate our model, we checked if we reproduce basic statistical observables as the line versus dust luminosities (Sect.\,\ref{sect:CO_scale}), the line luminosity functions (Sect.\,\ref{sect:CO_LF}, \ref{sect:CI_LF}, and \ref{sect:CII_LF}), or the measured average line ratios (Sect.\,\ref{sect:CO_SLED_valid}).

\subsection{Relation between the CO and infrared luminosity}

\label{sect:CO_scale}

The relation between the dust luminosity and the CO luminosity of the various transitions is one of the most basic test to check the reliability of the model. In the literature, the total infrared luminosity between 8 and 1000\,$\mu$m (L$_{\rm IR}$) is rarely used and instead the authors prefer the far-infrared luminosity L$_{\rm FIR}$. Using our SED templates, we pre-computed the ratio between L$_{\rm IR}$ and L$_{\rm FIR}$ for the various values of $\langle U \rangle$. The wavelength range integrated to derive L$_{\rm FIR}$ varies marginally depending of the authors and we chose to use 40--400\,$\mu$m in this paper.

In Fig.\,\ref{fig:CO_scale}, we compare the relations found in SIDES and from the literature. We obviously expect to recover the relation between the CO(1-0) and the far-infrared luminosities, since we used it to build our model. In contrast, the other transitions are produced from SLED templates and comparing our results with the observations allow us to test our model. There is an overall excellent agreement between the results of \citet{Kamenetzky2016} and \citet{Liu2015} based on \textit{Herschel} observations of low-z galaxies. However, for the 4--3 and 5--4 transitions, the center of the SIDES relation is a factor of $\sim$2 below the observed relation at low CO luminosity (L$'_{\rm CO} < 10^8$\,K\,km/s\,pc$^2$) although still in the scatter. At these luminosities, observational constraints come only from the local Universe. If we consider only the z$<$0.2 objects in SIDES, the results agree with the observed relations (see appendix\,\ref{sect:lowz_scaling_CO}). There might thus be a small selection bias, but the nature of these data compilations do not produce a clear selection function, which can be applied to our simulation. Contrary to the two previous study, \citet{Greve2014} used mainly local (ultra-)luminous infrared galaxies ((U)LIRGs) and high-z dusty star-forming galaxies including lensed ones. The SIDES simulation overall agrees with their relation probing higher luminosities. However, they found a much flatter slope for the 8--7 transition. In this case, the difference could be due to the low number of objects used in the study of \citet{Greve2014}.\\

\subsection{CO luminosity functions}

\label{sect:CO_LF}

The CO luminosity functions, i.e. the comoving volume density of galaxies at a given redshift as a function of their CO luminosity, also provide important constraints to test simulations and have already been used extensively to test theoretical models \citep[e.g.,][]{Lagos2012,Popping2014,Vallini2016}. A new generation of deep spectral scans with the Jansky very large array (JVLA) and ALMA provided important measurements at high redshift. The CO(1-0) luminosity at z$\sim$2.4 has been measured by \citet{Riechers2019} using the JVLA COLDz survey (51 arcmin$^2$ in GOODS-N and 9 arcmin$^2$ in COSMOS). As shown in Fig.\,\ref{fig:CO10_LF}, SIDES reproduces these observations very well.

We have much richer constraints on the higher-J transitions thanks to millimeter interferometers. A pilot deep survey has been performed by \citet{Walter2014} using the Plateau de Bure interferometer. This survey covered a single pointing ($\sim$1\,arcmin diameter) in the 79--115\,GHz range. The ALMA spectroscopic survey (ASPECS) started by a first pilot survey with a similar size covering the band 3 (84--115\,GHz) and 6 (212--272\,GHz) with an improved depth \citep{Walter2016, Decarli2016}. Finally, a large program extended the coverage to a 4.6\,arcmin$^2$ region \citep{Decarli2019,Decarli2020}. In Fig.\,\ref{fig:CO_LF}, we compare the luminosity function in our simulation with the measurements from these various surveys.

The band 3 and the band 6 windows correspond to different redshift range for each line. This offers a wide variety of constraints on the various transitions and redshifts. The SIDES luminosity functions are derived from the simulated catalog using a volume corresponding to a [$z_{\rm cen}-0.1$, $z_{\rm cen}+0.1$] redshift range, where $z_{\rm cen}$ is the central redshift of the ASPECS measurement. We used the apparent luminosity in this computation for consistency, since the observations have not been corrected for lensing magnification. However, this effect is negligible in the luminosity regime probed by the observation. The SIDES simulation is always in the 1-$\sigma$ range of the observations. However, for the higher-J transitions (J$_{\rm upper} \ge 4$), the simulation tends to be systematically on the lower end of the confidence interval. It could be a field-to-field variance effect, since the field covered by the ASPECS large program was only 4.6\,arcmin$^2$. A companion paper will demonstrate that the field-to-field variance caused by large-scale structures is non negligible (Gkogkou et al. in prep.).

\begin{table}
\caption{\label{tab:stacked_CO} Comparison between the measured ALMA stacked fluxes of CO(2-1) of MUSE-selected 1$<$z$<$1.7 galaxies by \citet{Inami2020} and our results from SIDES (see a description of the computation method in Sect.\,\ref{sect:CO_SLED_valid}).}
\begin{tabular}{lcc}
\hline
\hline
Stellar mass bin & Mean SIDES flux & Measured flux\\
 & Jy\,km/s & Jy\,km/s \\
\hline
8$<$log(M$_\star$/M$_\odot$)$<$9 &  0.0047$\pm$0.0007 & $<$0.022\\
9$<$log(M$_\star$/M$_\odot$)$<$10 &  0.029$\pm$0.005 & $<$0.019\\
10$<$log(M$_\star$/M$_\odot$)$<$11 &  0.15$\pm$0.03 & 0.13$\pm$0.02\\
11$<$log(M$_\star$/M$_\odot$)$<$12 &  0.33$\pm$0.12 & 0.45$\pm$0.04\\
\hline
\end{tabular}
\end{table}

\begin{figure}
\centering
\includegraphics[width=9cm]{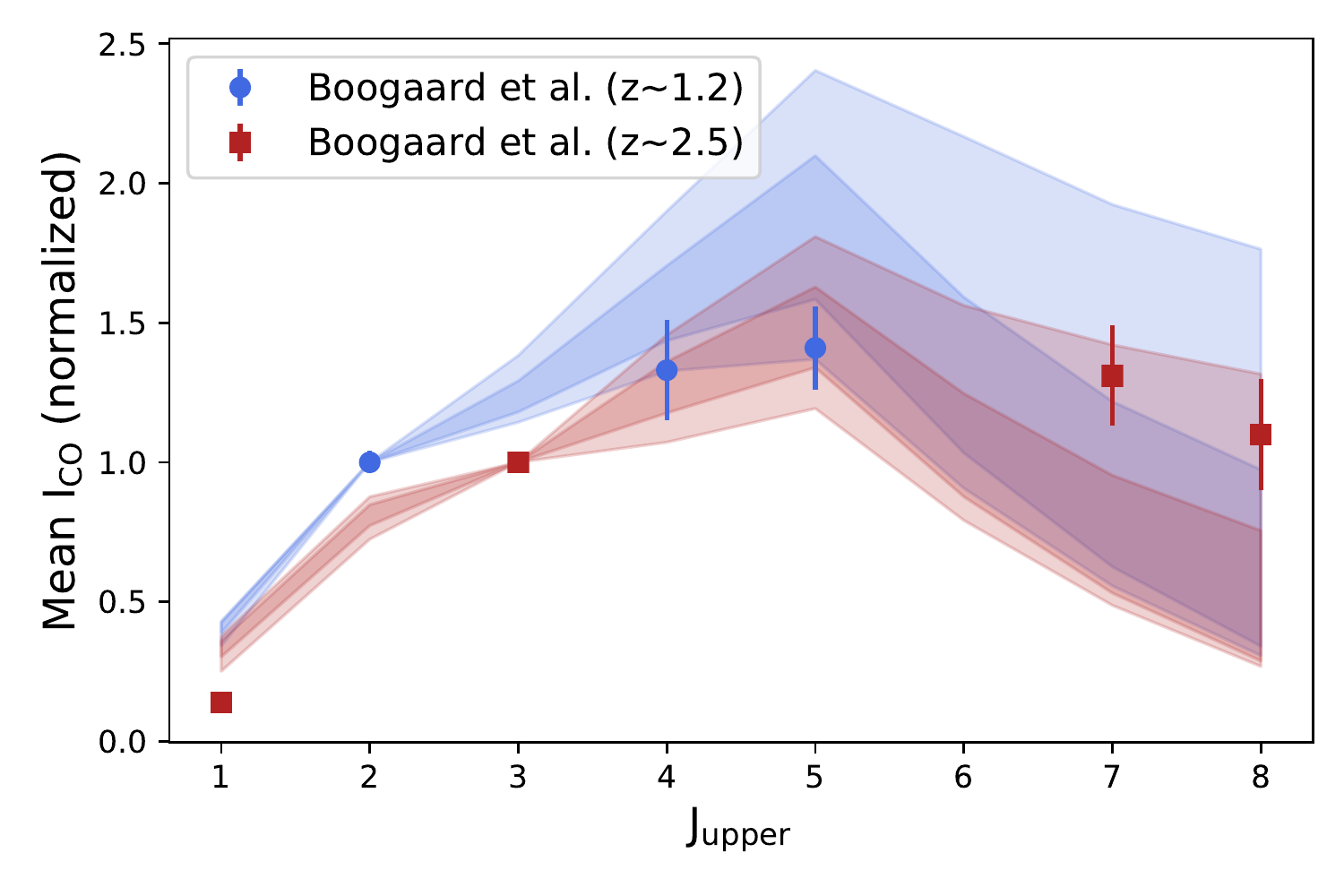}
\caption{\label{fig:sled} Comparison between the stacked SLEDs of \citet{Boogaard2020} and the results from SIDES. The blue red filled circles represent the SLED of z$\sim$1.2 objects detected in CO(2-1). The red filled squares correspond to the z$\sim$2.5 objects detected in CO(3-2). The SLED is normalized to unity for the transition used to select the sample to stack. The light and dark shaded areas show respectively the 1 and 2\,$\sigma$ confidence region of the mean SLED in our simulation taking into account the sample variance (see Sect.\,\ref{sect:CO_SLED_valid}).}
\end{figure}

\subsection{Comparison with CO line stacking}

\label{sect:CO_SLED_valid}

We can also check if our simulation agree with the results obtained using stacking, which usually probe fainter flux regime. \citet{Inami2020} stacked MUSE-selected galaxies with known spectroscopic redshifts and detected CO(2-1) in several stellar mass bins. Higher transitions were not detected. For their main-sequence sample, they measured the mean line flux in four stellar mass bins. To compare with our simulation, we computed the mean flux of the CO(2-1) in the same redshift range (1$<$z$<$1.7) and using the same stellar mass bins. Since they stacked a rather small number of objects (3--38 depending on the bin), we estimated the sample variance by drawing 10\,000 times a sample of the same size as \citet{Inami2020}. The results are presented in Table\,\ref{tab:stacked_CO}. Except for the 9$<$log(M$_\star$/M$_\odot$)$<$10 bin, the simulation agrees with the observations. The disagreement in the 9$<$log(M$_\star$/M$_\odot$)$<$10 bin could come from a small underestimation of this upper limit, since \citet{Inami2020} explicitly assume that their stacked galaxies are point sources to compute their upper limits but they could be marginally resolved.

To measure the CO SLED of galaxies in the ASPECS field, \citet{Boogaard2020} selected objects for which at least one line was detected and stacked the other transitions to measure their mean flux. In Fig.\,\ref{fig:sled}, we show their SLED for CO(2-1)-selected galaxies at z$\sim$1.2 and CO(3-2)-selected galaxies at z$\sim$2.5. To compare their results with our simulation, we performed a similar selection in our simulation. 

Simulating the full process of selecting sources from a single-line detections in the noisy cubes and then stack the other transitions is beyond the scope of this paper. We thus used a simplified selection, which should be roughly equivalent. We selected sources with redshifts corresponding to frequency range probed by ASPECS and with I$_{\rm CO}>$0.2\,Jy\,km/s corresponding typical flux of their faintest detections. We then computed the ratio between the mean line flux of a given transition and the reference transition (CO(2-1) or CO(3-2)) to obtain the same normalization. \citet{Boogaard2020} samples are small and only 5 objects are stacked to derive some ratios. We estimated the sample variance from our simulation by drawing 10\,000 samples of the same size in our simulation and recomputing the mean SLED. In our simulation, the variance comes only from the scatter on $\langle U \rangle$, which correlates with the CO excitation by construction and the presence of a few starbursts with a different SLED. We cannot exclude that the actual scatter is larger and our confidence region too small. 

These SLEDs derived from the simulation are shown in Fig.\,\ref{fig:sled}. Except the CO(1-0) at z$\sim$2.5, the observations are systematically in the 2-$\sigma$ region of the simulation. We can note that both CO(4-3) and CO(5-4) at z$\sim$1.2 in our simulation is systematically lower than the observations and both CO(7-6) and CO(8-7) at z$\sim$2.5 is higher. However, using our 10\,000 samples from the simulation, we determined that two consecutive transitions are strongly correlated with a Pearson correlation coefficient of 0.94 and 0.98 for CO(4-3) and CO(5-4) and CO(7-6) and CO(8-7), respectively. They are thus not independent. In contrast the CO(1-0) deficit at z$\sim$2.5 seems significant. However, it is hard to reconcile with the fact that both the CO(1-0) and CO(3-2) luminosity functions at z$\sim$2.5 are well reproduced (see Fig.\, \ref{fig:CO_LF} and \ref{fig:CO10_LF}).

\begin{figure}
\centering
\includegraphics[width=9cm]{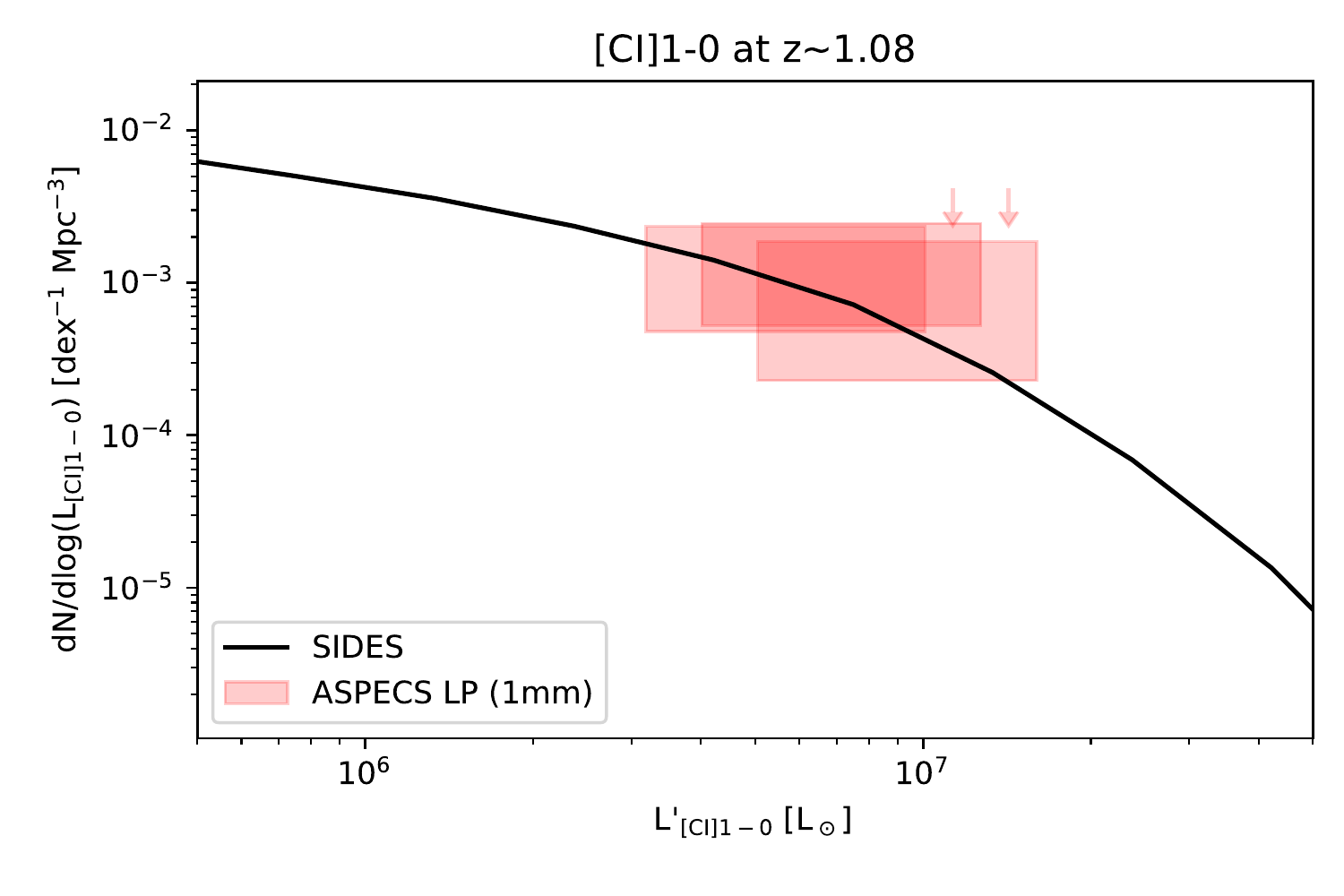}
\includegraphics[width=9cm]{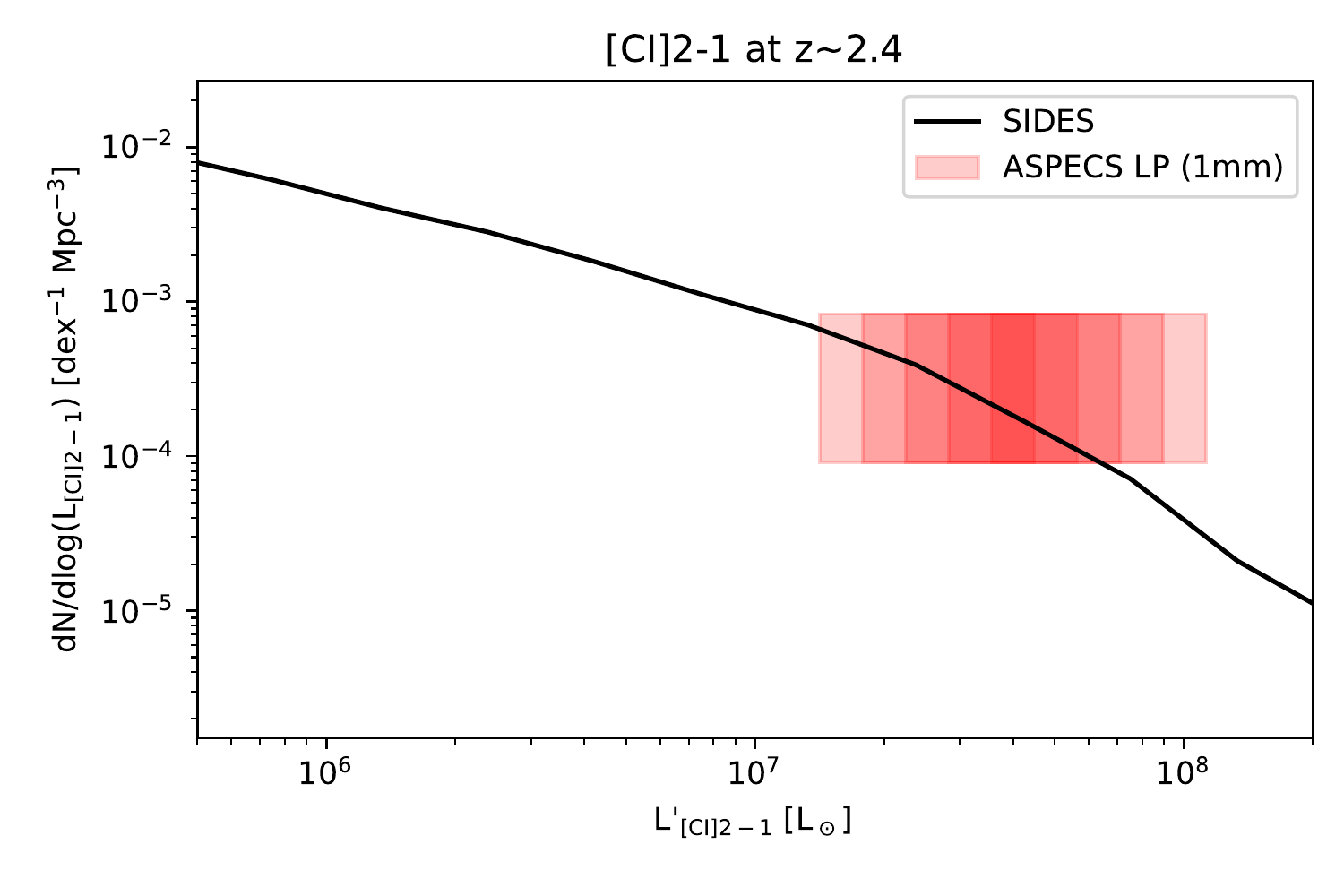}
\caption{\label{fig:CI_LF} Comparison between the [CI] luminosity functions from the SIDES simulation (black solid line) and the ASPECS survey (red areas, \citealt[4.6\,arcmin$^2$]{Decarli2020}).}
\end{figure}

\subsection{[CI] luminosity functions}

\label{sect:CI_LF}

Since there are fewer transitions of [CI] and they are usually slightly fainter than CO making them harder to detect, we have much less constraints on the [CI] luminosity functions. However, the ASPECS survey obtained first constraints around the knee of the luminosity function \citep{Decarli2020}, where the bulk of the [CI] integrated emission comes from. In Fig.\,\ref{fig:CI_LF}, we compare our simulation with their results. The simulation is always in the 1\,$\sigma$ range of the observations. The method used to generate [CI] line fluxes and presented in Sect.\,\ref{sect:recipe_CI} is thus sufficient to reproduce the current observational constraints.

\begin{figure}
\centering
\includegraphics[width=9cm]{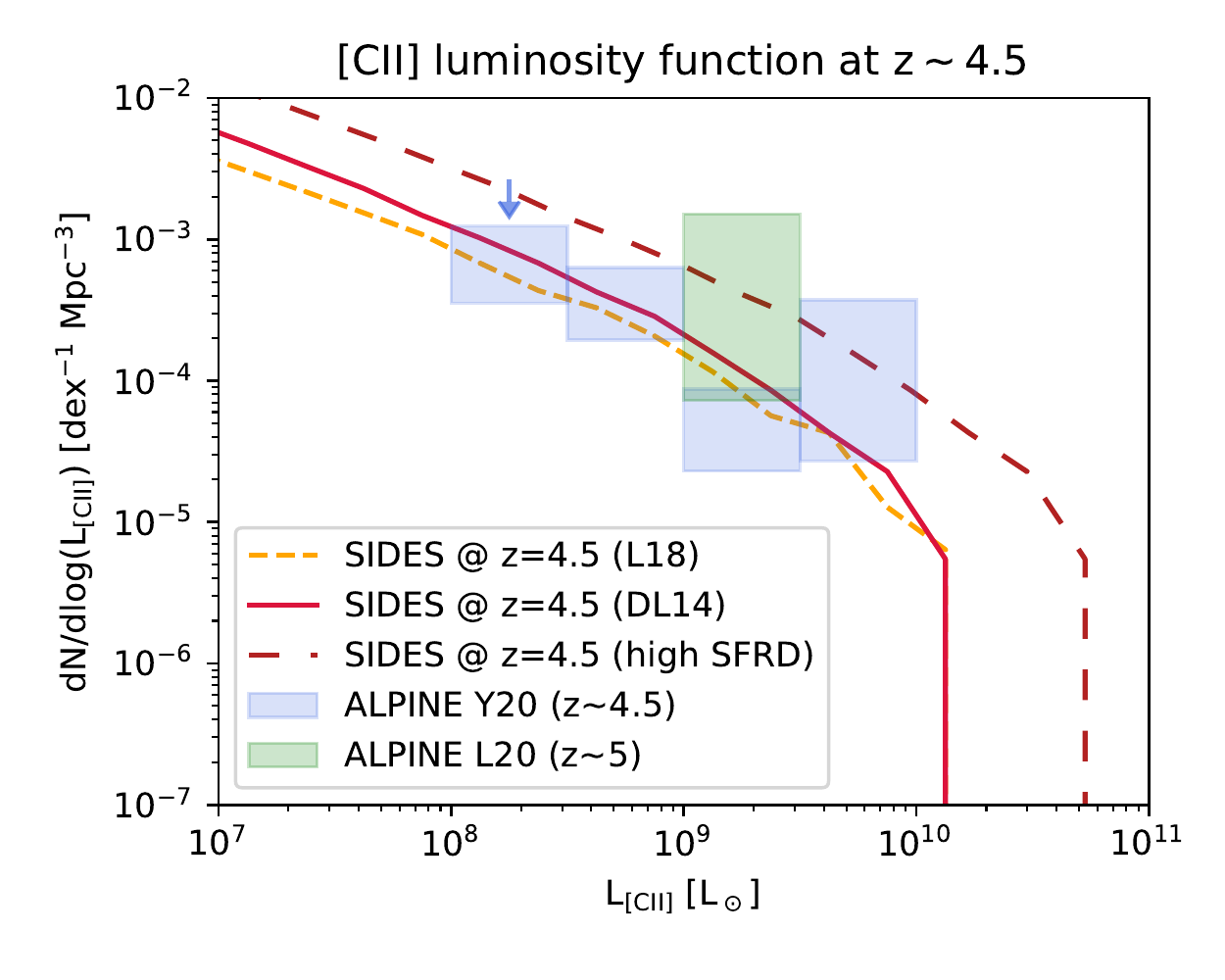}
\includegraphics[width=9cm]{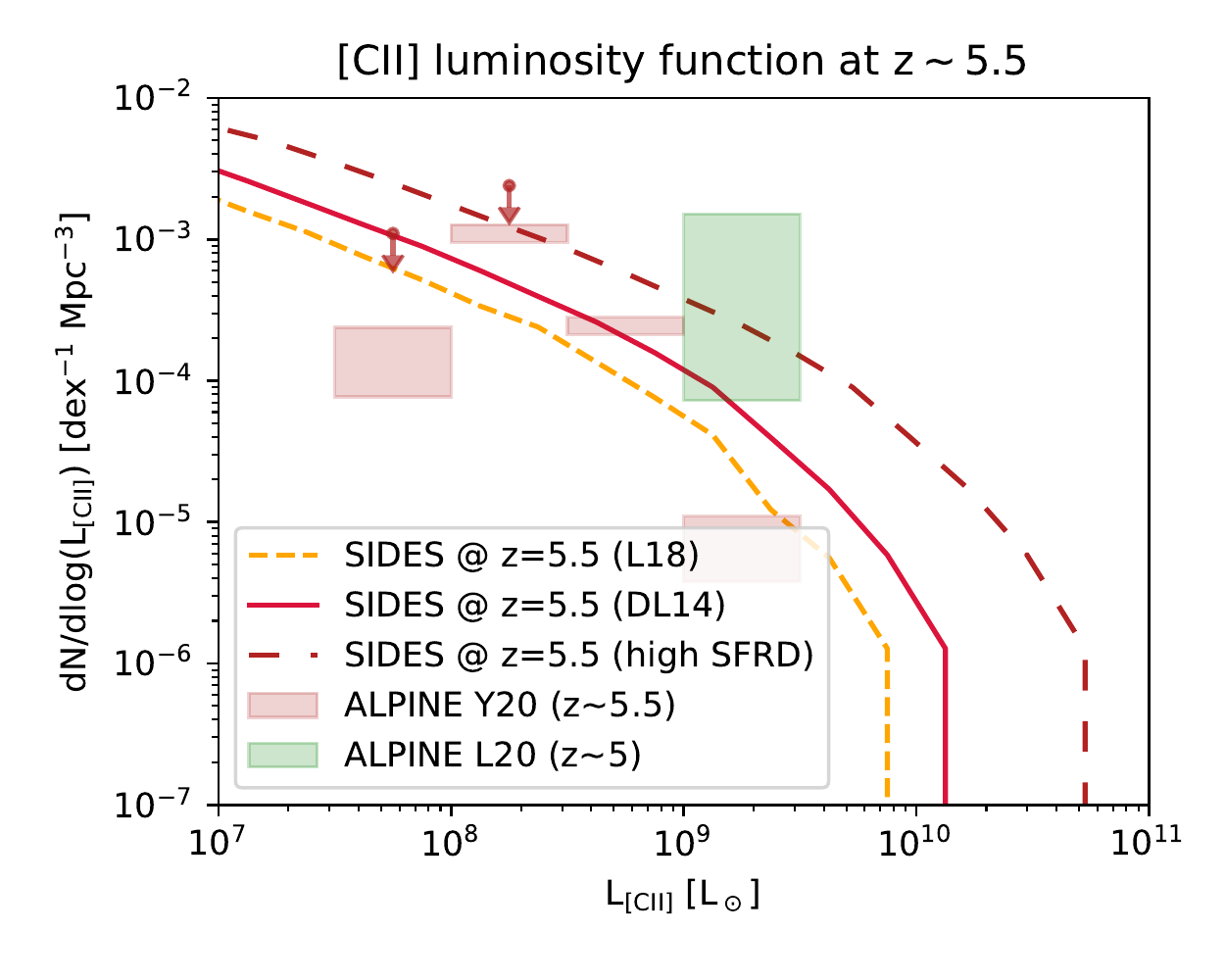}
\caption{\label{fig:CII_LF} Current constraints on the [CII] luminosity function at z$\sim$4.5 (upper panel) and z$\sim$5.5 (lower panel) and comparison with our SIDES simulation. The blue (z$\sim$4.5) and the red (z$\sim$5.5) filled rectangles show the reconstruction of the luminosity function from the ALPINE sample  \citep{Yan2020}. The two data points at the lowest luminosity could be impacted by non detections and the upper limits estimated by \citet{Yan2020} is shown as a downward-facing arrow. The constraint at z$\sim$5 using the other sideband of ALPINE is shown with green rectangles \citep{Loiacono2021}. The red solid lines and the orange dashed lines are the SIDES results assuming the DL14 or the L18 relation, respectively. The brown long-dashed line is the flat high-z SFRD version of SIDES.} 
\end{figure}

\subsection{[CII] luminosity function}

\label{sect:CII_LF}

The [CII] luminosity function is more difficult to measure. Contrary to CO and [CI] source, which are mainly below z=3, [CII] sources observables at ALMA frequency are at higher redshift. Consequently, they rarely have a good photometric coverage to firmly identify the line using a photometric redshift and the follow up of another line can be very difficult. \citet{Aravena2016c} performed a first attempt of measurement at z$>$6 with the ASPECS pilot survey. However, most of their detections were not found by deeper observations and were likely caused by noise. The most recent analysis of the full ASPECS survey provides only upper limits \citep{Uzgil2021}. The ALPINE survey \citep{Le_Fevre2020,Bethermin2020,Faisst2020} targeted the [CII] line with ALMA in 118 4.4$<$z$<$5.9 normal star-forming galaxies. Constraints on the [CII] luminosity function were obtained using two different methods. The first methods use the sideband where the ALPINE targets are not located (12\,GHz appart) to get 118 small blank fields \citep{Loiacono2021}. The second method is more indirect and uses both the properties of ALPINE sources (UV luminosity, redshift, and [CII] luminosity) and the luminosity function of the UV parent sample \citet{Yan2020}.

In Fig.\,\ref{fig:CII_LF}, we compare the SIDES [CII] luminosity function with the ALPINE constraints. To compare with the \citet{Yan2020} results, we computed the SIDES luminosity functions in the same $4.4<z<4.6$ (upper panel) and $5.1<z<5.9$ (lower panel) redshift bins as them. The SIDES results agree at $\sim$1\,$\sigma$ with these measurements at z$\sim$4.5 (in blue) independently of whether the DL14 or L18 relation is used to derive [CII]. The agreement is not as good at z$\sim$5.5 (in red). While we found an overall agreement between 10$^8$ and 10$^9$\,L$_\odot$ using the DL14 relation, the highest and the lowest luminosity points are lower than the simulation by 2\,$\sigma$. However, as discussed in \citet{Yan2020}, the faintest point could be affected by incompleteness of the detection. They propose a robust upper limit (downward arrow), which agrees with our simulation. In contrast, the simulation using the L18 relation is systematically low in the 10$^8$ and 10$^9$\,L$_\odot$ range at z$\sim$5.5 and the one using the flat high-z SFRD and DL14 is too high. Finally, our simulation agrees at $\sim$1\,$\sigma$ with the \citet{Loiacono2021} measurement, which is less accurate but less sensitive to assumptions or systematic effects. 

Our simulation thus roughly agrees with the early measurements of [CII] luminosity function.  While there are minor tensions with the indirect measurement of \citet{Yan2020}, it is hard to know if this is really a problem of the simulation or a problem with the measurement, since the tension between \citet{Loiacono2021} and \citet{Yan2020} at z$\sim$5.5 suggests that some significant systematic effects may impact these early measurements.


\section{Comparison with other models}

\label{sect:compare}

As shown in Sect.\,\ref{sect:LFs}, our new version of the SIDES simulation is compatible with the current constraints from the observations. In this section, we compare the results of this approach with other recent ($\ge$2018) models\footnote{The study of \citet{Karoumpis2021} was published after most of our analysis was completed and is therefore not included here.}. We study the [CII] luminosity function (Sect.\,\ref{sect:CIILF_model}), the cosmic [CII] background (Sect.\,\ref{sect:line_background}), and [CII] power spectrum at z=6 (Sect.\,\ref{sect:Pk3D}). Finally, we discuss the difference between the models and the strengths and weaknesses of the different approaches (Sect.\,\ref{sect:discussion_models}).

The models used for comparison are the following.
\begin{itemize}
\item \citet{Yue2019}: This model starts from the UV luminosity function and uses various empirical scaling relations to produce the [CII] luminosity. The UV luminosity is connected to the halo mass through abundance matching. All the observables are derived from analytical formula. The [CII] power spectrum is derived using a halo model.
\item \citet{Chung2020}: They start from dark-matter simulation. The dark-matter halos are populated by galaxies using the {\sc universemachine} \citep{Behroozi2019} approach. The [CII] luminosity is derived from the SFR using the \citet{Lagache2018} relation. The observables as background and power spectra are derived from the simulated cubes.
\item \citet{Yang2021}: The authors calibrated the relation between the [CII] luminosity using the \citet{Popping2019} semi-analytical model. The intensity mapping observables are derived using an analytical halo model.
\end{itemize}

\begin{figure}
\centering
\includegraphics[width=9cm]{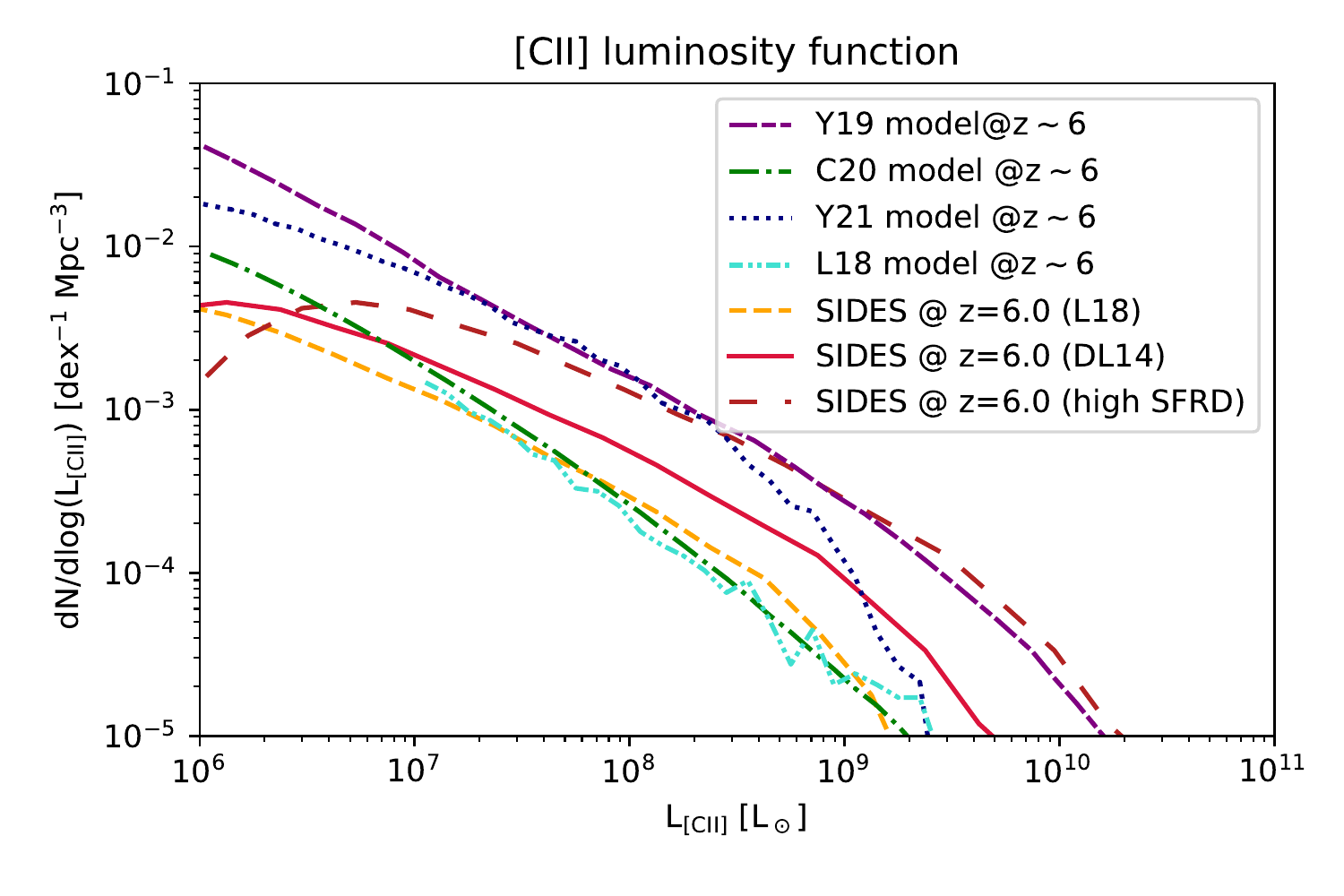}
\caption{\label{fig:CII_LF_model} Comparison between the luminosity functions produced by various models at z=6. The red solid lines and the orange dashed lines represents the SIDES luminosity function using the DL14 or the L18 relation, respectively. The brown long-dash line is the high SFRD version of SIDES. Ther models shown for comparison are \citet[][turquoise two-dot-dash line]{Lagache2018}, \citet[][blue dotted line]{Yang2021}, \citet[][purple short-long-dash line]{Yue2019}, \citet[][green dot-dash line]{Chung2020}.}
\end{figure}

\subsection{[CII] luminosity functions at z$\sim$6}

\label{sect:CIILF_model}

The [CII] luminosity function is one of the most basic observable to compare models. The [CII] background (see Sect.\,\ref{sect:line_background} and appendix \ref{app:CII_bkg}) and the shot noise of the [CII] power spectrum (see Sect.\,\ref{sect:Pk3D} and appendix \ref{app:CII_poisson}) derives directly from it. The link with the correlated fluctuations is less direct, since it also depends on the relation between the halo properties and the [CII] luminosity. In this paper, we choose to focus on the luminosity function at z$\sim$6, which is the aim of most of the first generation experiments (CONCERTO, TIME, FYST) and has been studied by all the models in our compilation. 

In Fig.\,\ref{fig:CII_LF_model}, we show the [CII] luminosity function predicted by these various models. The version of SIDES using the L18 relation and the \citet{Chung2020} model using the same relation have quite similar [CII] luminosity functions. However, we note that SIDES has a slightly shallower slope. \citet{Chung2020} is really close from the full semi-analytical L18 model, which indicates that their SFR distribution are very similar since they use the same SFR-L$_{\rm [CII]}$ relation. They are both very close from a power law, while the L18 version of SIDES has a knee around 10$^{9}$\,L$_\odot$. The DL14 version of SIDES is higher than the L18 version by a factor of $\sim$1.5 around 10$^{7}$\,L$_\odot$ and $\sim$3 around 10$^{9}$\,L$_\odot$. It also has an even shallower slope than the previous models.  This can be explained by the higher [CII] luminosity predicted by the DL14 and the non-linear slope of the L18 relation. Overall, these four models (SIDES L18, SIDES L14, \citealt{Chung2020}, and the original L18) have relatively similar luminosity functions with a significantly higher luminous end for the SIDES DL14.

The \citet{Yue2019}, \citet{Yang2021}, and the high-SFRD version of SIDES have overall much higher luminosity functions than the previously discussed models. Below 10$^{8.5}$\,L$_\odot$, these models are factor of $\sim$3 higher than SIDES assuming the DL14 relation. We can also note that the \citet{Yang2021} model has a steep slope around 10$^{8.5}$\,L$_\odot$, while the two other high models exhibit a more discrete knee. 

We remark that the high-SFRD version of SIDES has a sharp cutoff below 10$^7$\,L$_\odot$, which is caused by the halo mass limit of our underlying dark-matter simulation (see Sect.\,\ref{sect:discussion_models}). Since this model has a higher SFR and [CII] luminosity at fixed halo mass, this cutoff is at a higher luminosity than in the other versions of the model. This cutoff is around 10$^6$\,L$_ \odot$ for the DL14 version of SIDES. 

\begin{figure}
\centering
\includegraphics[width=9cm]{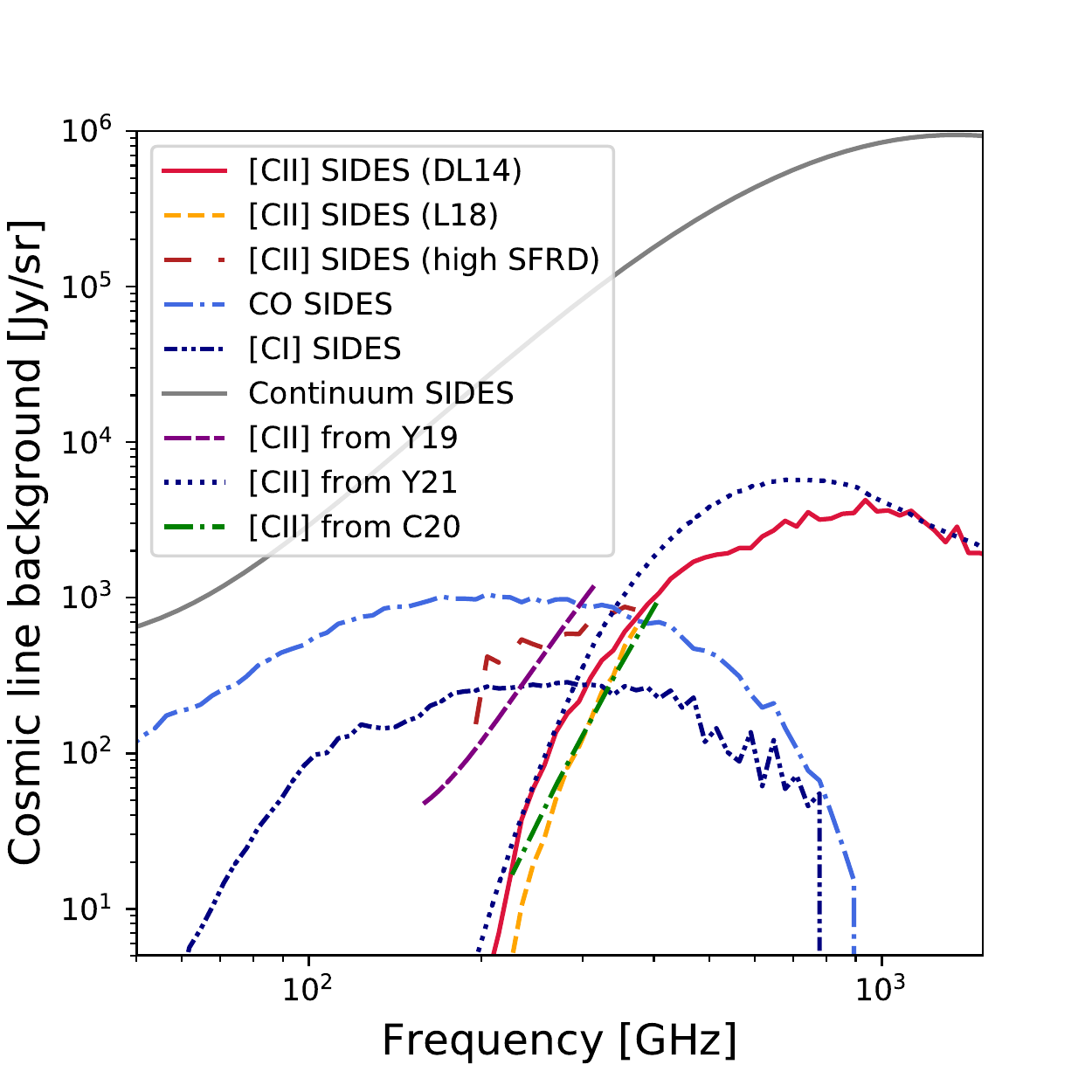}
\caption{\label{fig:line_background} Cosmic line background as a function of frequency.  The red solid, the yellow dashed, and the brown long-dashed lines shows the predictions of our simulation assuming the DL14 relation, the L18 relation, and the flat z$>$4 SFRD (see Sect.\,\ref{sect:SFRD}), respectively. The CO, [CI], and continuum background predicted by SIDES are shown as a blue dot-dashed, a dark blue two-dot-dashed, and a solid grey line, respectively. The green dot-dash, the dash-long-dash purple line, and the blue dotted line are the [CII] forecast from the \citet{Chung2020} model, the \citet{Yue2019}, and the \citet{Yang2021} models, respectively.}
\end{figure}

\subsection{Cosmic [CII] background}

\label{sect:line_background}

One of the most simple global quantities that we can derive from the simulated cubes, is the line background, i.e. the total surface brightness density from all the lines emitted by all the galaxies. Each galaxy contributes to a couple of specific observed frequencies corresponding to its various redshifted lines, but the total background is smooth since galaxies are distributed continuously over a wide range of redshift. While this quantity is very useful to compare models, it is very hard to measure it in practice, since it would require an absolute photometer and an extremely accurate subtraction of the CIB and the CMB, which are much brighter. For instance, the CIB is $\sim$100 times brighter than the [CII] background at 300\,GHz (see below). However, lower limits could be obtained from deep line spectral scans such as ASPECS, but a direct comparison with the luminosity functions is then more informative (see Sect.\,\ref{sect:compare}).

The [CII] background at a frequency $\nu_{\rm obs}$ is directly connected to the [CII] luminosity function ($\frac{d^2 N}{dL_{\rm [CII]} dV}$) at the associated redshift ($z = \nu_{\rm rest} / \nu_{\rm obs} - 1$) by the following equation:
\begin{equation}
B_\nu^{\rm [CII]} (\nu_{\rm obs}) = \frac{1}{C \, (1+z) \, \nu_{\rm obs} \, H(z)} \int L_{\rm [CII]} \, \frac{d^2 N}{dL_{\rm [CII]} dV} \, dL_{\rm [CII]},
\end{equation}
where $C$ is a constant conversion factor and $H(z)$ the Hubble parameter at a redshift $z$ (see appendix\,\ref{app:CII_bkg} for the full computation).

To check the consistency between our input catalogs and cubes, we derived the background using two different methods. The first approach uses the catalogs. We defined small bins in frequencies and sum the flux (in Jy\,km/s) of all the lines falling in each bin. We then divide this quantity by the solid angle in the sky associated to the catalog and the width of the velocity associated to each bin ($c \Delta \nu_{\rm obs} / \nu_{\rm obs}$). For the second approach, we averaged the cubes in the two spatial dimensions and obtained the mean spectra of the sky, which is exactly the line background. This task was performed on the cubes corresponding to each line to obtain their individual contribution. The results of the two methods agree at better than the percent demonstrating that no major artifact is created during the cube making.

In Fig.\,\ref{fig:line_background}, we present the line background derived from SIDES. The CIB (continuum) is always brighter than the lines. It is a factor of 5 higher than the line background at 100\,GHz and a factor of 200 at 1000\,GHz. If we consider only the lines, the [CII] background dominates at high frequencies, while CO dominates at low frequency. The crossing frequency slightly varies depending on the version of the [CII] model: 365\,GHz for the DL14 relation, 371\,GHz for the L18 relation, and 345\,GHz for the high SFDR scenario (see Sect.\,\ref{sect:SFRD}). The [CI] background is never dominant.

We can also compare our predictions with other models. The L18 version of our simulation agrees well with the \citet{Chung2020} work based on the same SFR-L$_{\rm [CII]}$ relation but using the {\sc universemachine} \citet{Behroozi2019} approach to populate dark-matter halos. The other versions of our model produce stronger [CII] background. The \citet{Yue2019} is much higher than the standard versions of our model whatever the assumed SFR-L$_{[CII]}$ relation, but it is similar to our high SFRD model. Their model is higher than SIDES (DL14 version) at z$\sim$6 by a factor of $\sim$ 3 (see Sect.\,\ref{sect:CIILF_model} and Fig.\,\ref{fig:CII_LF_model}) around 10$^9$\,L$_\odot$ and exhibits an even stronger excess at fainter fluxes. It is thus not surprising that this model produce a stronger background. Finally, the \citet{Yang2021} model is higher than SIDES between 250\,GHz and 1\,THz without being as extreme as the \citet{Yue2019} model. This is consistent with their luminosity function being overall higher than SIDES (except at the very bright end), but having a shallower faint-end slope and a sharper high-luminosity cutoff than the \citet{Yue2019} model.

\begin{figure}
\centering
\includegraphics[width=8cm]{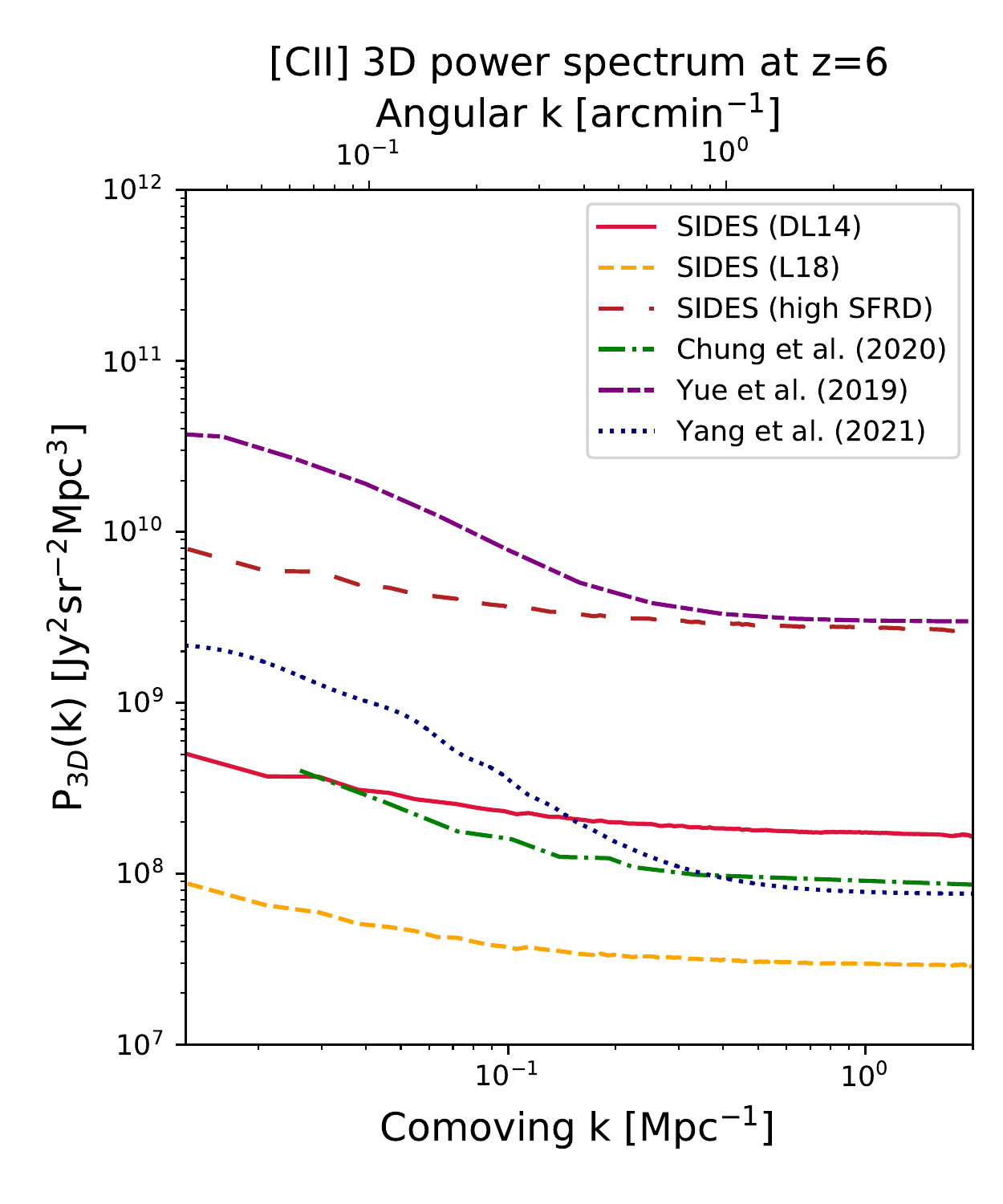}
\caption{\label{fig:Pk3D} Comparison of the 3-dimensional [CII] power spectra at z=6 from various models. The lengths are in comoving units. We also indicate the corresponding projected angular wavenumber at $z=6$ as upper x axis. The x-axis range corresponds approximately to the scales, which will be probed by CONCERTO.} The red solid and orange dashed lines are from the SIDES simulation assuming the DL14 and L18 relation, respectively. The brown long-dash line is the high-SFRD version of SIDES (see Sect.\,\ref{sect:SFRD}). We also compare with the models of \citet[][purple two-dot-dash line]{Yue2019}, \citet[][green dot-dash line]{Chung2020}, and \citet[][blue dotted line]{Yang2021}.
\end{figure}

\subsection{Computation of the [CII] 3D power spectrum at z$\sim$6 and comparison with other models}

\label{sect:Pk3D}

While our simulation produced spectral cubes, most of theoretical models forecasted only the 3D power spectra at some specific redshift. To perform a more direct comparison, we also produced a 3D cube from our catalog. We first defined a 3D grid in comoving units centered on z=6 (D$_C$ = 8435\,Mpc) and the middle of the SIDES field. This grid has 512 elements of 0.4\,Mpc in each direction to cover the full simulated field. For each of these voxels, we can associate a central redshift and redshift width by converting the depth coordinate into a redshift. Our cube covers the 5.76$<$z$<$6.24 range (8332\,Mpc$<$D$_C <$8538\,Mpc). Each source in this redshift range is associated to a voxel based on its redshift and sky coordinates. To derive the surface brightness density in Jy/sr of a voxel ($S_\nu$), we computed the sum of the [CII] line fluxes from all the sources associated to a given voxel and divided it by the velocity width $\Delta \upsilon_{\rm voxel}$ and the solid angle associated to the voxel $\Omega_{\rm voxel}$ ($= \Delta x \, \Delta y  / D_c^2$):
\begin{equation}
S_\nu = \frac{ \sum_{\rm sources}  I_{\rm [CII]}} {\Omega_{\rm voxel} \, \Delta \upsilon_{\rm voxel}} = \frac{(1+z_{\rm voxel}) \, \sum_{\rm sources}  I_{\rm [CII]}} {\Omega_{\rm voxel} \, \Delta z_{\rm voxel} \,c},
\end{equation}
where $z_{\rm voxel}$ and $\Delta z_{\rm voxel}$ are the center and the width of the voxel after converting the radial distances into redshifts, respectively. $I_{\rm [CII]}$ is the line flux in Jy\,km/s. Finally, we computed the 3D power spectrum of this cube. This quantity is independent from the choice of the voxel size (see appendix\,\ref{app:CII_poisson}).

Our results are presented in Fig.\ref{fig:Pk3D}. At small scale (high $k$), we can see a plateau corresponding to the shot noise (also called Poisson component), which we would still have in absence of any clustering and depends only on the luminosity functions (see appendix\,\ref{app:CII_poisson}). At large scale ($k \lesssim 0.5$\,arcmin$^{-1}$), we can observe an excess compared to the shot noise corresponding to the clustering of [CII]-emitting galaxies. This behavior is similar to what has already been observed for the cosmic infrared background \citep{Lagache2007,Viero2009,Planck_CIB2011,Amblard2011,Penin2012a,Planck_CIB2013,Bethermin2013,Viero2013}.

The various versions of SIDES produce power spectra with similar shapes but very different normalizations. The version using the L18 relation is lower than the one using the DL14 relation by a factor of $\sim$5. The difference is stronger than for the cosmic [CII] background. This is expected, since the power spectrum is proportional to the emissivity squared. In addition, because of the different shapes of the L$_{\rm [CII]}$-SFR relation, there are more luminous objects in the DL14 version. These rare luminous sources have a stronger relative contribution to the shot noise than the background (proportional to the luminosity squared instead of the luminosity, see appendix\,\ref{app:CII_bkg} and \ref{app:CII_poisson}). They also contribute more to correlated anisotropies at larger scales relatively to their flux since they tend to live on more massive halos, which are also more clustered. Finally, the high-SFRD version corresponds to a simple rescaling of the [CII] fluxes from DL14 version by a constant, which leads to a power spectrum higher by this constant squared.

The \citet{Chung2020} model has a lower shot noise at small scales than the DL14 version of SIDES, but a steeper slope at large scales where the clustering dominates. The \citet{Yang2021} model has a lower shot noise than SIDES DL14 and a similar one as the \citet{Chung2020} model, but a much stronger correlated signal. While their luminosity function is relatively high compared to SIDES DL14 (Sect.\,\ref{sect:CIILF_model}), \citet{Yang2021} have fewer objects than SIDES at the luminous end, which dominate the shot noise and cause their shot noise to be higher. In contrast, the larger amount of correlated signal is likely coming from the excess of 10$^7$-10$^8$\,L$_\odot$ emitters by a factor of $\sim$3 compared to SIDES DL14. Finally, the \citet{Yue2019} model, which has the highest luminosity function at all luminosities, has naturally a higher power spectrum than all the previously-cited models by at least an order of magnitude. It has a similar shot noise as the SIDES high SFRD model, but a stronger correlated signal. 

\begin{figure*}

\centering

\begin{tabular}{cc}
\includegraphics[width=8.5cm]{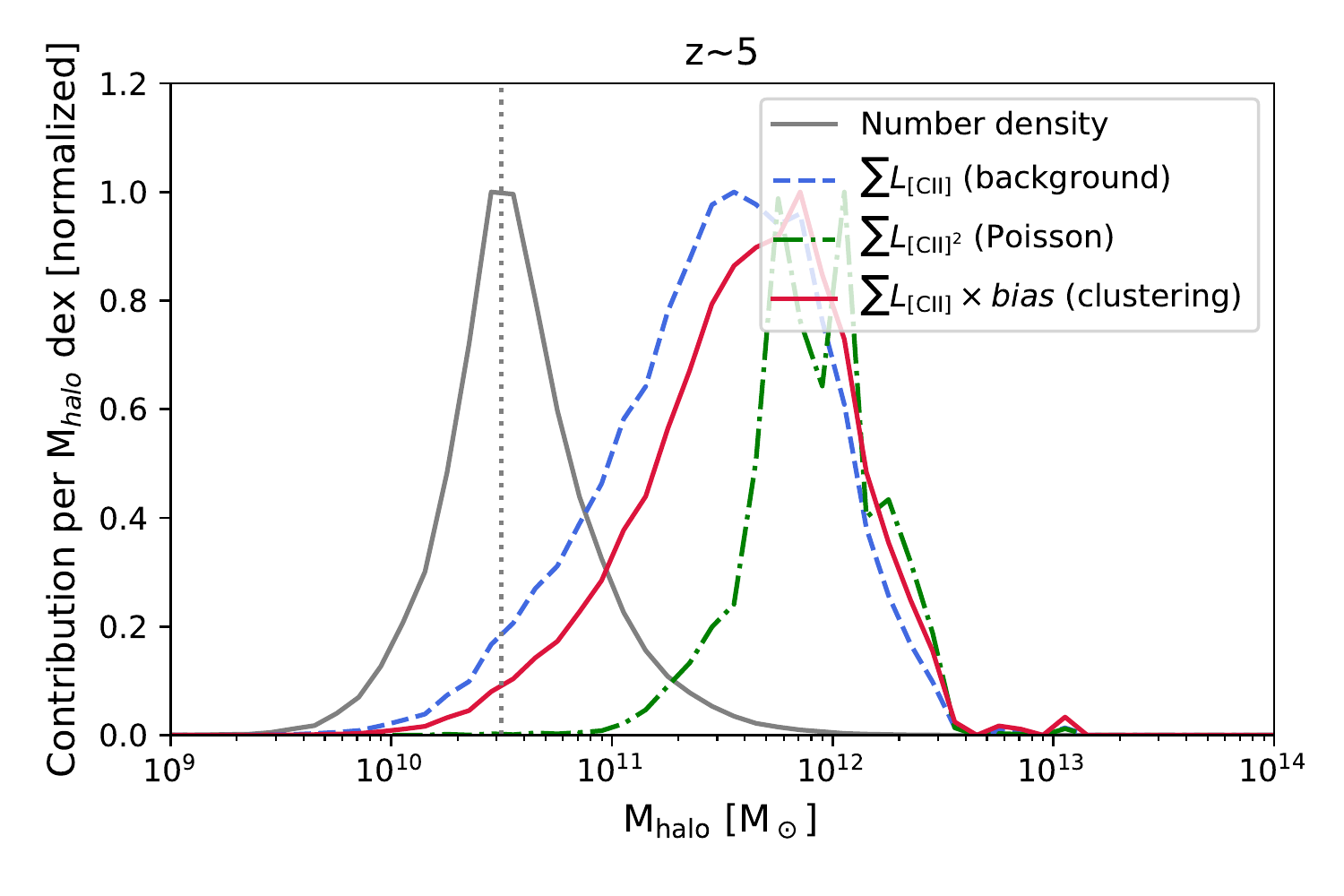} & \includegraphics[width=8.5cm]{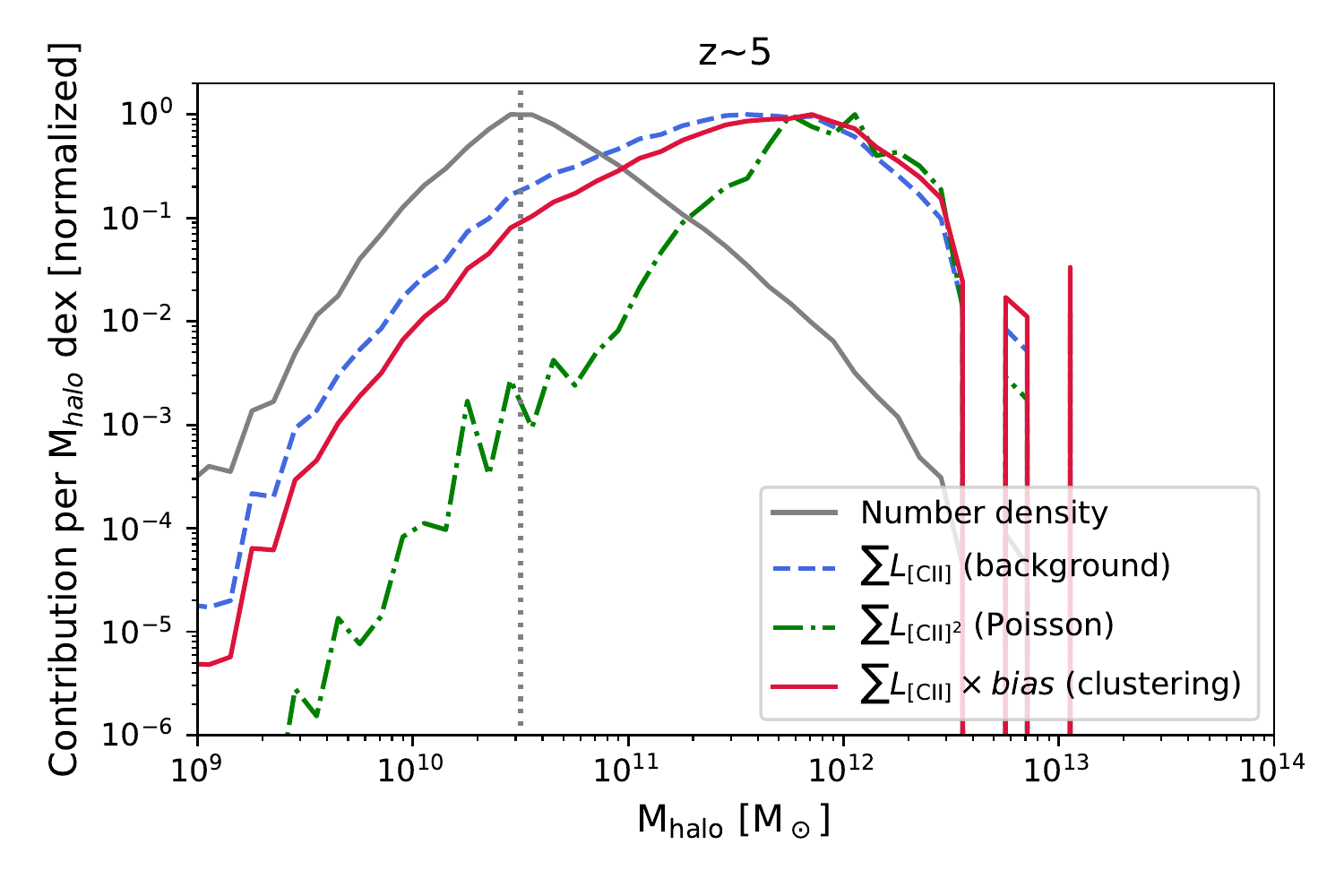}\\

\includegraphics[width=8.5cm]{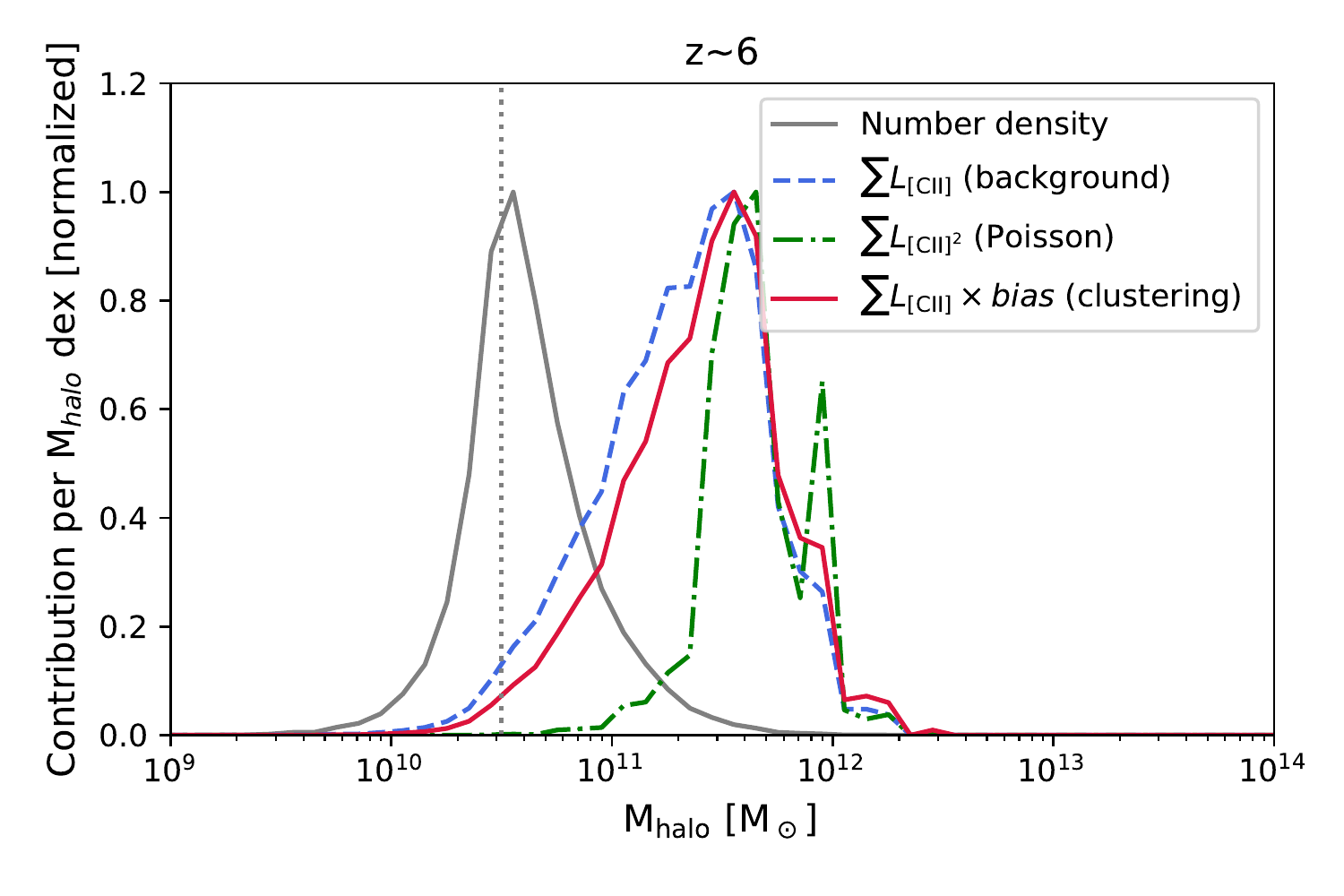} & \includegraphics[width=8.5cm]{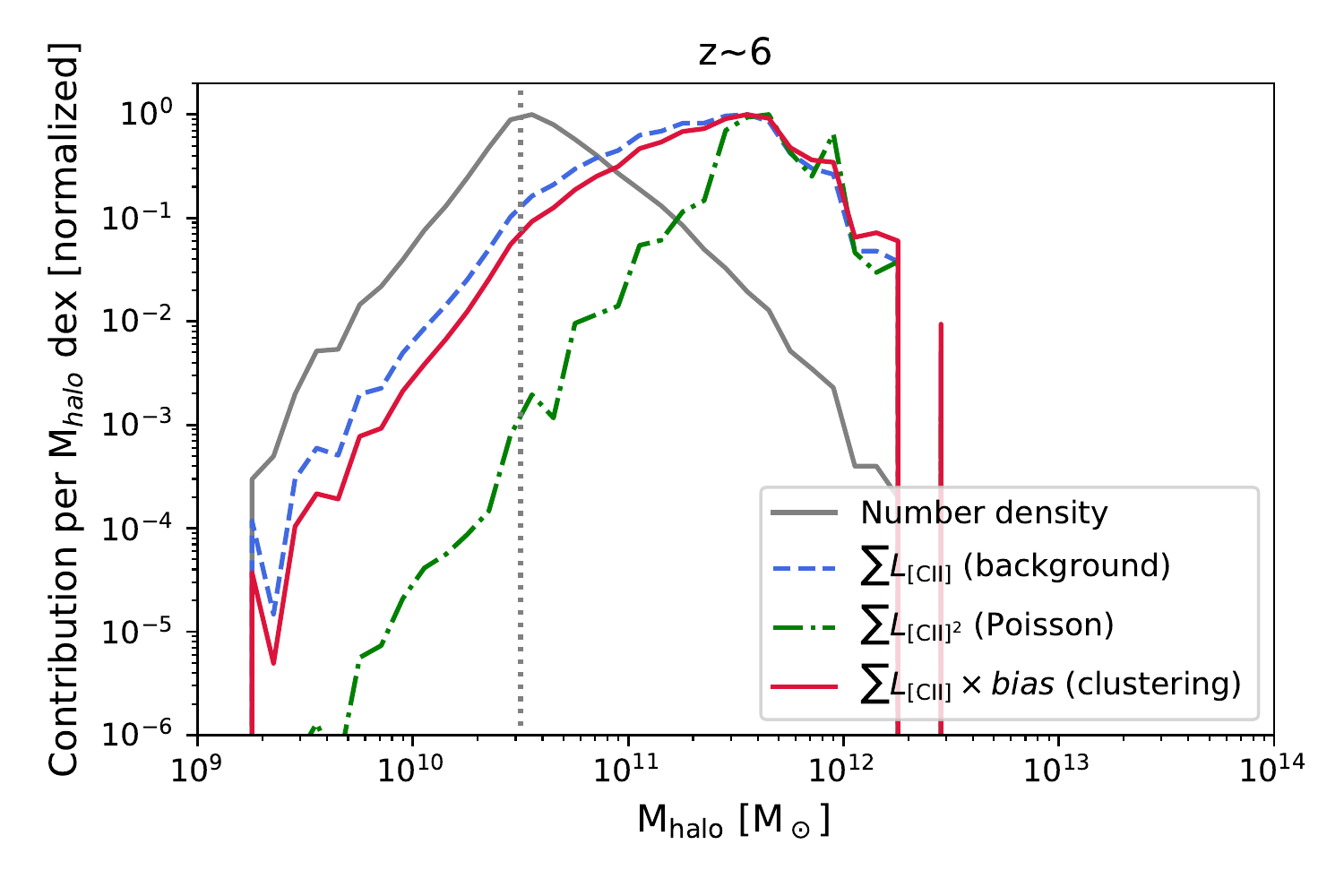}\\

\includegraphics[width=8.5cm]{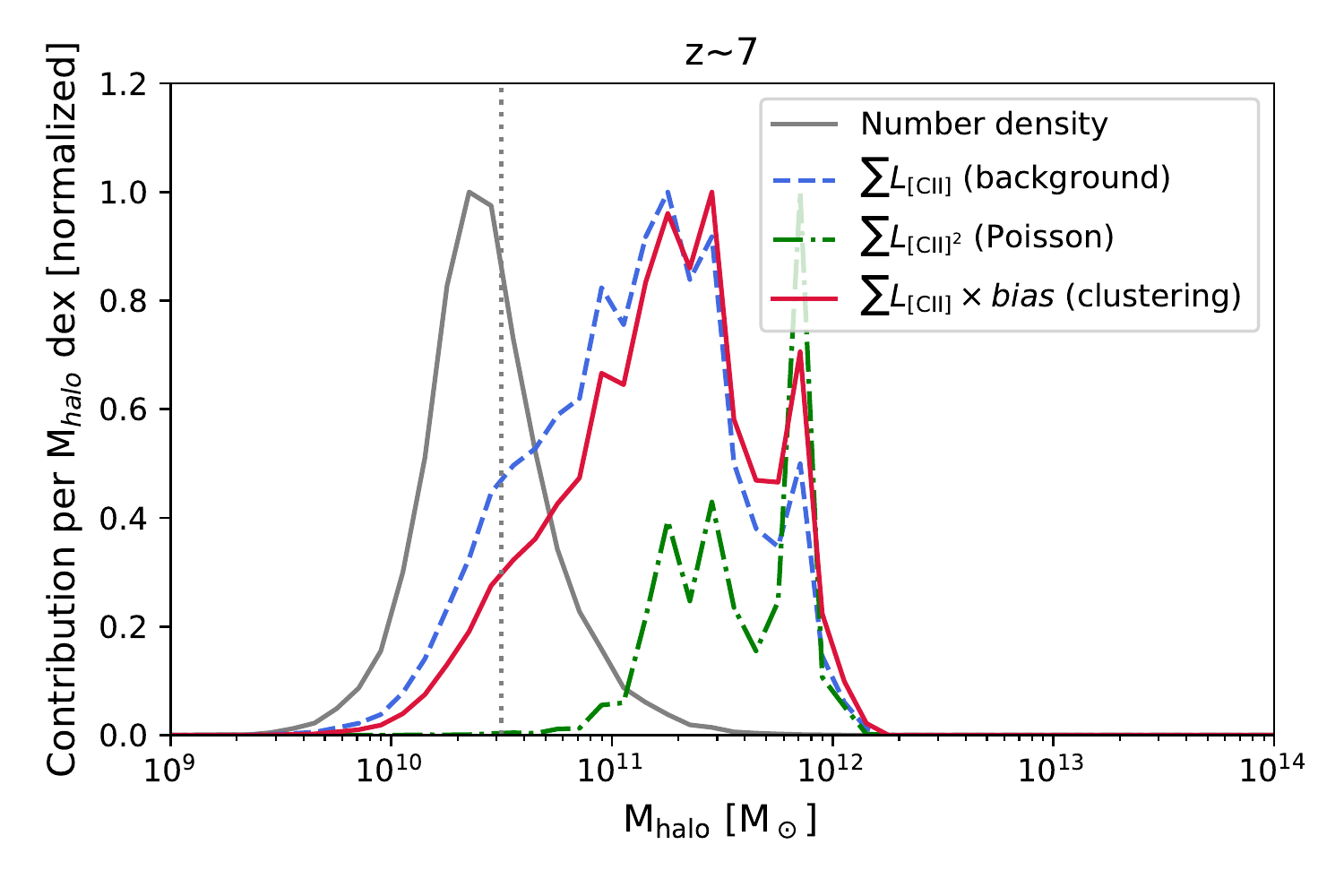} & \includegraphics[width=8.5cm]{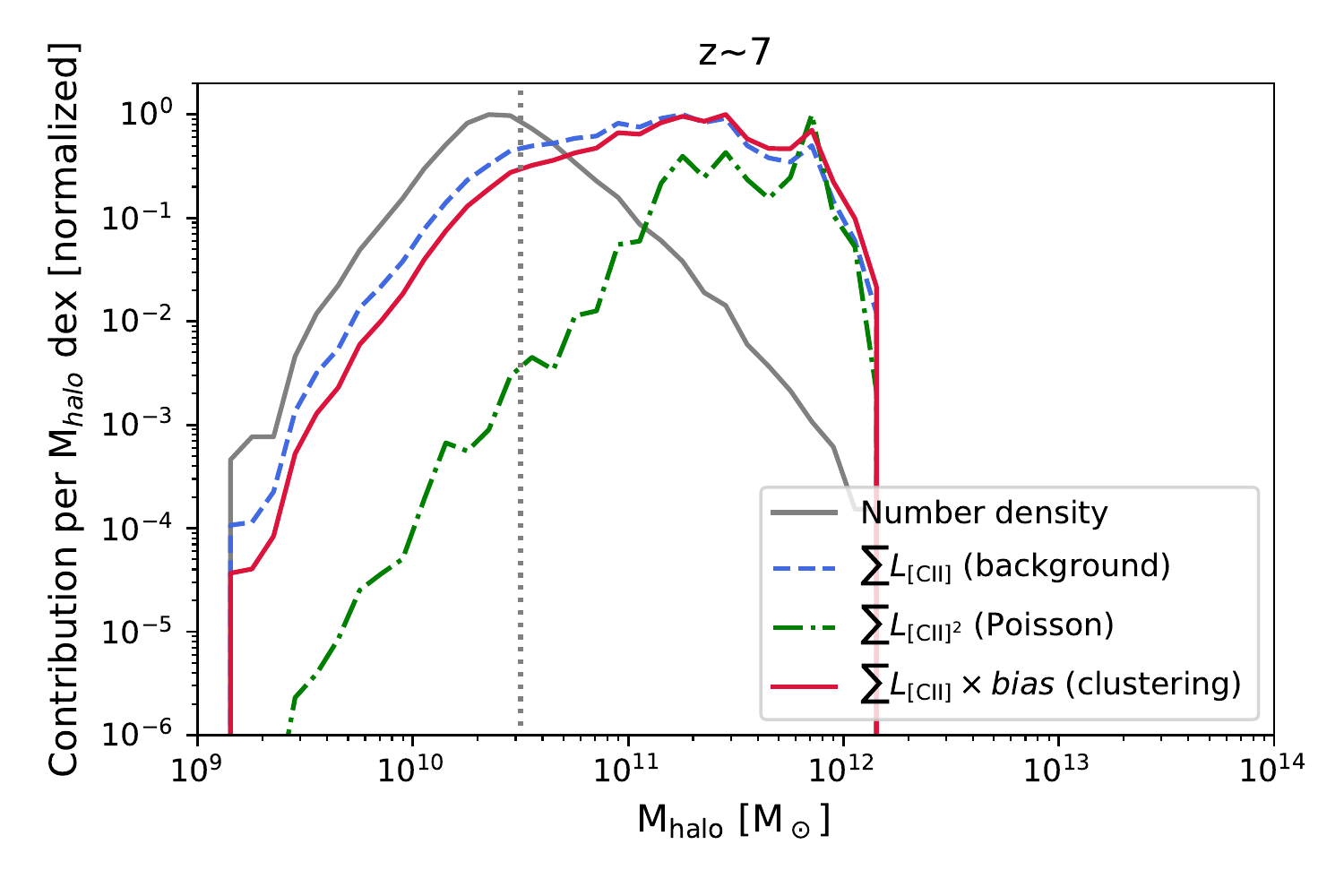}\\
\end{tabular}

\caption{\label{fig:halo_lim} Effect of the halo mass limit on various [CII] observables (see Sect.\,\ref{sect:discussion_models}). The curves show the total contribution of galaxies to these observables per logarithmic interval of host halo masses (dex). All the curves are normalized to have their maximum equal to one. The top, central, and bottom panels correspond to z$\sim$5, z$\sim$6, and z$\sim$7, respectively. The left panel have a y-axis in linear units allowing us to visualize naturally the area under the curves and the right panels are in logarithmic scale to better visualize the behavior at low mass. The grey curve is the number density of the halos and the dotted vertical line show the halo mass limit of 10$^{10.5}$\,M$_\odot$. The dashed blue line shows the contribution of galaxies to the [CII] background, which directly derives from the integral of the [CII] line luminosities (see appendix\,\ref{app:CII_bkg}). The green dot-dashed line is the contribution to the shot-noise, which is directly connected to the the integral of the [CII] luminosity squared (see appendix\,\ref{app:CII_poisson}). Finally, the solid red curve presents the integral of the luminosity of the galaxies multiplied by the linear bias corresponding to their halo mass. This term is directly connected to the amplitude of the correlated fluctuation of the [CII] (see Sect.\,\ref{sect:discussion_models}).}

\end{figure*}

\subsection{Discussion about the different models}

\label{sect:discussion_models}

The z$\sim$6 [CII] luminosity functions forecasts from the various models discussed in this section vary by an order of magnitude and the power spectra by more than two orders of magnitude. These models are all built on very reasonable assumptions and these disagreements demonstrate that it is still hard to predict the number density of [CII] galaxies and and how they are distributed in the dark-matter halos even in the ALMA era. On the positive side, having all these models available allows us to know in which range we can expect the real Universe to be and thus better prepare ongoing and future experiments. They will be key to better understand this early phase of galaxy evolution.

The shot-noise level can vary strongly from one model to another. It is really sensitive to the bright end of the luminosity function. There is no systematic difference between the family of models using a halo model \citep{Yue2019,Yang2021} and the ones using simulated cubes \citep[][SIDES]{Chung2020}. As discussed by \citet{Murmu2021}, the scatter on the L$_{\rm [CII]}$ can impact significantly the shot noise. However, it is hard to compare the various models since they do not connect quantities in the same way. \citet{Yue2019} connect the UV to the [CII] luminosity assuming a scatter, \citet{Chung2020} assume a scatter on the SFR-[CII] relation, and finally \citet{Yang2021} directly parametrize the scatter in the relation between the halo mass and the [CII] luminosity. In our model, we have a cascade of the scatters, when we connect successively the halo mass to the stellar mass to the SFR to the [CII] luminosity.

The relative level of the large-scale correlated fluctuations compared to the shot noise can also vary between models. Overall, the models based on simulations tend to have a lower ratio of correlated versus Poisson fluctuations (see Fig.\,\ref{fig:Pk3D}). The \citet{Yang2021} model has the largest ratio. This is not surprising, since this model has a large number of faint sources, but very few bright sources (L$_{\rm [CII]}>10^9$\,L$_\odot$, see Fig.\,\ref{fig:CII_LF_model}). The bright sources have a strong contribution to the Poisson term, since the contribution to it is proportional to the luminosity squared (see appendix\,\ref{app:CII_poisson}). In contrast, the contribution to the correlated fluctuations is linked to the luminosity weighted  by the linear bias of the host halos (see, e.g., Eq.\,17 of \citealt{Yue2019} or Eq.\,6 and 7 of \citealt{Yang2021}). Even if the low luminosity sources are usually hosted by lower mass halos with a lower bias, this usually does not compensate for the fact that they are much more numerous than the very bright objects.

A potential explanation for the lowest power spectra of the models based on dark-matter simulation could come from their halo mass limit. Since the very low mass halos are missing in the simulation, a significant fraction of the signal coming from the hosted galaxies may be lacking. \citet{Chung2020} discussed this effect and estimated that it could have an impact up to a factor of 3.

To estimate the actual effect of the halo-mass limit on the prediction of our simulation, we computed the relative contribution of galaxies to various observables as a function of their host halo mass. We divided our simulated catalog in halo bins with a 0.1\,dex width. We then computed the number density, the sum of [CII] luminosities, the sum of [CII] luminosities squared, and the sum of the [CII] luminosities weighted by the bias of the host halo from \citet{Tinker2010}. We normalize all the results, since we are only interested in the relative contribution. The results are shown in Fig.\,\ref{fig:halo_lim}. We see a strong turnover in the number density (grey solid line) at 10$^{10.5}$\,M$_\odot$ corresponding to the halo mass limit of the dark-matter simulation.

We first looked at the sum of the [CII] luminosities (blue dashed line), which is proportional to the contribution to the [CII] background (see appendix\,\ref{app:CII_bkg}). We can already notice visually that the contribution at the mass limit is small. To quantify more accurately the missing background, we fitted the curve between 10$^{10.5}$\,M$_\odot$ and 10$^{11}$\,M$_\odot$ by a power law. This function is used to extrapolate the contributions of the low-mass halos. We then integrated the curve above the limit and compared it with the integral down to 10$^5$\,M$_\odot$ using our power-law extrapolation below the mass limit. We found that the halos in our simulation contribute to 96\,\% of the total background at z$\sim$5 and z$\sim$6, but only 81\,\% at z$\sim$7.

As shown in appendix\,\ref{app:CII_poisson}, the contribution to the shot-noise is proportional to the luminosity squared (green dot-dash line). Thus, massive halos hosting luminous galaxies have a stronger contribution. Consequently, at z$\sim$5, the signal is mainly coming from galaxies hosted by $\sim$10$^{12}$\,M$_\odot$ halos. At higher redshift, these halos are rarer and the maximal contribution drifts to slightly lower masses. However, we estimated the contribution of the halos below the mass cut to be below 1\,\%. 

Finally, we estimated the contribution of the various halos to the correlated fluctuations from the product of the luminosity by the linear bias (red solid line). We find a peak contribution around 10$^{12}$\,M$_\odot$ at z$\sim$5 and 2$\times$10$^{11}$\,M$_\odot$ at z$\sim$7. This agrees with the estimate of \citet{Yue2015} at z$\sim$5 using a similar approach. At z=5 and z=6, we estimated that 98\,\% of the integral is coming from halos above the mass limit, but it is only 92\,\% at z=7. These values have to be squared to evaluate the impact on the power spectra. Our simulation should thus be reliable at a 20\,\% level up to z=7. At z$>$7, the results of our simulation should be taken with caution, since the mass resolution of the underlying dark-matter simulation may be not sufficient to be reliable at the very early stages of the structure formation. The sharp drop of the SFRD at z$>7$ (see Fig.\,\ref{fig:SFRD}) could be caused by the same problem. In contrast, our simulation could overestimate the correlated [CII] fluctuations, if the SFR-L$_{\rm [CII]}$ breaks in low mass galaxies due to the low metallicity. So far, the observations of lensed low-SFR galaxies obtained contrasted results on this question. Some studies found a clear deficit \citep{Knudsen2016,Bradac2017}, while \citet{Fujimoto2021} did not find any evidence for it.


\section{Contribution of the various astrophysical components to CONCERTO power spectra}

\label{sect:power_spectra}

Since our simulation reproduces the observed line luminosity functions (Sect.\,\ref{sect:LFs}), the observed dust continuum statistical properties \citep{Bethermin2017}, and the anisotropies of the cosmic infrared background \citep[][Gkogkou et al. in prep.]{Bethermin2017}, we use it to forecast the power spectra of the various lines at various redshifts.

\subsection{Computation of power spectra per frequency slices and link with 3D power spectra}

\label{sect:Pk_2d_3d}

In Sect.\,\ref{sect:Pk3D}, we compared the 3-dimensional power spectra of various models. In this section, we will instead use angular power spectra of 1\,GHz frequency slices. We simply calculate the angular power spectrum of each 1\,GHz spectral cube slice using the \textit{powspec} package\footnote{Public code by Alexandre Beelen hosted at \url{https://zenodo.org/record/4507624}}. Working in angular wavenumber and frequency is closer from the data that the instruments will produce.  It also allows us to produce figures, which are independent of the choice of a reference line for the projection into the 3-dimensional physical space. Finally, the low spectral resolution of CONCERTO will not allow us to probe the small scales in the radial direction. Projected to the physical space, the angular resolution (22\,arcsec at 305\,GHz) and the frequency resolution are very different. For instance, for [CII] at z=6, the transverse resolution is 0.9 comoving Mpc, while it is only 11 comoving Mpc in the radial direction.

However, we note that formalisms were developed to reproject another line acting as a contaminant on the 3-dimensional physical power spectrum of a given reference line \citep{Gong2014,Lidz2016,Breysse2021}. For two different lines, the same frequency interval corresponds to different width in the radial direction. Similarly, the same angular scale is associated to different physical scales in the transverse direction. However, for the interlopers, the radial and transverse distorsions have no reason to be the same and the power spectrum is thus anisotropic. This effect could be used to separate the signal from line interlopers from the targeted line. However, \citet{Lidz2016} showed that this decomposition requires high signal-to-noise ratios, which will not be achieved by first-generation experiments as CONCERTO. For simplicity, we do not consider anisotropies in this paper focused on CONCERTO.

In practice, the angular power spectrum $P_{\rm 2D}$ is approximately linked to the 3-dimensional power spectrum $P_{\rm 3D}$ by the following formula \citep{Neben2017,Yue2019}:
\begin{equation}
\label{eq:2d_to_3d}
P_{\rm 2D}(k_\theta) = \frac{1}{D_C^2 \, \Delta D_C} P_{\rm 3D} \left ( k_C = \frac{k_\theta}{D_C} \right ),
\end{equation} 
where $P_{\rm 2D}(k_\theta)$ is the angular power spectrum of the frequency slice, $D_C$ is the distance to the slice in comoving units, $P_{\rm 3D}(k_C = \frac{k_\theta}{D_C})$ is the 3-dimension power spectrum at the redshift associated to the frequency slice, and $\Delta D_C$ is the thickness of the slice in the same units:
\begin{equation}
\Delta D_C = \frac{c \,(1+z)\,\Delta \nu_{\rm obs}}{H(z) \nu_{\rm obs}},
\end{equation}
where H(z) is the Hubble parameter at a redshift $z$. Using this formula, we assume implicitly that  $\Delta D_C \ll D_C$ and that $P_{\rm 3D}$ does not vary significantly across the slice. To evaluate the accuracy of this approximation, we compared the 3-dimensional power spectrum derived at z=6  from the comoving cubes (see Sect.\,\ref{sect:Pk3D}) with the angular power spectrum derived from the simulated observed cubes. We found a maximal difference of 10\,\%, which is negligible compared to the various calibration uncertainties on the model (e.g., CO and [CII] scaling relations, stellar mass function, evolution of the main sequence and its scatter).

\begin{figure*}
\centering
\begin{tabular}{cc}
\includegraphics[width=9cm]{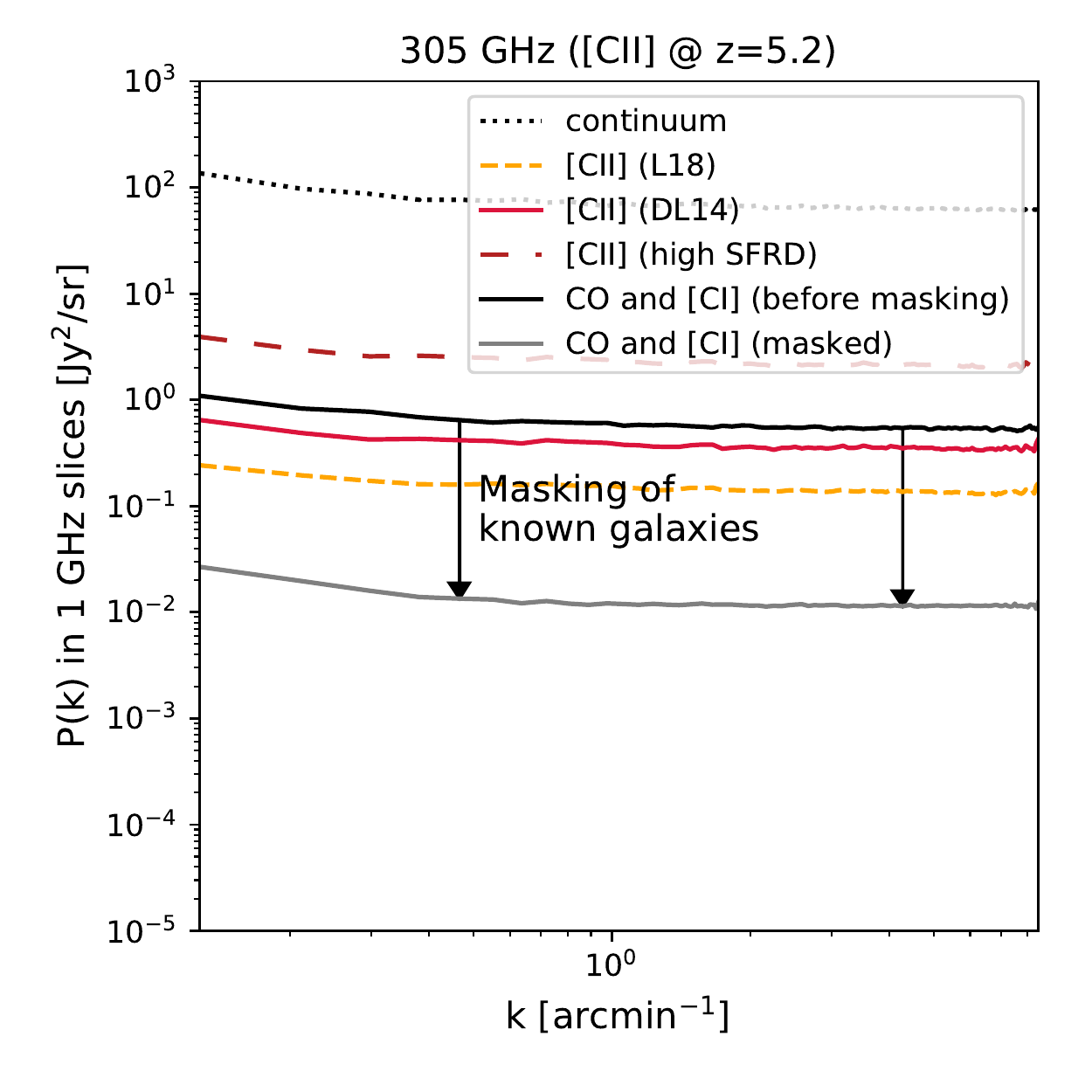} & \includegraphics[width=9cm]{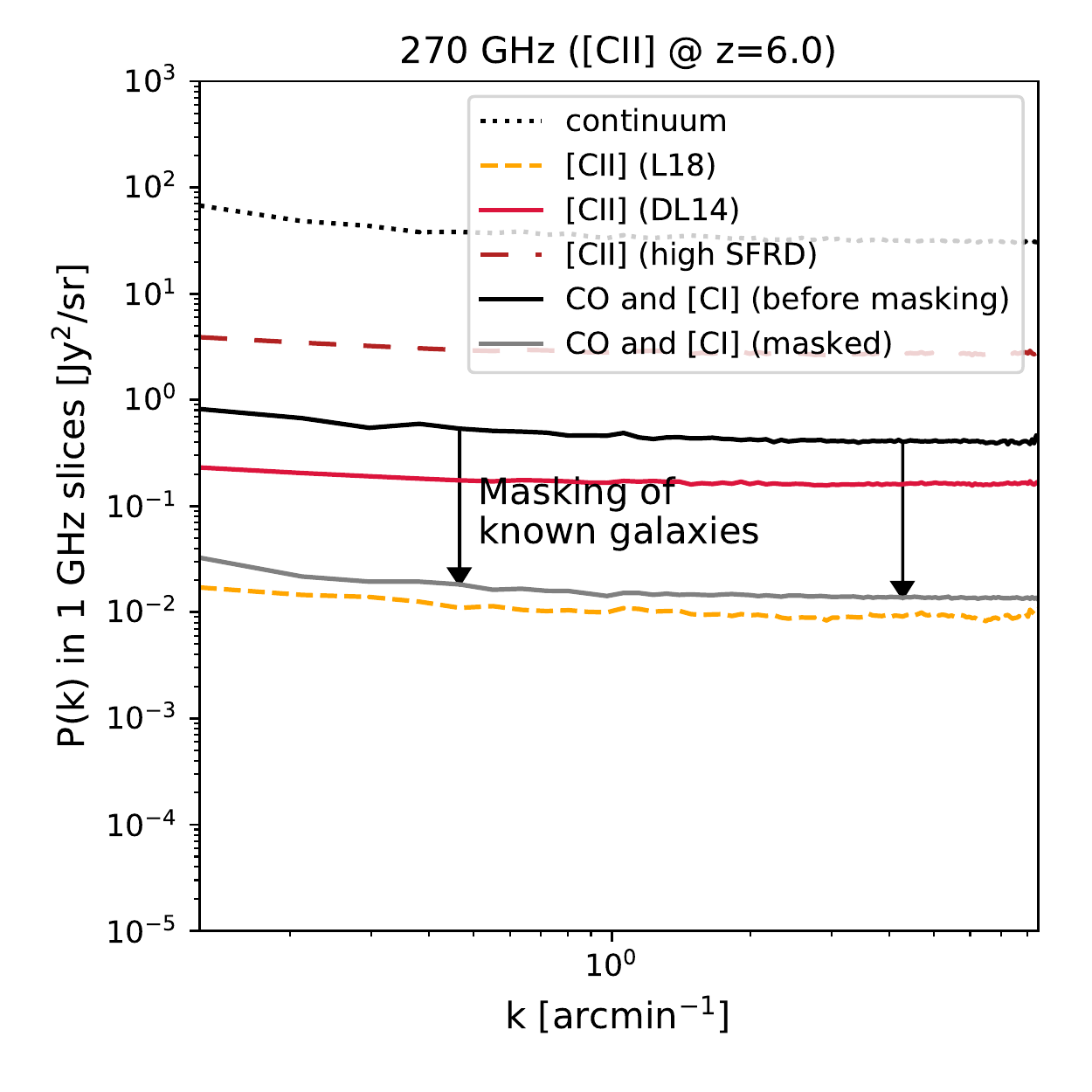}\\
\includegraphics[width=9cm]{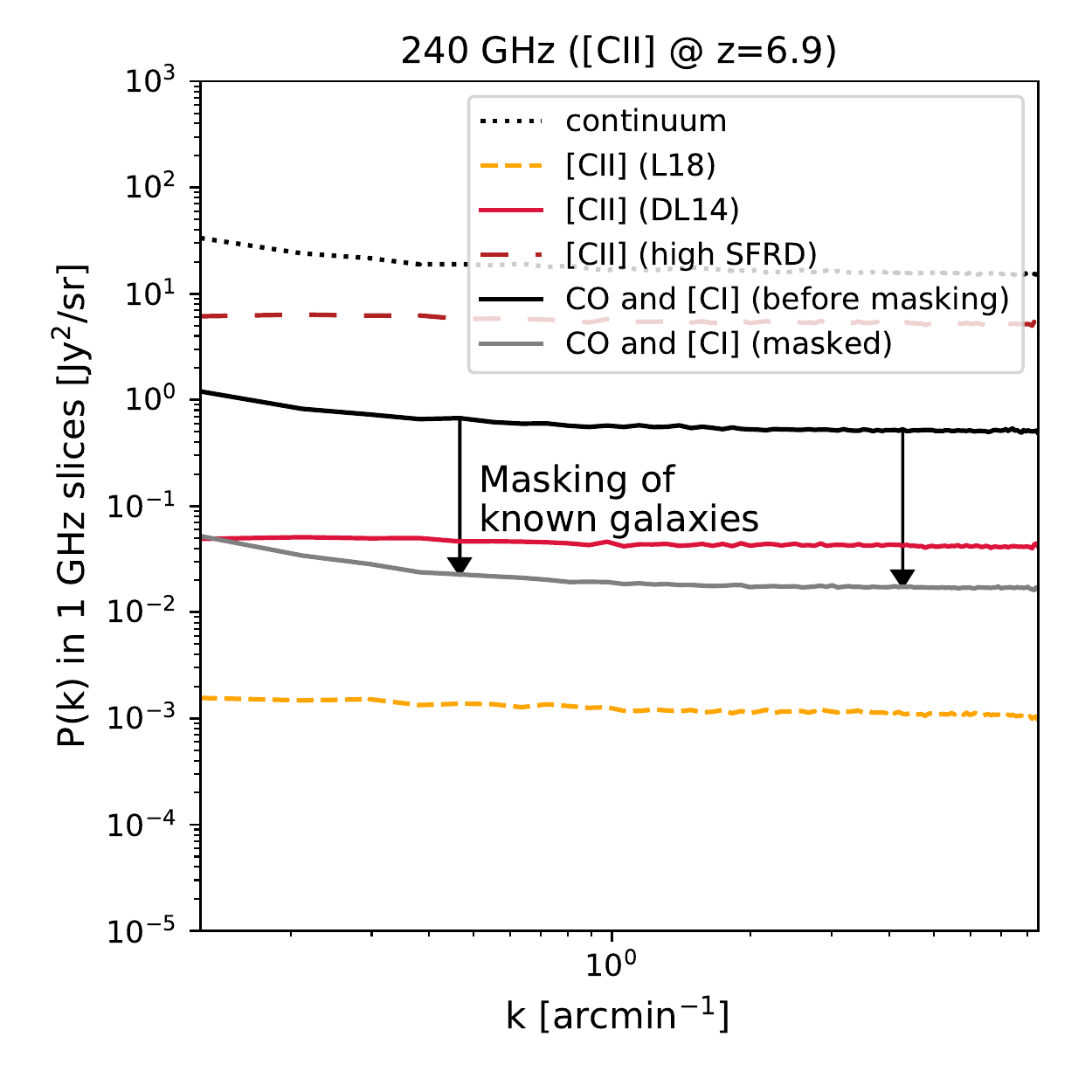} & \includegraphics[width=9cm]{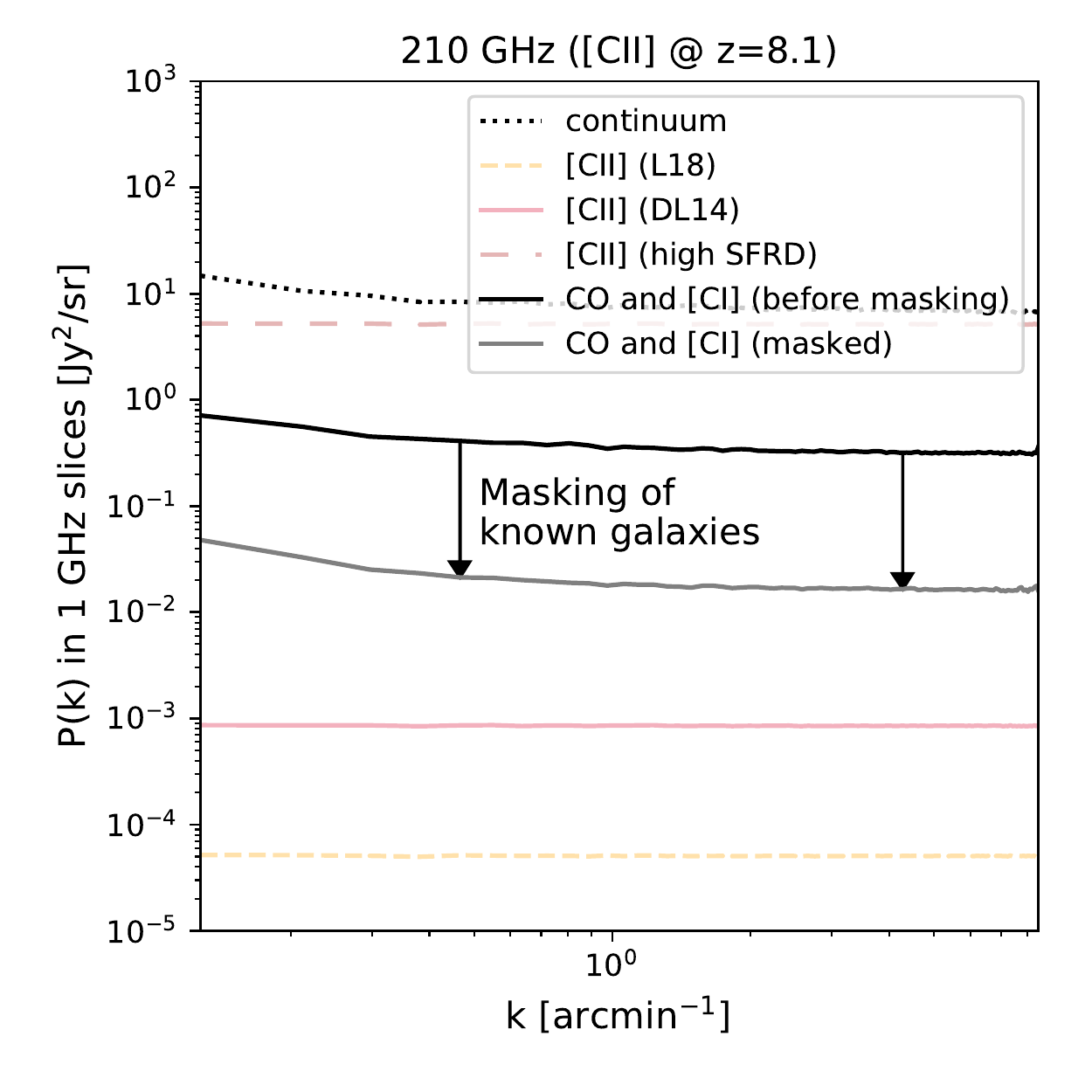}\\
\end{tabular}
\caption{\label{fig:Pk} Contribution of the various extragalactic components to the CONCERTO angular power spectra at 305\,GHz (upper left panel), 270\,GHz (upper right panel), 240\,GHz (lower left panel), and 210\,GHz (lower right panel) predicted by SIDES. As discussed in Sect.\,\ref{sect:discussion_models}, the [CII] power spectra at 210\,GHz (z=8.1) could be significantly underestimated by our model. To minimize the effect of the field-to-field variance, we computed the average of the 11 power power spectra corresponding to the 1\,GHz-wide slices placed between 5\,GHz below and above the central slice (except at 305\,GHz for which we have no slice at higher frequency). The continuum is represented by dotted line. The solid red and orange dashed lines show the [CII] contribution in SIDES assuming the DL14 and and the L18 relations, respectively. The long-dash brown lines represents the high SFRD variant of the model (see Sect.\,\ref{sect:SFRD}). The black and grey lines show the contribution from both CO and [CI] before and after removing the known galaxies from surveys (see Sect.\,\ref{sect:masking}).} 
\end{figure*}

\subsection{Power spectra at various frequencies}

\label{sect:Pk}

We computed the angular power spectrum of various astrophysical components at 305\,GHz, 270\,GHz, 240\,GHz, amd 210\,GHz, which corresponds to [CII] at a redshift of 5.2, 6.0, 6.9, and 8.1, respectively. At 210\,GHz, the [CII] forecast from our model should be taken with caution as discussed in Sect.\,\ref{sect:discussion_models}. The results are presented in Fig.\,\ref{fig:Pk}. Similarly to the 3-dimensional power spectra (Sect.\,\ref{sect:Pk3D}), we can see for all the components a Poisson plateau at small scales (large k) and an excess of power at large scales (small k) coming from the galaxy clustering. However, the relative contribution from the clustering decreases with increasing $z$ for [CII] and is very weak at z=8.1 (210\,GHz).

At all frequencies, the continuum (CIB) is the dominant component by more than one order of magnitude. However, this component should not be a too serious problem for intensity mapping experiment, since it should be easy to subtract due to its smooth dependence with frequency. Techniques were already developed for the 21\,cm survey \citep[e.g.,][]{Wang2006,Jelic2008,Alonso2014} and could be adapted to [CII] intensity mapping. To analyse the results of the mmIME experiment (see Sect.\,\ref{sect:mmIME}), \citet{Keating2020} ignored the modes affected by the continuum, since the continuum have a very different behavior in the radial and traverse direction because of its smooth frequency dependance. The continuum can also be subtracted directly in the coordinates-frequency cubes taking also advantage of the smoothness versus frequency (Van Cuyck et al. in prep.). The continuum component has thus very convenient properties, which should allow us to subtract it with minimal residuals. However, because of its intrinsic brightness, any systematic effects or residuals in its subtraction could have a significant impact on the line measurements.

At 305\,GHz (the highest frequency reachable by CONCERTO), the power spectrum of CO and [CI] combined (black solid line) is slightly higher than the DL14 version of [CII]. The L18 version is lower than the DL14 version by a factor of $\sim$2 and the high-SFRD model is higher by a factor of $\sim$5. At lower frequency (higher [CII] redshift), the level of the [CII] power spectrum is lower than the sum of the CO and [CI] contributions. The level of the L18 version decreases quicker with increasing redshift, since the normalization of the L$_{\rm [CII]}$-SFR relation from L18 decreases with increasing redshift.

To visualize the variation with frequency of each of these contributions, we computed for each slice the mean level of the power spectra between 0.15 and 0.35\, arcmin$^{-1}$ for the large scales and between 5 and 7\, arcmin$^{-1}$ for the Poisson. These results are interpreted in Sect.\,\ref{sect:Pk_freq}, \ref{sect:co_contrib}, and \ref{sect:ci_contrib}. The angular power spectrum is dependent on the choice of the spectral resolution (see appendix\,\ref{app:CII_poisson}). In the following sections, we normalize the Poisson component by the frequency width of the slice following Eq.\,\ref{eq:norm_poisson} to make it independent on the spectral grid. The corresponding unit is the Jy$^2$sr$^{-1}$GHz. This does not apply to large scales, since we cannot assume that several frequency slices are fully independent due to the large scale clustering.

\begin{figure*}
\centering
\begin{tabular}{cc}
\includegraphics[width=9cm]{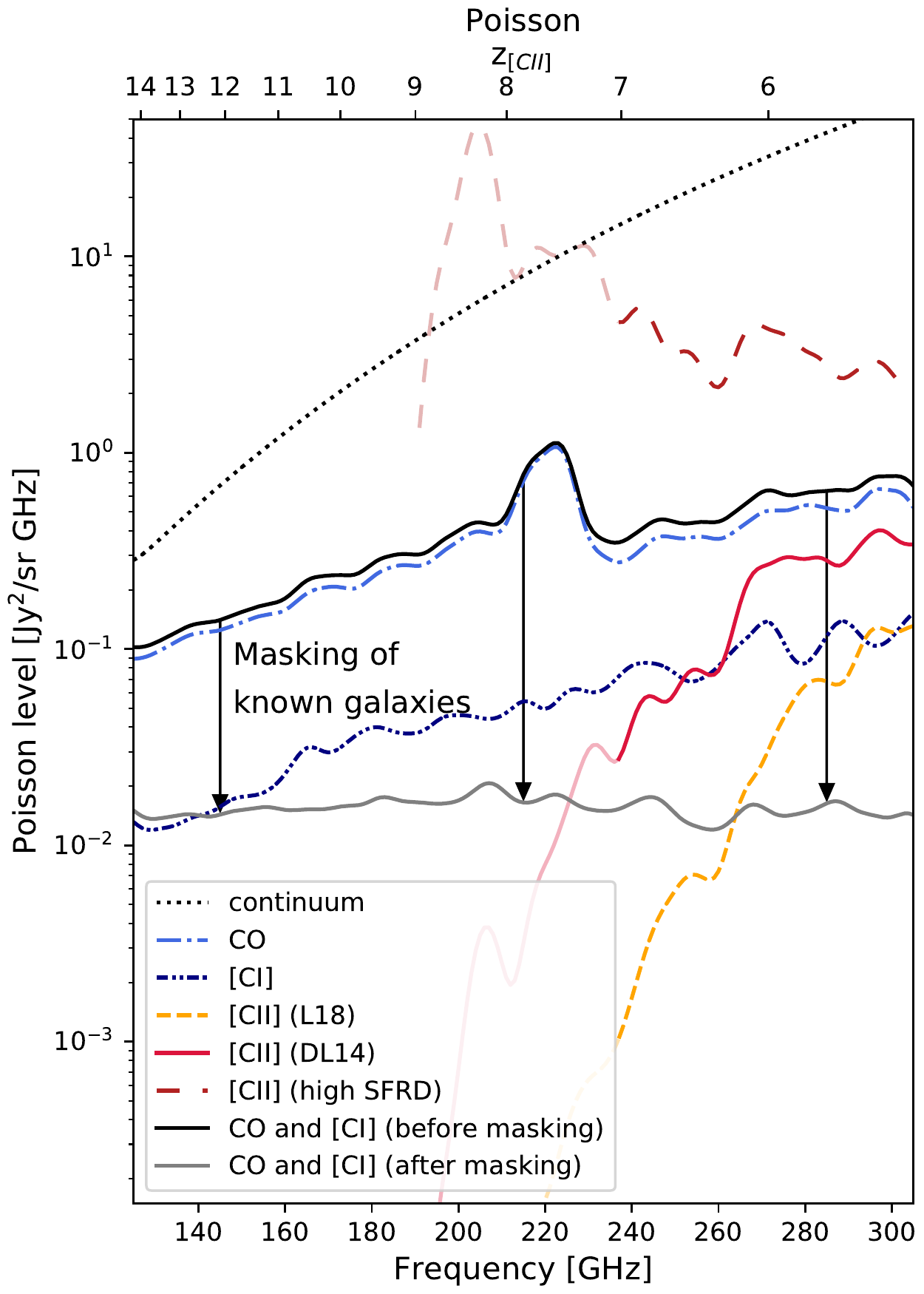} & \includegraphics[width=9cm]{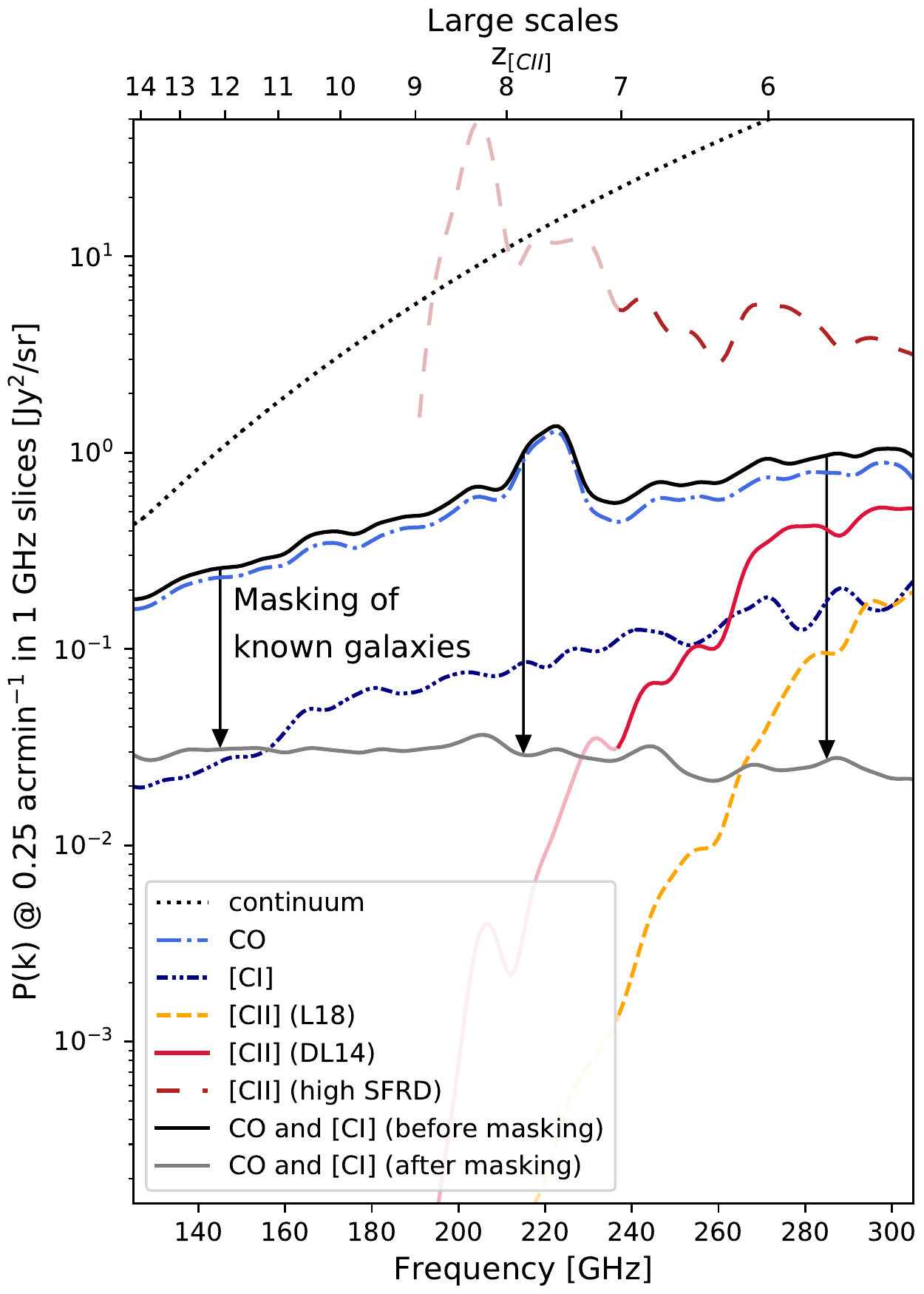}\\
\end{tabular}
\caption{\label{fig:Pk_freq} Level of the power spectrum of the various components in SIDES as a function of the frequency. The left panel shows the small scales dominated by the shot noise and the right panel shows the large scales. The black dotted line is the continuum computed using the standard model (low SFRD).  The red solid, the brown long-dashed  and the orange short-dashed  lines are the SIDES predictions for [CII] using the DL18, high SFRD, and L18 versions, respectively. As discussed in Sect.\,\ref{sect:discussion_models}, the [CII] power spectra at z$>$7 may be underestimated. We thus used a paler color to illustrate it. The light blue dot-dashed and the dark blue dash-dot-dot lines corresponds to all the CO and [CI] lines, respectively. The solid black is the total contribution of all the CO and [CI] lines, while the solid grey line is the same after masking galaxies known from galaxy surveys (see Sect.\,\ref{sect:masking}). For a better visualization, we smoothed the curves by a Gaussian kernel with a $\sigma$ of 3\,GHz.}
\end{figure*}

\subsection{Contribution of the main extragalactic lines as a function of frequency}

\label{sect:Pk_freq}

In Fig.\,\ref{fig:Pk_freq}, we present the contribution of the various astrophysical components as a function of frequency. Because of the small comoving volume associated to a frequency slice ($\sim 60^3$\,Mpc$^3$), there are large fluctuations between neighboring spectral slices. To obtain a better visualization, we smoothed the curves by a Gaussian kernel with a $\sigma$ of 3\,GHz. As expected, the continuum dominates at all frequencies, except if we have an unlikely flat SFRD scenario up to z$\sim$8 (high SFRD model). However, the continuum should not be too problematic to subtract from the cubes because of its smoothness versus frequencies \citep[e.g.,][]{Yue2015}. The CO-versus-continuum ratio varies with frequency with a higher ratio at lower frequency. This is expected, since the continuum increases strongly with increasing frequency, while the CO line flux does not increase as much from low-J to high-J transitions. 

[CI] contributes 5-10 times less to the power spectra than CO. It is not a surprise, since there are less [CI] transitions and they are usually slightly fainter. Both species have a shallow frequency dependance with more contribution at higher frequency. Finally, we can observe a strong spike for CO just below 230\,GHz. These frequencies correspond to low-z galaxies seen in CO(2-1) (see Sect.\,\ref{sect:co_contrib}). Since the Poisson fluctuations are proportional to the flux squared (see appendix\,\ref{app:CII_poisson}), a couple of bright nearby sources can have a major contribution to the power spectrum at this frequency. We can also observe a weaker $\sim$230\,GHz bump at large scales, where we observe the sum of the correlated and the Poisson fluctuations. Its amplitude compared to the baseline (in linear units) is similar to what is observed for the Poisson, suggesting that it comes mainly from the shot noise.

The [CII] forecast depends strongly on the assumptions of the simulation. For the DL14 and L18 prescriptions, the signal increases strongly with frequencies, while it is rather flat for the high SFRD version assuming a flat SFRD at z$>$4. There is virtually no signal below $\sim$200\,GHz, since this corresponds to z$>$8.5 and thus very small SFRDs in the standard version of our model. However, these results should be taken with caution, since the halo mass limit of our simulation can have a strong impact on our results, as discussed in Sect.\,\ref{sect:discussion_models}. At 305\,GHz (z$=$5.2), the three versions diverge by less than a factor of 10. In contrast, at z$=8$, there are already four orders of magnitude between the L18 and the high SFRD model. This illustrates how uncertain the [CII] intensity mapping are at the highest redshift and how important observational constraints will be.

At 305\,GHz, for the DL14 version of SIDES, the [CII] is a factor of 2 lower than the sum of CO and [CI] at both small (Poisson) and large scales. The [CII] amplitude is about a factor of five lower than the sum of CO and [CI] for the L18 version. The [CI] is always an order of magnitude lower than the two other species. Below 270\,GHz (above z$=$6.0), the [CII] decreases rapidly with decreasing frequency (or increasing redshift). At 250\,GHz (z$=$6.6), [CII] is already an order of magnitude below the the CO in the DL14 version of the model. This is low but more optimistic for intensity mapping experiments than the $<$1\,\% contribution of [CII] to the line background at 250\,GHz estimated empirically by \citet{Decarli2020}. These results demonstrate how crucial the accurate cleaning of the CO contribution will be if the SFRD is not flat at high redshift.

\begin{figure*}
\centering
\begin{tabular}{cc}
\includegraphics[width=9cm]{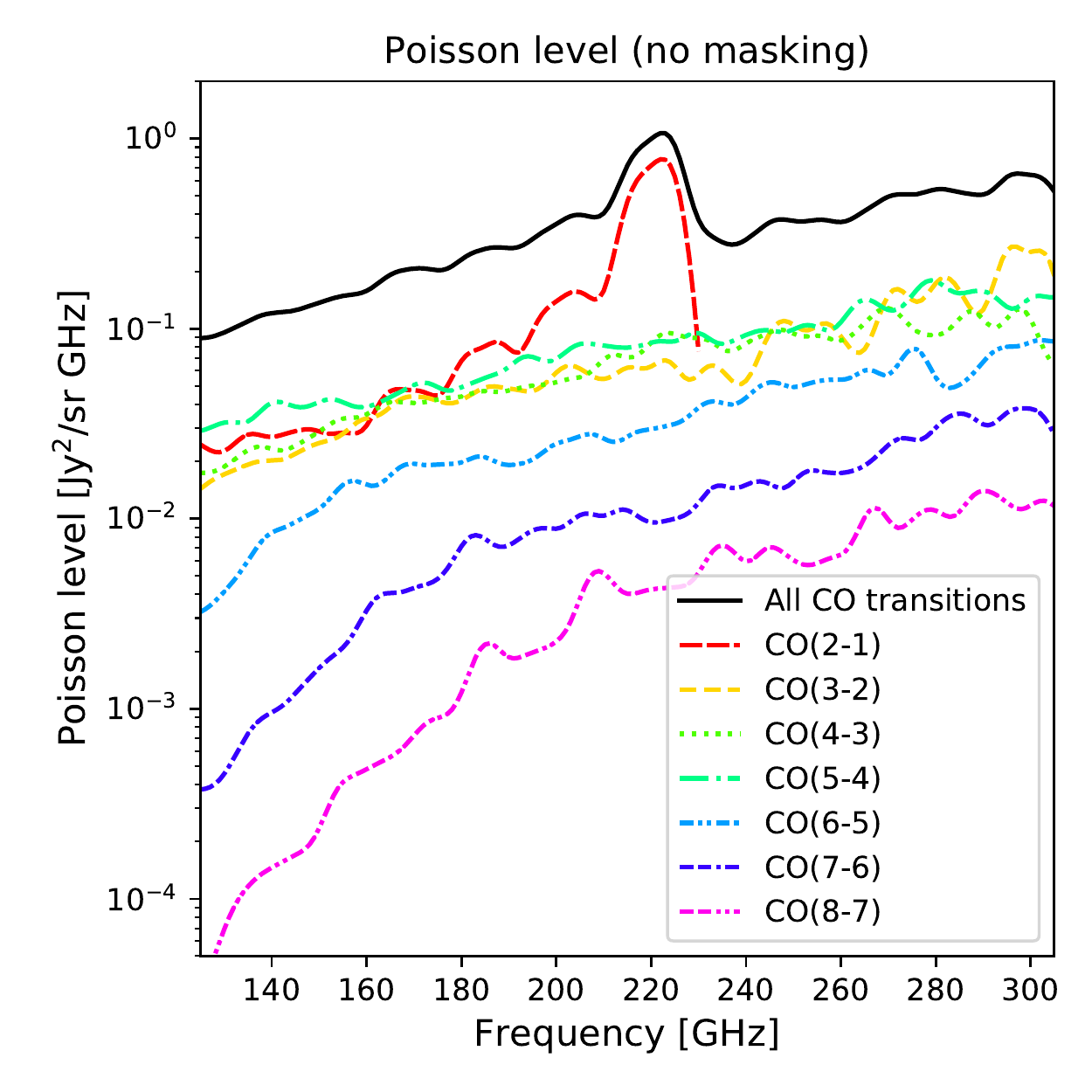} & \includegraphics[width=9cm]{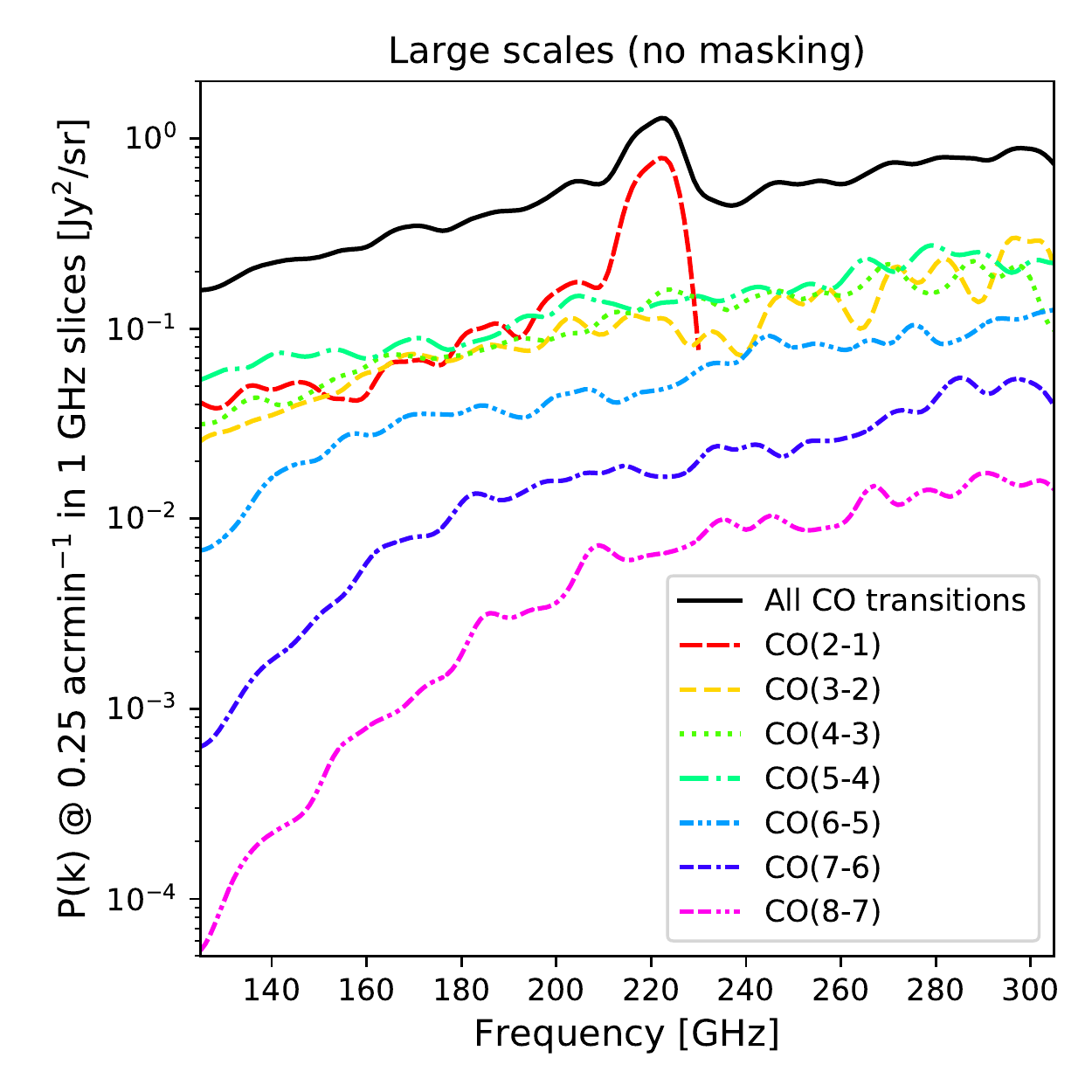}\\
\includegraphics[width=9cm]{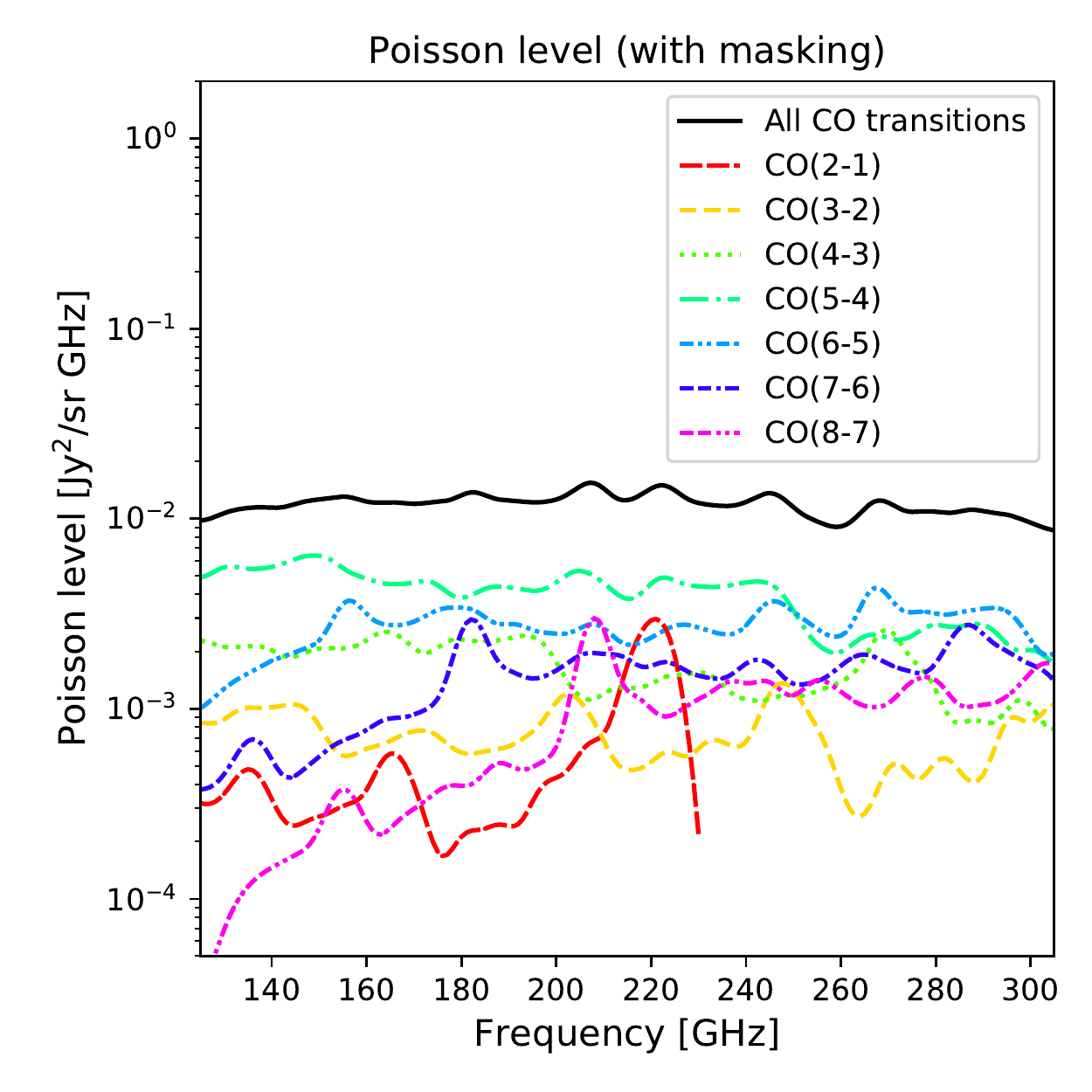} & \includegraphics[width=9cm]{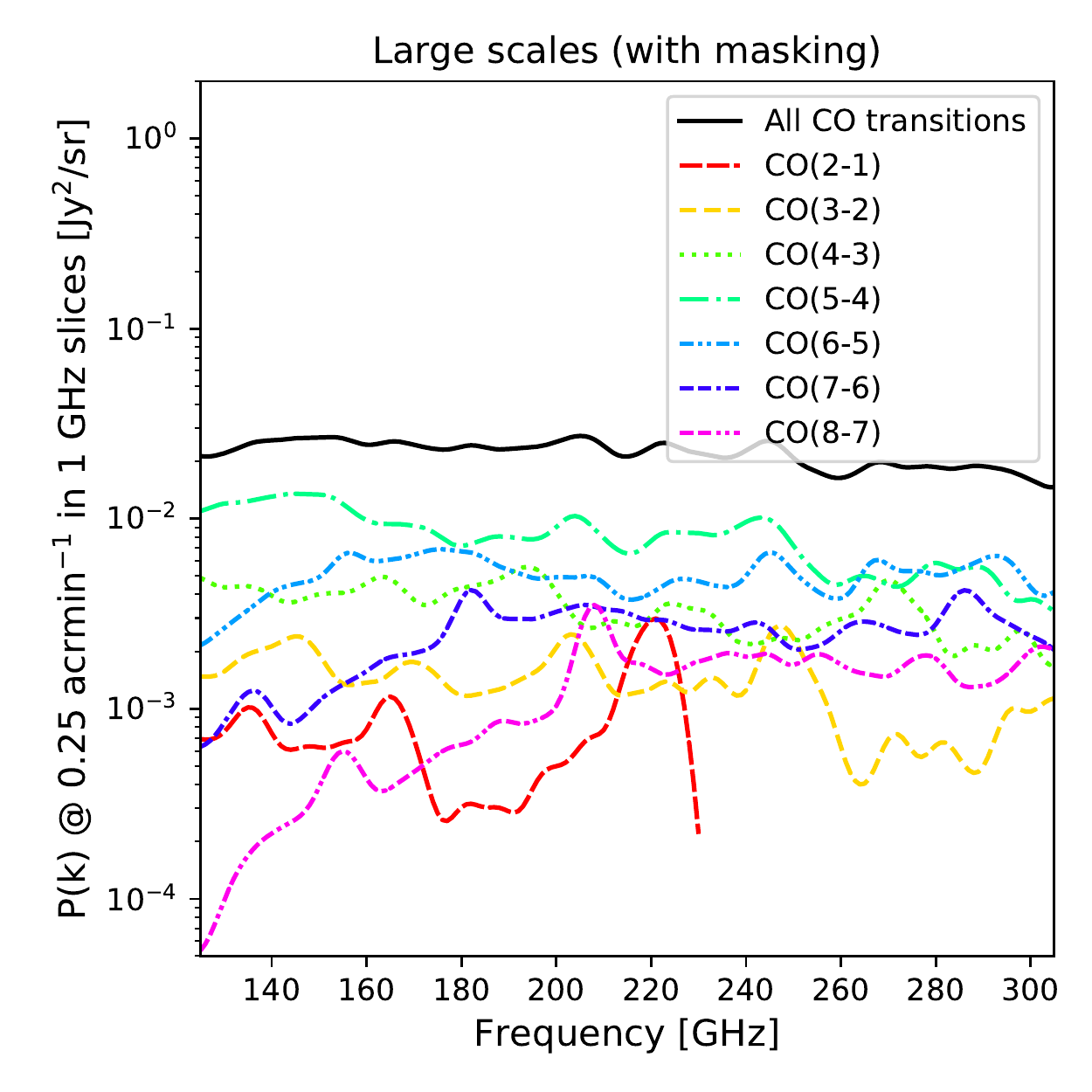}\\
\end{tabular}
\caption{\label{fig:co_contrib}Contribution of the various CO lines to the Poisson (left panels) and large-scale (right panels) to the power spectrum before (upper panels) and after (lower panels) masking the known galaxies from surveys. Together with the total CO (solid black line), we show the contribution of CO(2-1) (red long-dash line), CO(3-2) (orange short-dash line), CO(4-3) (green dotted line), CO(5-4) (turquoise long-dash-dot line), CO(6-5) (light blue two-dot-dash line), CO(7-6) (dark blue dot-two-dash line), and CO(8-7) (purple two-dash-two-dot line). For a better visualization, we smoothed the curves by a Gaussian kernel with a $\sigma$ of 3\,GHz.}
\end{figure*}

\begin{figure*}
\centering
\begin{tabular}{cc}
\includegraphics[width=9cm]{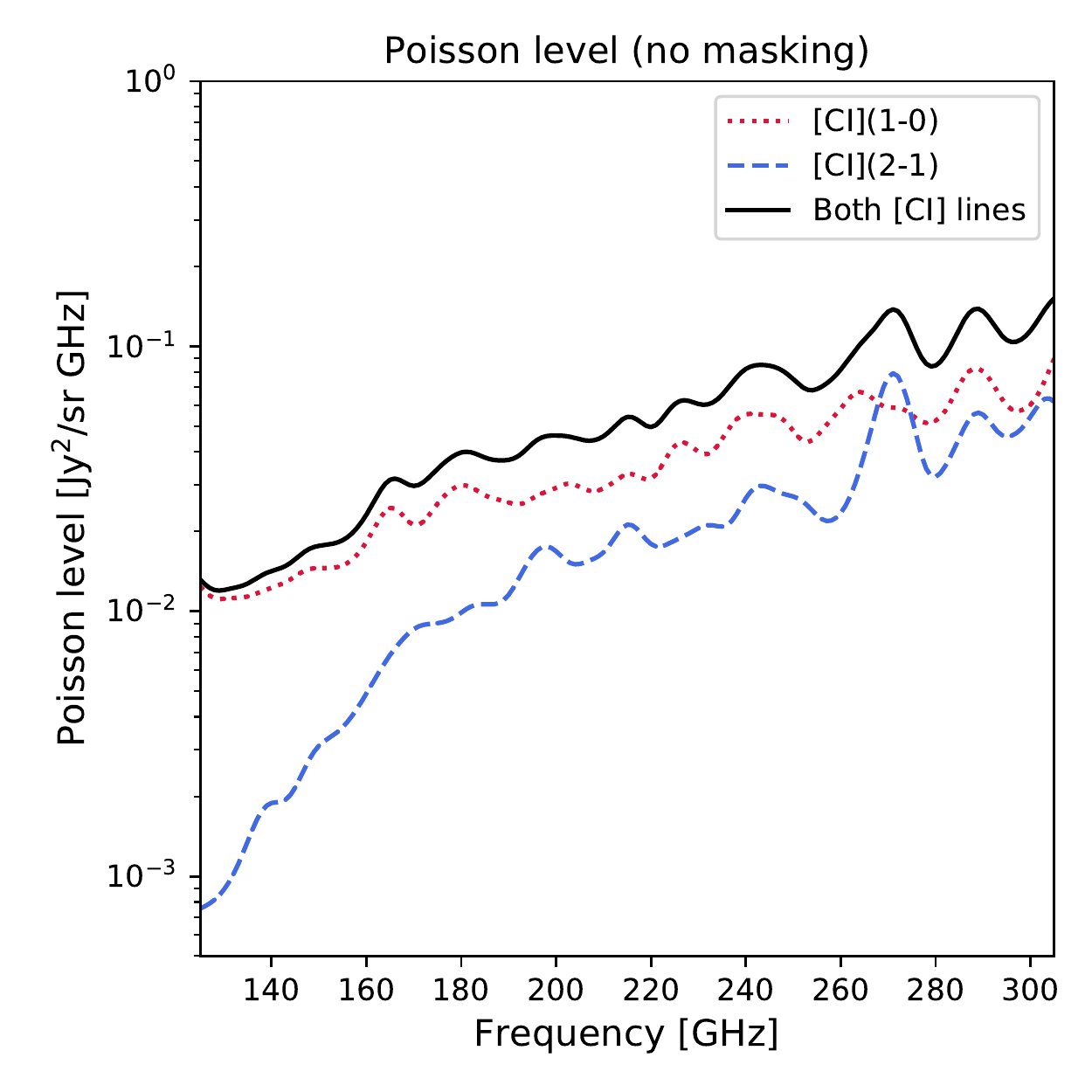} & \includegraphics[width=9cm]{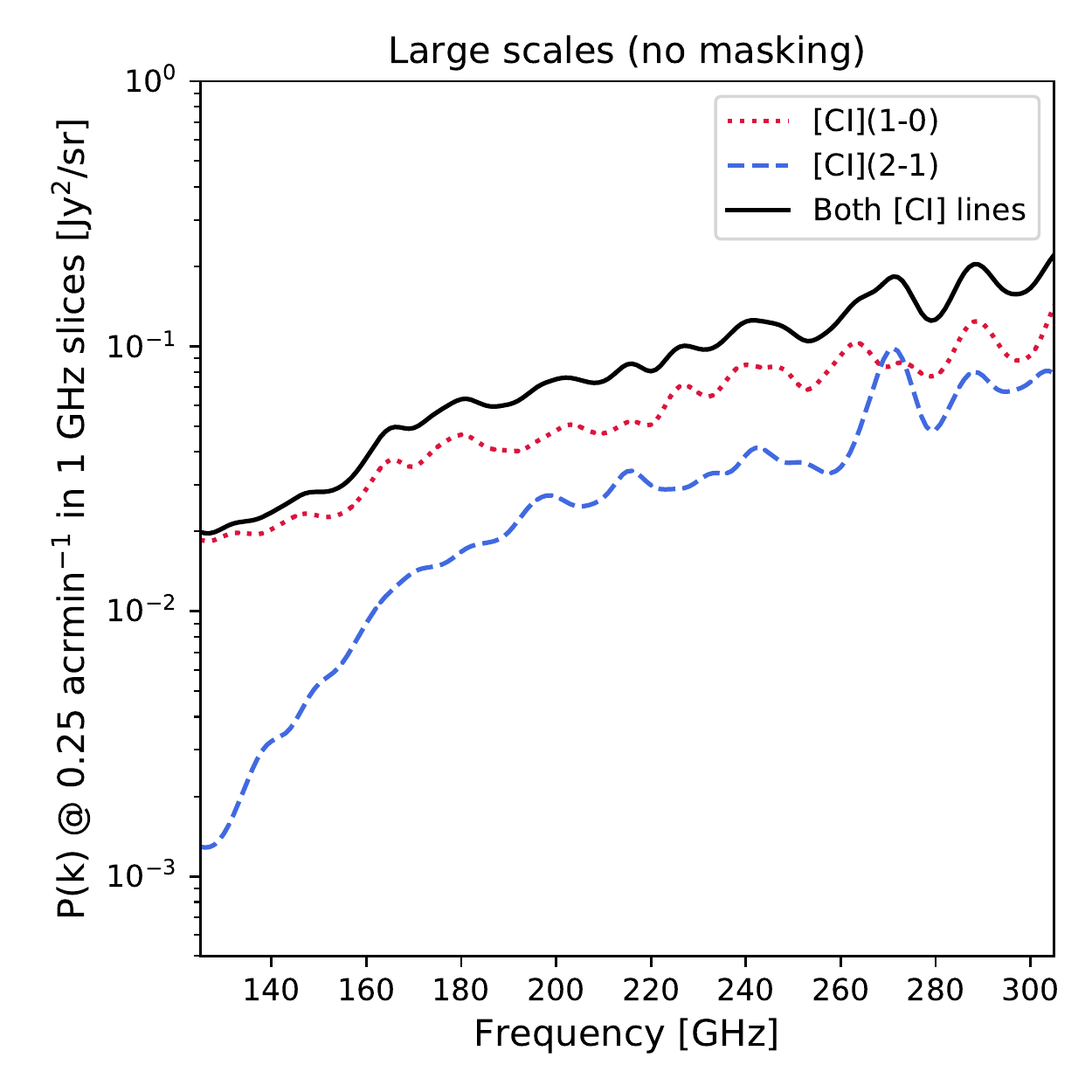}\\
\includegraphics[width=9cm]{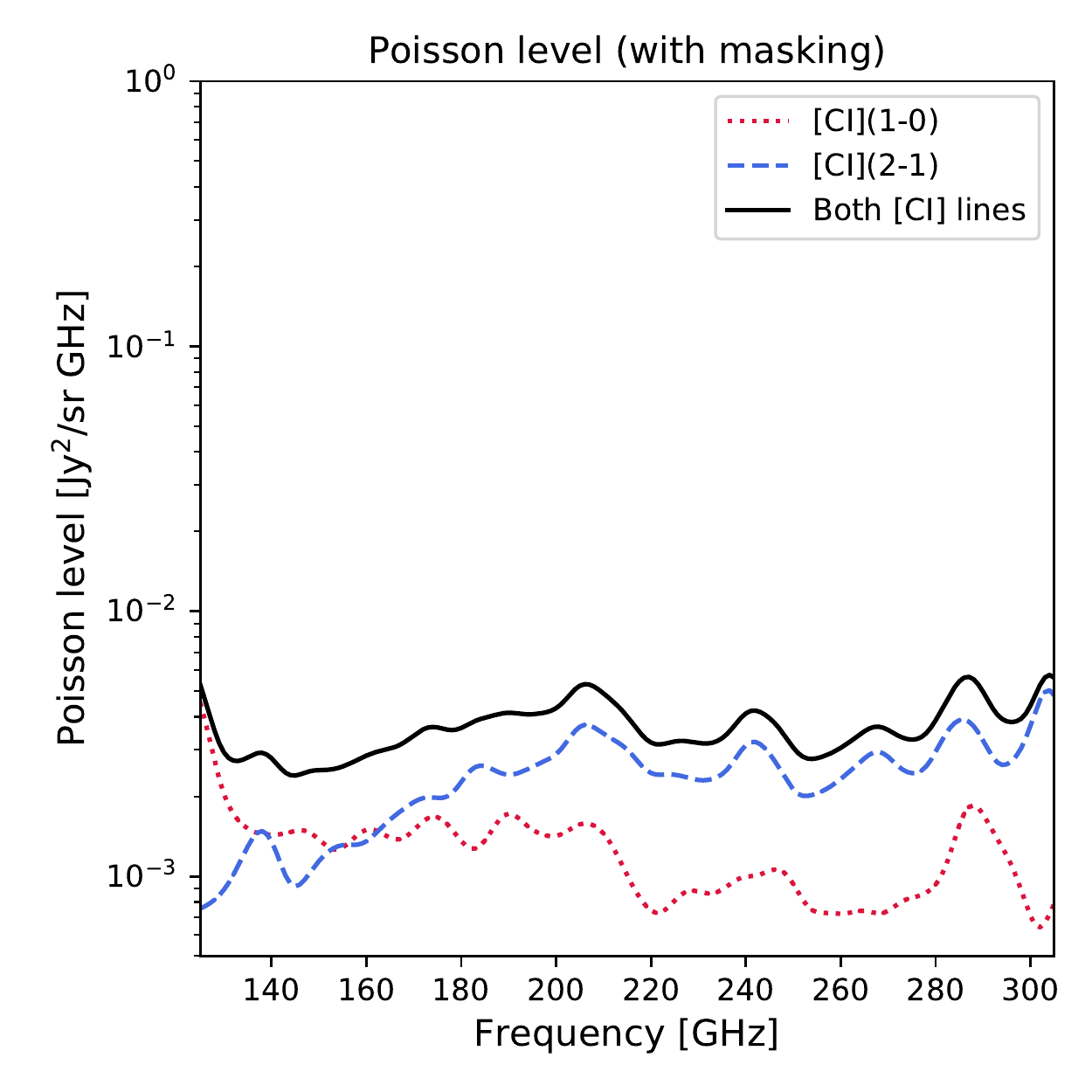} & \includegraphics[width=9cm]{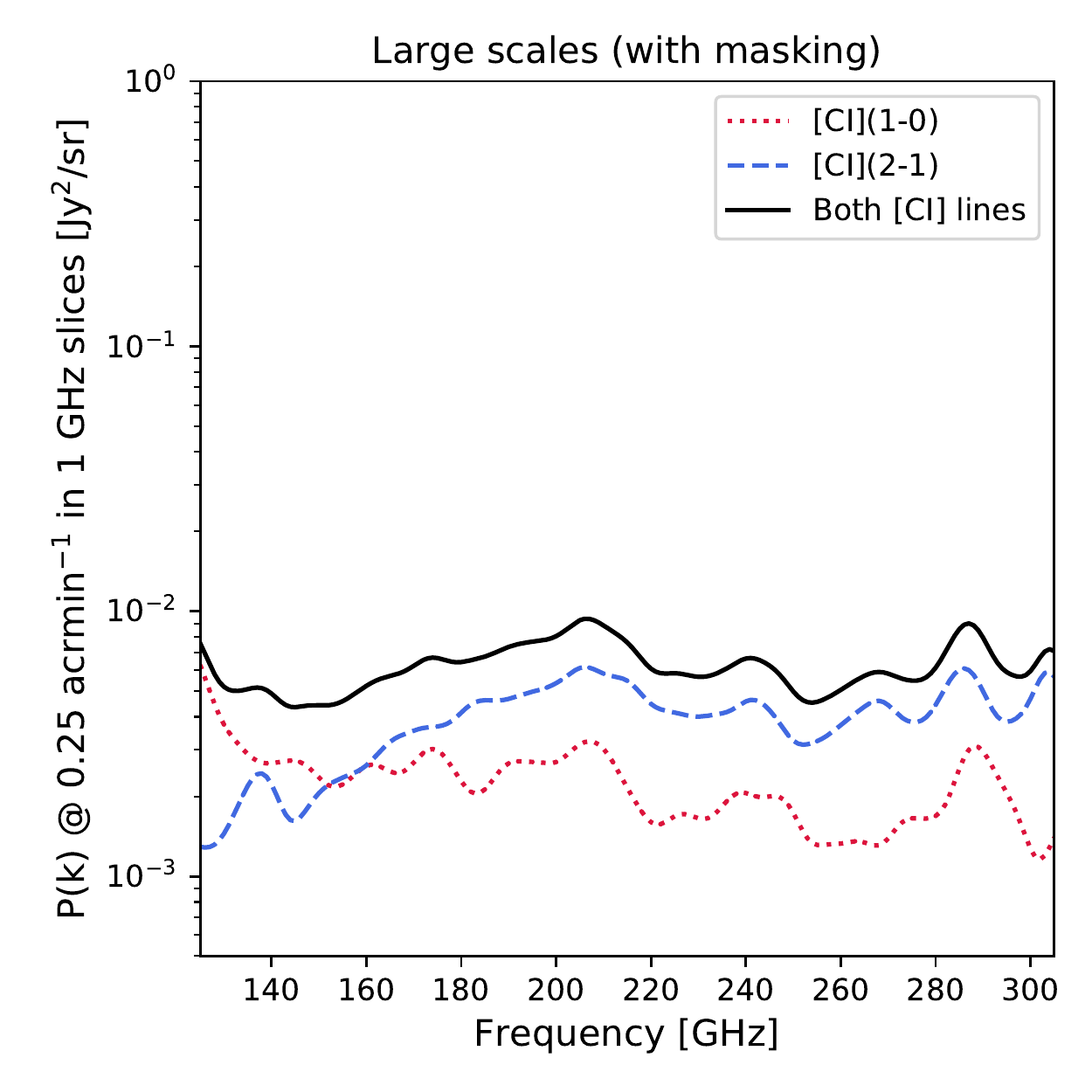}\\
\end{tabular}
\caption{\label{fig:ci_contrib} Contribution of the two [CI] transitions to the Poisson (left panels) and large-scale (right panels) to the power spectrum before (upper panels) and after (lower panels) masking the known galaxies from surveys. The black solid line represents the total, while the red dotted and the blue dashes lines corresponds to the [CI](1-0) and [CI](2-1) transitions, respectively. For a better visualization, we smoothed the curves by a Gaussian kernel with a $\sigma$ of 3\,GHz.}
\end{figure*}

\subsection{Effect of masking known sources}

\label{sect:masking}

\citet{Yue2015} and \citet{Sun2018} showed that the masking of the voxels associated to known galaxies from optical and near-infrared surveys could allow us to isolate the contribution of [CII]. To evaluate how powerful this technique could be, we produced new cubes injecting only galaxies, which would not be found by surveys at z$<$4. We use the stellar mass limits of the COSMOS catalog from \citet{Laigle2015} and consider that any galaxy above this stellar mass can be properly masked. This mass limit varies with redshift and we are removing sources down to much lower masses at lower redshift. This is indeed an approximation and it assumes that the redshift of the sources will be accurate enough to mask only the associated voxels and that no signal beyond the mask (PSF wings, map making artifacts) will contaminate the power spectra measured outside of the masked area. Our approach thus provides an upper limit on the efficiency of CO and [CI] masking technique. Detailed simulations of the full process will be presented in a future paper (Van Cuyck et al. in prep.).

As illustrated by Fig.\,\ref{fig:Pk_freq}, the masking reduces the level of the CO and [CI] contribution by more than one order of magnitude. At 305\,GHz (z$=$5.2), the residuals of CO and [CI] are a factor $\sim$25 below the [CII] signal predicted by DL14 version of SIDES (factor of $\sim$10 for the L18 version). The impact of these residuals increases rapidly with decreasing frequency. For the DL14 version, the CO and [CI] residuals reaches 20\,\% of the [CII] signal at 260\,GHz (z$=$6.3). The equality between [CII] and the residuals is reached at 230\,GHz (z$=$7.3). It suggests that the masking will not be sufficient beyond z$\sim$6.5 as already mentioned by \citet{Yue2015}, and more advanced techniques of decontamination will be necessary \citep[e.g.,][]{Cheng2020,CONCERTO2020}.

\subsection{Contribution of the different CO transitions}

\label{sect:co_contrib}

The CO is thus the main contaminant for [CII] intensity mapping. However, all the transitions do not contribute equally. In Fig.\,\ref{fig:co_contrib}, we show the contribution of each CO transition to the power spectrum at various frequencies. As discussed in Sect.\,\ref{sect:Pk_freq}, there is a large spike just below 230\,GHz. This figure confirms that it is caused by CO(2-1) at low redshift. If we do not mask the known galaxy populations, the CO power spectrum is dominated by the 2-1, 3-2, 4-3, and 5-4 transition above 230\,GHz. Between 200 and 230\,GHz, the signal comes mainly from CO(2-1) at low redshift. Below 200\,GHz, all the transitions between 2-1 and 5-4 have similar contributions. The contribution of higher-J transitions decreases strongly with increasing J.

The results are very different after masking. The main contributors are then the 5-4 and 6-5 transitions. Since the galaxy catalogs reach lower masses at lower redshift, most of the low-z signal is removed, while high-z intermediate mass galaxies are still contributing. The low-z CO(2-1) peak has also a much lower amplitude relatively to the other transition and does not impact significantly the total CO power spectra. We can also notice a small spike in CO(6-5) at 155\,GHz, CO(7-6) at 180\,GHz, and CO(8-7) at 210\,GHz. This is likely caused by the same overdensity around z$\sim$4.4 in the SIDES cube.

Finally, we can notice that the CO(8-7) transition has a non negligible contribution at high frequency after masking. High-J transitions could thus play an important roles after masking. These transitions are less known, especially in low-mass galaxies at high redshift, and an hypothetical population of low-mass galaxies with a particularly high CO excitation could have a significant impact on the CO residuals. However, this is disfavored by theoretical models, which predict a turnover of the SLED around J=7 for normal high-z galaxies \citep[e.g.,][]{Vallini2018}.

\subsection{Contribution of the [CI] lines}

\label{sect:ci_contrib}

The two [CI] transitions also contribute to the power spectrum. In Fig.\,\ref{fig:ci_contrib}, we present [CI] power spectra levels as a function of frequency. Before masking, the [CI]1-0 and 2-1 transitions have similar contributions above 270\,GHz. Between 180 and 270\,GHz, the ratio between [CI](2-1) and [CI](1-0) decreases mildly with decreasing frequency (increasing redshift). This is expected, since [CI]2-1 is probed between z=2 and z=3.5, where the SFRD is almost flat, while [CI]1-0 is probed between z=0.8 and 1.7, where it has a mild increase. Below 180\,GHz, [CI]2-1 drops sharply, since it corresponds to z$>$3.5, where SFRD decreases quickly with increasing redshift.

Similarly with CO, the relative contribution of the various transitions is strongly affected by the masking. The [CI]2-1 transition is dominant down to 150\,GHz. This can be easily explained by [CI]2-1 being emitted at higher redshift and thus less affected by the masking. 

\begin{figure}
\centering
\includegraphics[width=8cm]{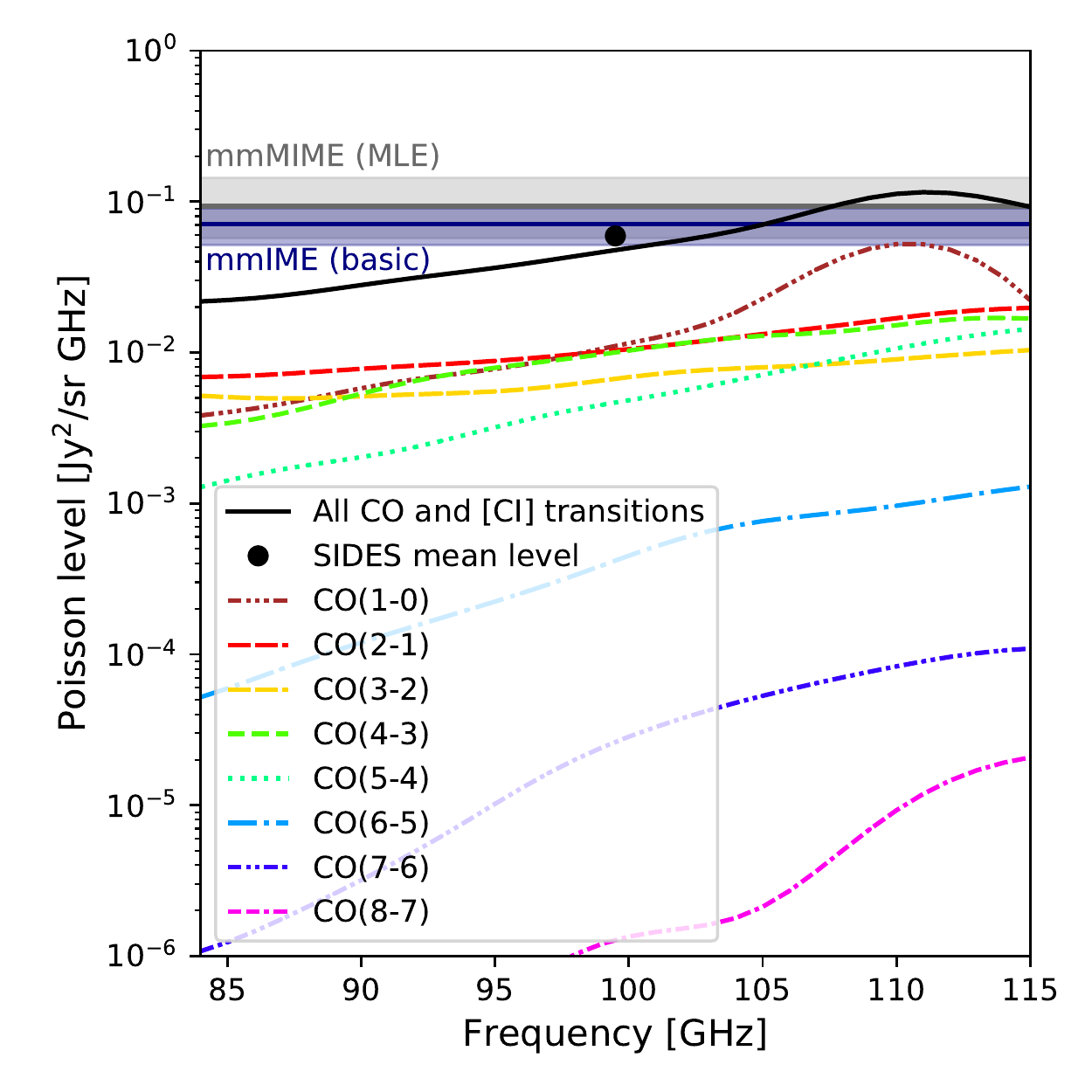}
\caption{\label{fig:mmIME} Shot noise level predicted by SIDES as a function of frequency and comparison with the mmIME measurements \citep[][20\,arcmin$^2$ in total, horizontal dark grey line and associated 1\,$\sigma$ area for the MLE method and similar dark blue line and area for the simpler instrumental noise correction]{Keating2020}. The frequency range shown in this figure correspond to what mmIME used. The black filled circle is the average shot noise predicted by SIDES in this range. The solid black line is the total of CO and [CI]. The various colored lines captioned directly in the figures indicate the contribution of the various CO transitions.}
\end{figure}

\subsection{Comparison with the mmIME measurements}

\label{sect:mmIME}

Millimeter intensity mapping is an emerging technique and so far very few measurements have been performed. First pioneering results have been obtained by the millimeter-wave intensity mapping experiment \citep[mmIME,][]{Keating2020} using both the results of the ASPECS survey and a dedicated Atacama compact array (ACA) survey. They managed to obtain a 2.5\,$\sigma$ tentative detection of the shot noise around 100\,GHz. They subtracted the continuum signal from galaxies (CIB) by discarding the modes corresponding to smooth components on the spectral axis. They found 770$_{-210}^{+210}$\,$\mu$K$^2$\,Hz\,sr correcting only from the instrumental noise and 1010$_{-390}^{+550}$\,$\mu$K$^2$\,Hz\,sr using more complex maximum likelihood estimation (MLE). Since the SIDES simulation is in Jy units, we converted the mmIME results to more convenient unit (see appendix\,\ref{sect:mmIME_conversion}). We obtained 0.07$_{-0.02}^{+0.02}$\,Jy$^2$\,sr$^{-1}$\,GHz$^{-1}$ for the simple noise correction and 0.09$_{-0.04}^{+0.05}$\,Jy$^2$\,sr$^{-1}$\,GHz$^{-1}$ for the MLE. 

To compare SIDES with mmIME, we produced simulated cubes using the method described in Sect.\,\ref{sect:data_cubes} between 84 and 115\,GHz (the frequency range probed by mmIME). For each 1\,GHz element, we then computed the angular power spectrum of the slice using the \textit{powspec} package. The Poisson level is estimated averaging the k$>$5\,arcmin$^{-1}$ scales. The results are presented in Fig.\,\ref{fig:mmIME}. 

Above 100\,GHz, the shot noise is dominated by CO(1-0) at low redshift. This is not surprising, since the flux of low-z sources can be very high and the shot-noise is particularly impacted by bright sources contrary to the mean background ($\int L^2 \frac{dN}{dL} dL$ instead of $\int L \frac{dL}{dS} dL$, see appendix\,\ref{app:CII_bkg} and \ref{app:CII_poisson}). A similar behavior has already been identified for the redshift distribution of the CIB shot noise with both a low-redshift and a z$\sim$2 peak (e.g., \citealt{Bethermin2013}, Fig.\,12). In \citet{Keating2020}, they computed the CO(1-0) shot noise predicted by their model only up to $\sim$75\,GHz and did not investigate the possibility of a low-z second peak. Below 100\,GHz, there is a similar contribution from the 4 lowest CO transition. 

The total contribution of the CO and [CI] averaged over the mmIME frequency range (filled circle in Fig.\,\ref{fig:mmIME}) is 0.059\,Jy$^2$\,sr$^{-1}$\,GHz$^{-1}$, which agrees with mmIME measurements. Only four lines contribute to 10\,\% or more of the total shot noise: CO(1-0) with 33\,\%, CO(2-1) with 19\,\%, CO(3-2) with 12\,\%, and CO(4-3) with 17\,\%. The signal is thus mainly coming from CO emitters below z=4. The two [CI] lines contribute only to 7.4\,\% of the total shot noise and [CI](1-0) produces 98\,\% of it. Since the CO(1-0) contributes only to approximately one third of the signal, the model presented in \citet{Keating2020} should remain in the 1-$\sigma$ confidence region of their measurements if they add this contribution.


\section{Conclusion}

We presented a new model and its associated public code to generate submillimeter and millimeter intensity mapping simulated cubes from dark-matter lightcones. This new model produce realistic dust continuum SEDs evolving with redshift and with an intrinsic scatter in radiation-field intensity and thus temperature. The various CO transitions are generated using on the combination of two SLEDs (clump and diffuse medium). The fraction of each component is also linked to the radiation field. The CO SLED of galaxies thus evolves with z and have an intrinsic scatter. [CI] is generated using empirical relations. Finally, [CII] is generated using several prescriptions based on recent [CII] studies and includes a scatter. Our model thus includes a certain level of complexity to test future analysis methods without being too computationally demanding.

Our model successfully reproduce the various observed line luminosity functions of CO and [CI] at z$<4$. It also reproduce correctly the CO-dominated shot-noise signal measured around 100\,GHz by the mmIME intensity mapping precursor program. Finally, the version of our model using the DL14 prescription for the SFR-[CII] relation is compatible with the first constraints on the [CII] luminosity function at z$>$4 from ALPINE.

The forecast of the [CII] power spectra at z$=$6 can vary significantly depending on the model. We compiled recent models and found differences of up to 2.5 orders of magnitude between them. These differences can mainly be explained by the very different luminosity functions. Models with high number densities of bright sources have high shot-noise levels, while models with numerous faint populations tend to produce higher levels of correlated fluctuations. However, these discrepancies between models highlight how uncertain is the emissivity and the spatial distribution of [CII] in the high redshift Universe. The first generation of experiments will thus have a key role to either confirm or rule out the optimistic models.

In addition, our new simulation provides a detailed view of the contribution of the various extragalactic components as a function of frequency. The continuum is by far the dominant component and will have to be subtracted with a precision higher than the percent. At 305\,GHz (z=5.2), the CO is higher than [CII] in the versions of our model with a standard star formation history at high redshift. 

However, by subtracting the known galaxies from surveys, this contribution becomes lower than 20\,\% of the [CII]. The masking method could thus be suitable at z$\lesssim$6.5. At higher redshift (lower frequency), the [CII] level drops sharply, while CO is almost constant. We will need to develop more advanced methods to isolate [CII]. Our simulation will be ideal to test these future methods, since it does not contain the classic but simplified assumption of component separation methods as fixed CO SLED or dust continuum templates. With our simulation in input, we will be able to perform end-to-end simulations of the full process and to identify and correct the main biases of the analysis pipelines.

The new version of the SIDES simulation is thus a powerful tool to prepare and interpret future intensity mapping experiments. It can also be used for interferometric spectral scans and photometric survey (e.g., forecast of the number of detections, properties of the detections, end-to-end simulations). A companion paper (Gkogkou et al., in prep) will present a larger simulation and discuss the field-to-field variance on these various probes of galaxy evolution. The code and its products are publicly available at \url{https://cesamsi.lam.fr/instance/sides/home}. The code has been developed to be flexible and easy to modify to adapt to the needs of the community. 

\begin{acknowledgements}
We thanks the referee for their insightful comments. We warmly thank Bin Yue, Dongwoo Chung, Marco Viero, Gergö Popping, and Shengqi Yang for sharing their model predictions and discussing them. We thank Gareth Keating for suggesting to test our simulation on the mmIME results and his explanations. This project has received funding from the European Research Council (ERC) under the European Union’s Horizon 2020 research and innovation program (grant agreement No 788212) and from the Excellence Initiative of Aix-Marseille University-A*Midex, a French “Investissements d’Avenir” program. MA acknowledges support from FONDECYT grant 1211951, ANID+PCI+INSTITUTO MAX PLANCK DE ASTRONOMIA MPG 190030, ANID+PCI+REDES 190194 and ANID BASAL project FB210003. AP acknowledges support from the ERC Advanced Grant INTERSTELLAR H2020/740120.

\end{acknowledgements}

\bibliographystyle{aa}

\bibliography{biblio}

\begin{thebibliography}{167}
\expandafter\ifx\csname natexlab\endcsname\relax\def\natexlab#1{#1}\fi

\bibitem[{{Alaghband-Zadeh} {et~al.}(2013){Alaghband-Zadeh}, {Chapman},
  {Swinbank}, {Smail}, {Danielson}, {Decarli}, {Ivison}, {Meijerink}, {Weiss},
  \& {van der Werf}}]{Alaghband_Zadeh2013}
{Alaghband-Zadeh}, S., {Chapman}, S.~C., {Swinbank}, A.~M., {et~al.} 2013,
  \mnras, 435, 1493

\bibitem[{{Alonso} {et~al.}(2014){Alonso}, {Ferreira}, \&
  {Santos}}]{Alonso2014}
{Alonso}, D., {Ferreira}, P.~G., \& {Santos}, M.~G. 2014, \mnras, 444, 3183

\bibitem[{{Amblard} {et~al.}(2011){Amblard}, {Cooray}, {Serra}, {Altieri},
  {Arumugam}, {Aussel}, {Blain}, {Bock}, {Boselli}, {Buat},
  {Castro-Rodr{\'{\i}}guez}, {Cava}, {Chanial}, {Chapin}, {Clements}, {Conley},
  {Conversi}, {Dowell}, {Dwek}, {Eales}, {Elbaz}, {Farrah}, {Franceschini},
  {Gear}, {Glenn}, {Griffin}, {Halpern}, {Hatziminaoglou}, {Ibar}, {Isaak},
  {Ivison}, {Khostovan}, {Lagache}, {Levenson}, {Lu}, {Madden}, {Maffei},
  {Mainetti}, {Marchetti}, {Marsden}, {Mitchell-Wynne}, {Nguyen}, {O'Halloran},
  {Oliver}, {Omont}, {Page}, {Panuzzo}, {Papageorgiou}, {Pearson},
  {P{\'e}rez-Fournon}, {Pohlen}, {Rangwala}, {Roseboom}, {Rowan-Robinson},
  {Portal}, {Schulz}, {Scott}, {Seymour}, {Shupe}, {Smith}, {Stevens},
  {Symeonidis}, {Trichas}, {Tugwell}, {Vaccari}, {Valiante}, {Valtchanov},
  {Vieira}, {Vigroux}, {Wang}, {Ward}, {Wright}, {Xu}, \&
  {Zemcov}}]{Amblard2011}
{Amblard}, A., {Cooray}, A., {Serra}, P., {et~al.} 2011, \nat, 470, 510

\bibitem[{{Arata} {et~al.}(2020){Arata}, {Yajima}, {Nagamine}, {Abe}, \&
  {Khochfar}}]{Arata2020}
{Arata}, S., {Yajima}, H., {Nagamine}, K., {Abe}, M., \& {Khochfar}, S. 2020,
  \mnras, 498, 5541

\bibitem[{{Aravena} {et~al.}(2016{\natexlab{a}}){Aravena}, {Decarli}, {Walter},
  {Bouwens}, {Oesch}, {Carilli}, {Bauer}, {Da Cunha}, {Daddi},
  {G{\'o}nzalez-L{\'o}pez}, {Ivison}, {Riechers}, {Smail}, {Swinbank}, {Weiss},
  {Anguita}, {Bacon}, {Bell}, {Bertoldi}, {Cortes}, {Cox}, {Hodge}, {Ibar},
  {Inami}, {Infante}, {Karim}, {Magnelli}, {Ota}, {Popping}, {van der Werf},
  {Wagg}, \& {Fudamoto}}]{Aravena2016c}
{Aravena}, M., {Decarli}, R., {Walter}, F., {et~al.} 2016{\natexlab{a}}, \apj,
  833, 71

\bibitem[{{Aravena} {et~al.}(2016{\natexlab{b}}){Aravena}, {Spilker},
  {Bethermin}, {Bothwell}, {Chapman}, {de Breuck}, {Furstenau},
  {G{\'o}nzalez-L{\'o}pez}, {Greve}, {Litke}, {Ma}, {Malkan}, {Marrone},
  {Murphy}, {Stark}, {Strandet}, {Vieira}, {Weiss}, {Welikala}, {Wong}, \&
  {Collier}}]{Aravena2016}
{Aravena}, M., {Spilker}, J.~S., {Bethermin}, M., {et~al.} 2016{\natexlab{b}},
  Monthly Notices of the Royal Astronomical Society, 457, 4406

\bibitem[{{Astropy Collaboration} {et~al.}(2013){Astropy Collaboration},
  {Robitaille}, {Tollerud}, {Greenfield}, {Droettboom}, {Bray}, {Aldcroft},
  {Davis}, {Ginsburg}, {Price-Whelan}, {Kerzendorf}, {Conley}, {Crighton},
  {Barbary}, {Muna}, {Ferguson}, {Grollier}, {Parikh}, {Nair}, {Unther},
  {Deil}, {Woillez}, {Conseil}, {Kramer}, {Turner}, {Singer}, {Fox}, {Weaver},
  {Zabalza}, {Edwards}, {Azalee Bostroem}, {Burke}, {Casey}, {Crawford},
  {Dencheva}, {Ely}, {Jenness}, {Labrie}, {Lim}, {Pierfederici}, {Pontzen},
  {Ptak}, {Refsdal}, {Servillat}, \& {Streicher}}]{astropy:2013}
{Astropy Collaboration}, {Robitaille}, T.~P., {Tollerud}, E.~J., {et~al.} 2013,
  \aap, 558, A33

\bibitem[{{Behroozi} {et~al.}(2019){Behroozi}, {Wechsler}, {Hearin}, \&
  {Conroy}}]{Behroozi2019}
{Behroozi}, P., {Wechsler}, R.~H., {Hearin}, A.~P., \& {Conroy}, C. 2019,
  \mnras, 488, 3143

\bibitem[{{Behroozi} {et~al.}(2013){Behroozi}, {Wechsler}, \&
  {Conroy}}]{Behroozi2013}
{Behroozi}, P.~S., {Wechsler}, R.~H., \& {Conroy}, C. 2013, \apj, 770, 57

\bibitem[{{B{\'e}thermin} {et~al.}(2012){B{\'e}thermin}, {Dor{\'e}}, \&
  {Lagache}}]{Bethermin2012a}
{B{\'e}thermin}, M., {Dor{\'e}}, O., \& {Lagache}, G. 2012, \aap, 537, L5

\bibitem[{{B{\'e}thermin} {et~al.}(2020){B{\'e}thermin}, {Fudamoto}, {Ginolfi},
  {Loiacono}, {Khusanova}, {Capak}, {Cassata}, {Faisst}, {Le F{\`e}vre},
  {Schaerer}, {Silverman}, {Yan}, {Amorin}, {Bardelli}, {Boquien}, {Cimatti},
  {Davidzon}, {Dessauges-Zavadsky}, {Fujimoto}, {Gruppioni}, {Hathi}, {Ibar},
  {Jones}, {Koekemoer}, {Lagache}, {Lemaux}, {Moreau}, {Oesch}, {Pozzi},
  {Riechers}, {Talia}, {Toft}, {Vallini}, {Vergani}, {Zamorani}, \&
  {Zucca}}]{Bethermin2020}
{B{\'e}thermin}, M., {Fudamoto}, Y., {Ginolfi}, M., {et~al.} 2020, \aap, 643,
  A2

\bibitem[{{B{\'e}thermin} {et~al.}(2014){B{\'e}thermin}, {Kilbinger}, {Daddi},
  {Gabor}, {Finoguenov}, {McCracken}, {Wolk}, {Aussel}, {Strazzulo}, {Le
  Floc'h}, {Gobat}, {Rodighiero}, {Dickinson}, {Wang}, {Lutz}, \&
  {Heinis}}]{Bethermin2014}
{B{\'e}thermin}, M., {Kilbinger}, M., {Daddi}, E., {et~al.} 2014, \aap, 567,
  A103

\bibitem[{{B{\'e}thermin} {et~al.}(2013){B{\'e}thermin}, {Wang}, {Dor{\'e}},
  {Lagache}, {Sargent}, {Daddi}, {Cousin}, \& {Aussel}}]{Bethermin2013}
{B{\'e}thermin}, M., {Wang}, L., {Dor{\'e}}, O., {et~al.} 2013, \aap, 557, A66

\bibitem[{{B{\'e}thermin} {et~al.}(2017){B{\'e}thermin}, {Wu}, {Lagache},
  {Davidzon}, {Ponthieu}, {Cousin}, {Wang}, {Dor{\'e}}, {Daddi}, \&
  {Lapi}}]{Bethermin2017}
{B{\'e}thermin}, M., {Wu}, H.-Y., {Lagache}, G., {et~al.} 2017, \aap, 607, A89

\bibitem[{{Birkin} {et~al.}(2021){Birkin}, {Weiss}, {Wardlow}, {Smail},
  {Swinbank}, {Dudzevi{\v{c}}i{\={u}}t{\.{e}}}, {An}, {Ao}, {Chapman}, {Chen},
  {da Cunha}, {Dannerbauer}, {Gullberg}, {Hodge}, {Ikarashi}, {Ivison},
  {Matsuda}, {Stach}, {Walter}, {Wang}, \& {van der Werf}}]{Birkin2021}
{Birkin}, J.~E., {Weiss}, A., {Wardlow}, J.~L., {et~al.} 2021, \mnras, 501,
  3926

\bibitem[{{Bisbas} {et~al.}(2021){Bisbas}, {Tan}, \& {Tanaka}}]{Bisbas2021}
{Bisbas}, T.~G., {Tan}, J.~C., \& {Tanaka}, K. E.~I. 2021, \mnras, 502, 2701

\bibitem[{{Boogaard} {et~al.}(2020){Boogaard}, {van der Werf}, {Weiss},
  {Popping}, {Decarli}, {Walter}, {Aravena}, {Bouwens}, {Riechers},
  {Gonz{\'a}lez-L{\'o}pez}, {Smail}, {Carilli}, {Kaasinen}, {Daddi}, {Cox},
  {D{\'\i}az-Santos}, {Inami}, {Cortes}, \& {Wagg}}]{Boogaard2020}
{Boogaard}, L.~A., {van der Werf}, P., {Weiss}, A., {et~al.} 2020, \apj, 902,
  109

\bibitem[{{Bothwell} {et~al.}(2016){Bothwell}, {Maiolino}, {Peng}, {Cicone},
  {Griffith}, \& {Wagg}}]{Bothwell2016}
{Bothwell}, M.~S., {Maiolino}, R., {Peng}, Y., {et~al.} 2016, \mnras, 455, 1156

\bibitem[{{Bournaud} {et~al.}(2015){Bournaud}, {Daddi}, {Wei{\ss}}, {Renaud},
  {Mastropietro}, \& {Teyssier}}]{Bournaud2015}
{Bournaud}, F., {Daddi}, E., {Wei{\ss}}, A., {et~al.} 2015, \aap, 575, A56

\bibitem[{{Bouwens} {et~al.}(2007){Bouwens}, {Illingworth}, {Franx}, \&
  {Ford}}]{Bouwens2007}
{Bouwens}, R.~J., {Illingworth}, G.~D., {Franx}, M., \& {Ford}, H. 2007, \apj,
  670, 928

\bibitem[{{Bouwens} {et~al.}(2015){Bouwens}, {Illingworth}, {Oesch}, {Trenti},
  {Labb{\'e}}, {Bradley}, {Carollo}, {van Dokkum}, {Gonzalez}, {Holwerda},
  {Franx}, {Spitler}, {Smit}, \& {Magee}}]{Bouwens2015}
{Bouwens}, R.~J., {Illingworth}, G.~D., {Oesch}, P.~A., {et~al.} 2015, \apj,
  803, 34

\bibitem[{{Bouwens} {et~al.}(2012){Bouwens}, {Illingworth}, {Oesch}, {Trenti},
  {Labb{\'e}}, {Franx}, {Stiavelli}, {Carollo}, {van Dokkum}, \&
  {Magee}}]{Bouwens2012}
{Bouwens}, R.~J., {Illingworth}, G.~D., {Oesch}, P.~A., {et~al.} 2012, \apjl,
  752, L5

\bibitem[{{Brada{\v{c}}} {et~al.}(2017){Brada{\v{c}}}, {Garcia-Appadoo},
  {Huang}, {Vallini}, {Quinn Finney}, {Hoag}, {Lemaux}, {Borello Schmidt},
  {Treu}, {Carilli}, {Dijkstra}, {Ferrara}, {Fontana}, {Jones}, {Ryan}, {Wagg},
  \& {Gonzalez}}]{Bradac2017}
{Brada{\v{c}}}, M., {Garcia-Appadoo}, D., {Huang}, K.-H., {et~al.} 2017, \apjl,
  836, L2

\bibitem[{{Breysse} {et~al.}(2021){Breysse}, {Chung}, {Cleary}, {Ihle},
  {Padmanabhan}, {Silva}, {Bond}, {Borowska}, {Catha}, {Church}, {Dunne},
  {Eriksen}, {Foss}, {Gaier}, {Ott Gundersen}, {Harris}, {Hobbs}, {Keating},
  {Lamb}, {Lawrence}, {Lunde}, {Murray}, {Pearson}, {Philip}, {Rasmussen},
  {Readhead}, {Rennie}, {Stutzer}, {Viero}, {Watts}, {Wehus}, \&
  {Woody}}]{Breysse2021}
{Breysse}, P.~C., {Chung}, D.~T., {Cleary}, K.~A., {et~al.} 2021, arXiv
  e-prints, arXiv:2111.05933

\bibitem[{{Ca{\~n}ameras} {et~al.}(2018){Ca{\~n}ameras}, {Yang}, {Nesvadba},
  {Beelen}, {Kneissl}, {Koenig}, {Le Floc'h}, {Limousin}, {Malhotra}, {Omont},
  \& {Scott}}]{Canameras2018}
{Ca{\~n}ameras}, R., {Yang}, C., {Nesvadba}, N.~P.~H., {et~al.} 2018, \aap,
  620, A61

\bibitem[{{Capak} {et~al.}(2015){Capak}, {Carilli}, {Jones}, {Casey},
  {Riechers}, {Sheth}, {Carollo}, {Ilbert}, {Karim}, {Lefevre}, {Lilly},
  {Scoville}, {Smolcic}, \& {Yan}}]{Capak2015}
{Capak}, P.~L., {Carilli}, C., {Jones}, G., {et~al.} 2015, \nat, 522, 455

\bibitem[{{Carilli} \& {Walter}(2013)}]{Carilli2013}
{Carilli}, C.~L. \& {Walter}, F. 2013, \araa, 51, 105

\bibitem[{{Carniani} {et~al.}(2020){Carniani}, {Ferrara}, {Maiolino},
  {Castellano}, {Gallerani}, {Fontana}, {Kohandel}, {Lupi}, {Pallottini},
  {Pentericci}, {Vallini}, \& {Vanzella}}]{Carniani2020}
{Carniani}, S., {Ferrara}, A., {Maiolino}, R., {et~al.} 2020, \mnras, 499, 5136

\bibitem[{{Casey} {et~al.}(2018){Casey}, {Zavala}, {Spilker}, {da Cunha},
  {Hodge}, {Hung}, {Staguhn}, {Finkelstein}, \& {Drew}}]{Casey2018}
{Casey}, C.~M., {Zavala}, J.~A., {Spilker}, J., {et~al.} 2018, \apj, 862, 77

\bibitem[{{Chabrier}(2003)}]{Chabrier2003}
{Chabrier}, G. 2003, \pasp, 115, 763

\bibitem[{{Cheng} {et~al.}(2016){Cheng}, {Chang}, {Bock}, {Bradford}, \&
  {Cooray}}]{Cheng2016}
{Cheng}, Y.-T., {Chang}, T.-C., {Bock}, J., {Bradford}, C.~M., \& {Cooray}, A.
  2016, \apj, 832, 165

\bibitem[{{Cheng} {et~al.}(2020){Cheng}, {Chang}, \& {Bock}}]{Cheng2020}
{Cheng}, Y.-T., {Chang}, T.-C., \& {Bock}, J.~J. 2020, \apj, 901, 142

\bibitem[{{Chung} {et~al.}(2020){Chung}, {Viero}, {Church}, \&
  {Wechsler}}]{Chung2020}
{Chung}, D.~T., {Viero}, M.~P., {Church}, S.~E., \& {Wechsler}, R.~H. 2020,
  \apj, 892, 51

\bibitem[{{Concerto Collaboration} {et~al.}(2020){Concerto Collaboration},
  {Ade}, {Aravena}, {Barria}, {Beelen}, {Benoit}, {B{\'e}thermin}, {Bounmy},
  {Bourrion}, {Bres}, {De Breuck}, {Calvo}, {Cao}, {Catalano}, {D{\'e}sert},
  {Dur{\'a}n}, {Fasano}, {Fenouillet}, {Garcia}, {Garde}, {Goupy}, {Groppi},
  {Hoarau}, {Lagache}, {Lambert}, {Leggeri}, {Levy-Bertrand},
  {Mac{\'\i}as-P{\'e}rez}, {Mani}, {Marpaud}, {Mauskopf}, {Monfardini},
  {Pisano}, {Ponthieu}, {Prieur}, {Roni}, {Roudier}, {Tourres}, \&
  {Tucker}}]{CONCERTO2020}
{Concerto Collaboration}, {Ade}, P., {Aravena}, M., {et~al.} 2020, \aap, 642,
  A60

\bibitem[{{Cowley} {et~al.}(2018){Cowley}, {Caputi}, {Deshmukh}, {Ashby},
  {Fazio}, {Le F{\`e}vre}, {Fynbo}, {Ilbert}, {McCracken}, {Milvang-Jensen}, \&
  {Somerville}}]{Cowley2018}
{Cowley}, W.~I., {Caputi}, K.~I., {Deshmukh}, S., {et~al.} 2018, \apj, 853, 69

\bibitem[{{Cowley} {et~al.}(2017){Cowley}, {Lacey}, {Baugh}, {Cole}, \&
  {Wilkinson}}]{Cowley2017}
{Cowley}, W.~I., {Lacey}, C.~G., {Baugh}, C.~M., {Cole}, S., \& {Wilkinson}, A.
  2017, \mnras, 469, 3396

\bibitem[{{Crites} {et~al.}(2014){Crites}, {Bock}, {Bradford}, {Chang},
  {Cooray}, {Duband}, {Gong}, {Hailey-Dunsheath}, {Hunacek}, {Koch}, {Li},
  {O'Brient}, {Prouve}, {Shirokoff}, {Silva}, {Staniszewski}, {Uzgil}, \&
  {Zemcov}}]{Crites2014}
{Crites}, A.~T., {Bock}, J.~J., {Bradford}, C.~M., {et~al.} 2014, in Society of
  Photo-Optical Instrumentation Engineers (SPIE) Conference Series, Vol. 9153,
  Millimeter, Submillimeter, and Far-Infrared Detectors and Instrumentation for
  Astronomy VII, ed. W.~S. {Holland} \& J.~{Zmuidzinas}, 91531W

\bibitem[{{da Cunha} {et~al.}(2013){da Cunha}, {Groves}, {Walter}, {Decarli},
  {Weiss}, {Bertoldi}, {Carilli}, {Daddi}, {Elbaz}, {Ivison}, {Maiolino},
  {Riechers}, {Rix}, {Sargent}, \& {Smail}}]{Da_Cunha2013}
{da Cunha}, E., {Groves}, B., {Walter}, F., {et~al.} 2013, \apj, 766, 13

\bibitem[{{Daddi} {et~al.}(2015){Daddi}, {Dannerbauer}, {Liu}, {Aravena},
  {Bournaud}, {Walter}, {Riechers}, {Magdis}, {Sargent}, {B{\'e}thermin},
  {Carilli}, {Cibinel}, {Dickinson}, {Elbaz}, {Gao}, {Gobat}, {Hodge}, \&
  {Krips}}]{Daddi2015}
{Daddi}, E., {Dannerbauer}, H., {Liu}, D., {et~al.} 2015, \aap, 577, A46

\bibitem[{{Daddi} {et~al.}(2010){Daddi}, {Elbaz}, {Walter}, {Bournaud},
  {Salmi}, {Carilli}, {Dannerbauer}, {Dickinson}, {Monaco}, \&
  {Riechers}}]{Daddi2010b}
{Daddi}, E., {Elbaz}, D., {Walter}, F., {et~al.} 2010, \apjl, 714, L118

\bibitem[{{Davidzon} {et~al.}(2017){Davidzon}, {Ilbert}, {Laigle}, {Coupon},
  {McCracken}, {Delvecchio}, {Masters}, {Capak}, {Hsieh}, {Le F{\`e}vre},
  {Tresse}, {Bethermin}, {Chang}, {Faisst}, {Le Floc'h}, {Steinhardt}, {Toft},
  {Aussel}, {Dubois}, {Hasinger}, {Salvato}, {Sanders}, {Scoville}, \&
  {Silverman}}]{Davidzon2017}
{Davidzon}, I., {Ilbert}, O., {Laigle}, C., {et~al.} 2017, \aap, 605, A70

\bibitem[{{De Looze} {et~al.}(2014){De Looze}, {Cormier}, {Lebouteiller},
  {Madden}, {Baes}, {Bendo}, {Boquien}, {Boselli}, {Clements}, {Cortese},
  {Cooray}, {Galametz}, {Galliano}, {Graci{\'a}-Carpio}, {Isaak}, {Karczewski},
  {Parkin}, {Pellegrini}, {R{\'e}my-Ruyer}, {Spinoglio}, {Smith}, \&
  {Sturm}}]{De_Looze2014}
{De Looze}, I., {Cormier}, D., {Lebouteiller}, V., {et~al.} 2014, \aap, 568,
  A62

\bibitem[{{Decarli} {et~al.}(2020){Decarli}, {Aravena}, {Boogaard}, {Carilli},
  {Gonz{\'a}lez-L{\'o}pez}, {Walter}, {Cortes}, {Cox}, {da Cunha}, {Daddi},
  {D{\'\i}az-Santos}, {Hodge}, {Inami}, {Neeleman}, {Novak}, {Oesch},
  {Popping}, {Riechers}, {Smail}, {Uzgil}, {van der Werf}, {Wagg}, \&
  {Weiss}}]{Decarli2020}
{Decarli}, R., {Aravena}, M., {Boogaard}, L., {et~al.} 2020, \apj, 902, 110

\bibitem[{{Decarli} {et~al.}(2016){Decarli}, {Walter}, {Aravena}, {Carilli},
  {Bouwens}, {da Cunha}, {Daddi}, {Ivison}, {Popping}, {Riechers}, {Smail},
  {Swinbank}, {Weiss}, {Anguita}, {Assef}, {Bauer}, {Bell}, {Bertoldi},
  {Chapman}, {Colina}, {Cortes}, {Cox}, {Dickinson}, {Elbaz},
  {G{\'o}nzalez-L{\'o}pez}, {Ibar}, {Infante}, {Hodge}, {Karim}, {Le Fevre},
  {Magnelli}, {Neri}, {Oesch}, {Ota}, {Rix}, {Sargent}, {Sheth}, {van der Wel},
  {van der Werf}, \& {Wagg}}]{Decarli2016}
{Decarli}, R., {Walter}, F., {Aravena}, M., {et~al.} 2016, \apj, 833, 69

\bibitem[{{Decarli} {et~al.}(2019){Decarli}, {Walter},
  {G{\'o}nzalez-L{\'o}pez}, {Aravena}, {Boogaard}, {Carilli}, {Cox}, {Daddi},
  {Popping}, {Riechers}, {Uzgil}, {Weiss}, {Assef}, {Bacon}, {Bauer},
  {Bertoldi}, {Bouwens}, {Contini}, {Cortes}, {da Cunha}, {D{\'\i}az-Santos},
  {Elbaz}, {Inami}, {Hodge}, {Ivison}, {Le F{\`e}vre}, {Magnelli}, {Novak},
  {Oesch}, {Rix}, {Sargent}, {Smail}, {Swinbank}, {Somerville}, {van der Werf},
  {Wagg}, \& {Wisotzki}}]{Decarli2019}
{Decarli}, R., {Walter}, F., {G{\'o}nzalez-L{\'o}pez}, J., {et~al.} 2019, \apj,
  882, 138

\bibitem[{{Dessauges-Zavadsky} {et~al.}(2015){Dessauges-Zavadsky}, {Zamojski},
  {Schaerer}, {Combes}, {Egami}, {Swinbank}, {Richard}, {Sklias}, {Rawle},
  {Rex}, {Kneib}, {Boone}, \& {Blain}}]{Dessauges2015}
{Dessauges-Zavadsky}, M., {Zamojski}, M., {Schaerer}, D., {et~al.} 2015, \aap,
  577, A50

\bibitem[{{Downes} \& {Solomon}(1998)}]{Downes1998}
{Downes}, D. \& {Solomon}, P.~M. 1998, \apj, 507, 615

\bibitem[{{Draine} \& {Li}(2007)}]{Draine2007}
{Draine}, B.~T. \& {Li}, A. 2007, \apj, 657, 810

\bibitem[{{Faisst} {et~al.}(2020){Faisst}, {Schaerer}, {Lemaux}, {Oesch},
  {Fudamoto}, {Cassata}, {B{\'e}thermin}, {Capak}, {Le F{\`e}vre}, {Silverman},
  {Yan}, {Ginolfi}, {Koekemoer}, {Morselli}, {Amor{\'\i}n}, {Bardelli},
  {Boquien}, {Brammer}, {Cimatti}, {Dessauges-Zavadsky}, {Fujimoto},
  {Gruppioni}, {Hathi}, {Hemmati}, {Ibar}, {Jones}, {Khusanova}, {Loiacono},
  {Pozzi}, {Talia}, {Tasca}, {Riechers}, {Rodighiero}, {Romano}, {Scoville},
  {Toft}, {Vallini}, {Vergani}, {Zamorani}, \& {Zucca}}]{Faisst2020}
{Faisst}, A.~L., {Schaerer}, D., {Lemaux}, B.~C., {et~al.} 2020, \apjs, 247, 61

\bibitem[{{Ferrara} {et~al.}(2019){Ferrara}, {Vallini}, {Pallottini},
  {Gallerani}, {Carniani}, {Kohandel}, {Decataldo}, \& {Behrens}}]{Ferrara2019}
{Ferrara}, A., {Vallini}, L., {Pallottini}, A., {et~al.} 2019, \mnras, 489, 1

\bibitem[{{Freundlich} {et~al.}(2019){Freundlich}, {Combes}, {Tacconi},
  {Genzel}, {Garcia-Burillo}, {Neri}, {Contini}, {Bolatto}, {Lilly},
  {Salom{\'e}}, {Bicalho}, {Boissier}, {Boone}, {Bouch{\'e}}, {Bournaud},
  {Burkert}, {Carollo}, {Cooper}, {Cox}, {Feruglio}, {F{\"o}rster Schreiber},
  {Juneau}, {Lippa}, {Lutz}, {Naab}, {Renzini}, {Saintonge}, {Sternberg},
  {Walter}, {Weiner}, {Wei{\ss}}, \& {Wuyts}}]{Freundlich2019}
{Freundlich}, J., {Combes}, F., {Tacconi}, L.~J., {et~al.} 2019, \aap, 622,
  A105

\bibitem[{{Fudamoto} {et~al.}(2020){Fudamoto}, {Oesch}, {Faisst},
  {B{\'e}thermin}, {Ginolfi}, {Khusanova}, {Loiacono}, {Le F{\`e}vre}, {Capak},
  {Schaerer}, {Silverman}, {Cassata}, {Yan}, {Amorin}, {Bardelli}, {Boquien},
  {Cimatti}, {Dessauges-Zavadsky}, {Fujimoto}, {Gruppioni}, {Hathi}, {Ibar},
  {Jones}, {Koekemoer}, {Lagache}, {Lemaux}, {Maiolino}, {Narayanan}, {Pozzi},
  {Riechers}, {Rodighiero}, {Talia}, {Toft}, {Vallini}, {Vergani}, {Zamorani},
  \& {Zucca}}]{Fudamoto2020}
{Fudamoto}, Y., {Oesch}, P.~A., {Faisst}, A., {et~al.} 2020, \aap, 643, A4

\bibitem[{{Fudamoto} {et~al.}(2021){Fudamoto}, {Oesch}, {Schouws}, {Stefanon},
  {Smit}, {Bouwens}, {Bowler}, {Endsley}, {Gonzalez}, {Inami}, {Labbe},
  {Stark}, {Aravena}, {Barrufet}, {da Cunha}, {Dayal}, {Ferrara}, {Graziani},
  {Hodge}, {Hutter}, {Li}, {De Looze}, {Nanayakkara}, {Pallottini}, {Riechers},
  {Schneider}, {Ucci}, {van der Werf}, \& {White}}]{Fudamoto2021}
{Fudamoto}, Y., {Oesch}, P.~A., {Schouws}, S., {et~al.} 2021, \nat, 597, 489

\bibitem[{{Fujimoto} {et~al.}(2021){Fujimoto}, {Oguri}, {Brammer}, {Yoshimura},
  {Laporte}, {Gonz{\'a}lez-L{\'o}pez}, {Caminha}, {Kohno}, {Zitrin}, {Richard},
  {Ouchi}, {Bauer}, {Smail}, {Hatsukade}, {Ono}, {Kokorev}, {Umehata},
  {Schaerer}, {Knudsen}, {Sun}, {Magdis}, {Valentino}, {Ao}, {Toft},
  {Dessauges-Zavadsky}, {Shimasaku}, {Caputi}, {Kusakabe}, {Morokuma-Matsui},
  {Shotaro}, {Egami}, {Lee}, {Rawle}, \& {Espada}}]{Fujimoto2021}
{Fujimoto}, S., {Oguri}, M., {Brammer}, G., {et~al.} 2021, \apj, 911, 99

\bibitem[{{Genzel} {et~al.}(2010){Genzel}, {Tacconi}, {Gracia-Carpio},
  {Sternberg}, {Cooper}, {Shapiro}, {Bolatto}, {Bouch{\'e}}, {Bournaud},
  {Burkert}, {Combes}, {Comerford}, {Cox}, {Davis}, {Schreiber},
  {Garcia-Burillo}, {Lutz}, {Naab}, {Neri}, {Omont}, {Shapley}, \&
  {Weiner}}]{Genzel2010}
{Genzel}, R., {Tacconi}, L.~J., {Gracia-Carpio}, J., {et~al.} 2010, \mnras,
  407, 2091

\bibitem[{{Genzel} {et~al.}(2015){Genzel}, {Tacconi}, {Lutz}, {Saintonge},
  {Berta}, {Magnelli}, {Combes}, {Garc{\'\i}a-Burillo}, {Neri}, {Bolatto},
  {Contini}, {Lilly}, {Boissier}, {Boone}, {Bouch{\'e}}, {Bournaud}, {Burkert},
  {Carollo}, {Colina}, {Cooper}, {Cox}, {Feruglio}, {F{\"o}rster Schreiber},
  {Freundlich}, {Gracia-Carpio}, {Juneau}, {Kovac}, {Lippa}, {Naab}, {Salome},
  {Renzini}, {Sternberg}, {Walter}, {Weiner}, {Weiss}, \& {Wuyts}}]{Genzel2015}
{Genzel}, R., {Tacconi}, L.~J., {Lutz}, D., {et~al.} 2015, \apj, 800, 20

\bibitem[{{Gong} {et~al.}(2012){Gong}, {Cooray}, {Silva}, {Santos}, {Bock},
  {Bradford}, \& {Zemcov}}]{Gong2012}
{Gong}, Y., {Cooray}, A., {Silva}, M., {et~al.} 2012, \apj, 745, 49

\bibitem[{{Gong} {et~al.}(2014){Gong}, {Silva}, {Cooray}, \&
  {Santos}}]{Gong2014}
{Gong}, Y., {Silva}, M., {Cooray}, A., \& {Santos}, M.~G. 2014, \apj, 785, 72

\bibitem[{{Grazian} {et~al.}(2015){Grazian}, {Fontana}, {Santini}, {Dunlop},
  {Ferguson}, {Castellano}, {Amorin}, {Ashby}, {Barro}, {Behroozi}, {Boutsia},
  {Caputi}, {Chary}, {Dekel}, {Dickinson}, {Faber}, {Fazio}, {Finkelstein},
  {Galametz}, {Giallongo}, {Giavalisco}, {Grogin}, {Guo}, {Kocevski},
  {Koekemoer}, {Koo}, {Lee}, {Lu}, {Merlin}, {Mobasher}, {Nonino}, {Papovich},
  {Paris}, {Pentericci}, {Reddy}, {Renzini}, {Salmon}, {Salvato}, {Sommariva},
  {Song}, \& {Vanzella}}]{Grazian2015}
{Grazian}, A., {Fontana}, A., {Santini}, P., {et~al.} 2015, \aap, 575, A96

\bibitem[{{Greve} {et~al.}(2014){Greve}, {Leonidaki}, {Xilouris}, {Wei{\ss}},
  {Zhang}, {van der Werf}, {Aalto}, {Armus}, {D{\'{\i}}az-Santos}, {Evans},
  {Fischer}, {Gao}, {Gonz{\'a}lez-Alfonso}, {Harris}, {Henkel}, {Meijerink},
  {Naylor}, {Smith}, {Spaans}, {Stacey}, {Veilleux}, \& {Walter}}]{Greve2014}
{Greve}, T.~R., {Leonidaki}, I., {Xilouris}, E.~M., {et~al.} 2014, \apj, 794,
  142

\bibitem[{{Gruppioni} {et~al.}(2020){Gruppioni}, {B{\'e}thermin}, {Loiacono},
  {Le F{\`e}vre}, {Capak}, {Cassata}, {Faisst}, {Schaerer}, {Silverman}, {Yan},
  {Bardelli}, {Boquien}, {Carraro}, {Cimatti}, {Dessauges-Zavadsky}, {Ginolfi},
  {Fujimoto}, {Hathi}, {Jones}, {Khusanova}, {Koekemoer}, {Lagache}, {Lemaux},
  {Oesch}, {Pozzi}, {Riechers}, {Rodighiero}, {Romano}, {Talia}, {Vallini},
  {Vergani}, {Zamorani}, \& {Zucca}}]{Gruppioni2020}
{Gruppioni}, C., {B{\'e}thermin}, M., {Loiacono}, F., {et~al.} 2020, \aap, 643,
  A8

\bibitem[{{Gullberg} {et~al.}(2016){Gullberg}, {Lehnert}, {De Breuck},
  {Branchu}, {Dannerbauer}, {Drouart}, {Emonts}, {Guillard}, {Hatch},
  {Nesvadba}, {Omont}, {Seymour}, \& {Vernet}}]{Gullberg2016}
{Gullberg}, B., {Lehnert}, M.~D., {De Breuck}, C., {et~al.} 2016, \aap, 591,
  A73

\bibitem[{Harris {et~al.}(2020)Harris, Millman, van~der Walt, Gommers,
  Virtanen, Cournapeau, Wieser, Taylor, Berg, Smith, Kern, Picus, Hoyer, van
  Kerkwijk, Brett, Haldane, del R{'{\i}}o, Wiebe, Peterson,
  G{'{e}}rard-Marchant, Sheppard, Reddy, Weckesser, Abbasi, Gohlke, \&
  Oliphant}]{numpy}
Harris, C.~R., Millman, K.~J., van~der Walt, S.~J., {et~al.} 2020, Nature, 585,
  357

\bibitem[{{Hernandez-Monteagudo} {et~al.}(2017){Hernandez-Monteagudo}, {Maio},
  {Ciardi}, \& {Sunyaev}}]{Hernandez-Monteagudo2017}
{Hernandez-Monteagudo}, C., {Maio}, U., {Ciardi}, B., \& {Sunyaev}, R.~A. 2017,
  arXiv e-prints, arXiv:1707.01910

\bibitem[{{Herrera-Camus} {et~al.}(2015){Herrera-Camus}, {Bolatto}, {Wolfire},
  {Smith}, {Croxall}, {Kennicutt}, {Calzetti}, {Helou}, {Walter}, {Leroy},
  {Draine}, {Brandl}, {Armus}, {Sandstrom}, {Dale}, {Aniano}, {Meidt},
  {Boquien}, {Hunt}, {Galametz}, {Tabatabaei}, {Murphy}, {Appleton}, {Roussel},
  {Engelbracht}, \& {Beirao}}]{Herrera-Camus2015}
{Herrera-Camus}, R., {Bolatto}, A.~D., {Wolfire}, M.~G., {et~al.} 2015, \apj,
  800, 1

\bibitem[{{Hogg}(1999)}]{Hogg1999}
{Hogg}, D.~W. 1999, ArXiv Astrophysics e-prints
  [\eprint{arXiv:astro-ph/9905116}]

\bibitem[{{Ilbert} {et~al.}(2013){Ilbert}, {McCracken}, {Le F{\`e}vre},
  {Capak}, {Dunlop}, {Karim}, {Renzini}, {Caputi}, {Boissier}, {Arnouts},
  {Aussel}, {Comparat}, {Guo}, {Hudelot}, {Kartaltepe}, {Kneib}, {Krogager},
  {Le Floc'h}, {Lilly}, {Mellier}, {Milvang-Jensen}, {Moutard}, {Onodera},
  {Richard}, {Salvato}, {Sanders}, {Scoville}, {Silverman}, {Taniguchi},
  {Tasca}, {Thomas}, {Toft}, {Tresse}, {Vergani}, {Wolk}, \&
  {Zirm}}]{Ilbert2013}
{Ilbert}, O., {McCracken}, H.~J., {Le F{\`e}vre}, O., {et~al.} 2013, \aap, 556,
  A55

\bibitem[{{Inami} {et~al.}(2020){Inami}, {Decarli}, {Walter}, {Weiss},
  {Carilli}, {Aravena}, {Boogaard}, {Gonza{\'l}ez-L{\'o}pez}, {Popping}, {da
  Cunha}, {Bacon}, {Bauer}, {Contini}, {Cortes}, {Cox}, {Daddi},
  {D{\'\i}az-Santos}, {Kaasinen}, {Riechers}, {Wagg}, {van der Werf}, \&
  {Wisotzki}}]{Inami2020}
{Inami}, H., {Decarli}, R., {Walter}, F., {et~al.} 2020, \apj, 902, 113

\bibitem[{{Ishigaki} {et~al.}(2018){Ishigaki}, {Kawamata}, {Ouchi}, {Oguri},
  {Shimasaku}, \& {Ono}}]{Ishigaki2018}
{Ishigaki}, M., {Kawamata}, R., {Ouchi}, M., {et~al.} 2018, \apj, 854, 73

\bibitem[{{Ishikawa} {et~al.}(2016){Ishikawa}, {Kashikawa}, {Hamana},
  {Toshikawa}, \& {Onoue}}]{Ishikawa2016}
{Ishikawa}, S., {Kashikawa}, N., {Hamana}, T., {Toshikawa}, J., \& {Onoue}, M.
  2016, \mnras, 458, 747

\bibitem[{{Ishikawa} {et~al.}(2017){Ishikawa}, {Kashikawa}, {Toshikawa},
  {Tanaka}, {Hamana}, {Niino}, {Ichikawa}, \& {Uchiyama}}]{Ishikawa2017}
{Ishikawa}, S., {Kashikawa}, N., {Toshikawa}, J., {et~al.} 2017, \apj, 841, 8

\bibitem[{{Jeli{\'c}} {et~al.}(2008){Jeli{\'c}}, {Zaroubi}, {Labropoulos},
  {Thomas}, {Bernardi}, {Brentjens}, {de Bruyn}, {Ciardi}, {Harker},
  {Koopmans}, {Pandey}, {Schaye}, \& {Yatawatta}}]{Jelic2008}
{Jeli{\'c}}, V., {Zaroubi}, S., {Labropoulos}, P., {et~al.} 2008, \mnras, 389,
  1319

\bibitem[{{Kamenetzky} {et~al.}(2016){Kamenetzky}, {Rangwala}, {Glenn},
  {Maloney}, \& {Conley}}]{Kamenetzky2016}
{Kamenetzky}, J., {Rangwala}, N., {Glenn}, J., {Maloney}, P.~R., \& {Conley},
  A. 2016, \apj, 829, 93

\bibitem[{{Karoumpis} {et~al.}(2021){Karoumpis}, {Magnelli},
  {Romano-D{\'\i}az}, {Haslbauer}, \& {Bertoldi}}]{Karoumpis2021}
{Karoumpis}, C., {Magnelli}, B., {Romano-D{\'\i}az}, E., {Haslbauer}, M., \&
  {Bertoldi}, F. 2021, arXiv e-prints, arXiv:2111.12847

\bibitem[{{Kashikawa} {et~al.}(2006){Kashikawa}, {Yoshida}, {Shimasaku},
  {Nagashima}, {Yahagi}, {Ouchi}, {Matsuda}, {Malkan}, {Doi}, {Iye}, {Ajiki},
  {Akiyama}, {Ando}, {Aoki}, {Furusawa}, {Hayashino}, {Iwamuro}, {Karoji},
  {Kobayashi}, {Kodaira}, {Kodama}, {Komiyama}, {Miyazaki}, {Mizumoto},
  {Morokuma}, {Motohara}, {Murayama}, {Nagao}, {Nariai}, {Ohta}, {Okamura},
  {Sasaki}, {Sato}, {Sekiguchi}, {Shioya}, {Tamura}, {Taniguchi}, {Umemura},
  {Yamada}, \& {Yasuda}}]{Kashikawa2006}
{Kashikawa}, N., {Yoshida}, M., {Shimasaku}, K., {et~al.} 2006, \apj, 637, 631

\bibitem[{{Katz} {et~al.}(2017){Katz}, {Kimm}, {Sijacki}, \&
  {Haehnelt}}]{Katz2017}
{Katz}, H., {Kimm}, T., {Sijacki}, D., \& {Haehnelt}, M.~G. 2017, \mnras, 468,
  4831

\bibitem[{{Keating} {et~al.}(2020){Keating}, {Marrone}, {Bower}, \&
  {Keenan}}]{Keating2020}
{Keating}, G.~K., {Marrone}, D.~P., {Bower}, G.~C., \& {Keenan}, R.~P. 2020,
  \apj, 901, 141

\bibitem[{{Khusanova} {et~al.}(2021){Khusanova}, {Bethermin}, {Le F{\`e}vre},
  {Capak}, {Faisst}, {Schaerer}, {Silverman}, {Cassata}, {Yan}, {Ginolfi},
  {Fudamoto}, {Loiacono}, {Amorin}, {Bardelli}, {Boquien}, {Cimatti},
  {Dessauges-Zavadsky}, {Gruppioni}, {Hathi}, {Jones}, {Koekemoer}, {Lagache},
  {Maiolino}, {Lemaux}, {Oesch}, {Pozzi}, {Riechers}, {Romano}, {Talia},
  {Toft}, {Vergani}, {Zamorani}, \& {Zucca}}]{Khusanova2021}
{Khusanova}, Y., {Bethermin}, M., {Le F{\`e}vre}, O., {et~al.} 2021, \aap, 649,
  A152

\bibitem[{{Knudsen} {et~al.}(2016){Knudsen}, {Richard}, {Kneib}, {Jauzac},
  {Cl{\'e}ment}, {Drouart}, {Egami}, \& {Lindroos}}]{Knudsen2016}
{Knudsen}, K.~K., {Richard}, J., {Kneib}, J.-P., {et~al.} 2016, \mnras, 462, L6

\bibitem[{{Kovetz} {et~al.}(2017){Kovetz}, {Viero}, {Lidz}, {Newburgh},
  {Rahman}, {Switzer}, {Kamionkowski}, {Aguirre}, {Alvarez}, {Bock}, {Bond},
  {Bower}, {Bradford}, {Breysse}, {Bull}, {Chang}, {Cheng}, {Chung}, {Cleary},
  {Corray}, {Crites}, {Croft}, {Dor{\'e}}, {Eastwood}, {Ferrara}, {Fonseca},
  {Jacobs}, {Keating}, {Lagache}, {Lakhlani}, {Liu}, {Moodley}, {Murray},
  {P{\'e}nin}, {Popping}, {Pullen}, {Reichers}, {Saito}, {Saliwanchik},
  {Santos}, {Somerville}, {Stacey}, {Stein}, {Villaescusa-Navarro}, {Visbal},
  {Weltman}, {Wolz}, \& {Zemcov}}]{Kovetz2017}
{Kovetz}, E.~D., {Viero}, M.~P., {Lidz}, A., {et~al.} 2017, arXiv e-prints,
  arXiv:1709.09066

\bibitem[{{Lagache} {et~al.}(2007){Lagache}, {Bavouzet}, {Fernandez-Conde},
  {Ponthieu}, {Rodet}, {Dole}, {Miville-Desch{\^e}nes}, \&
  {Puget}}]{Lagache2007}
{Lagache}, G., {Bavouzet}, N., {Fernandez-Conde}, N., {et~al.} 2007, \apjl,
  665, L89

\bibitem[{{Lagache} {et~al.}(2018){Lagache}, {Cousin}, \&
  {Chatzikos}}]{Lagache2018}
{Lagache}, G., {Cousin}, M., \& {Chatzikos}, M. 2018, Astronomy and
  Astrophysics, 609, A130

\bibitem[{{Lagache} {et~al.}(2003){Lagache}, {Dole}, \& {Puget}}]{Lagache2003}
{Lagache}, G., {Dole}, H., \& {Puget}, J.-L. 2003, \mnras, 338, 555

\bibitem[{{Lagos} {et~al.}(2012){Lagos}, {Bayet}, {Baugh}, {Lacey}, {Bell},
  {Fanidakis}, \& {Geach}}]{Lagos2012}
{Lagos}, C.~d.~P., {Bayet}, E., {Baugh}, C.~M., {et~al.} 2012, \mnras, 426,
  2142

\bibitem[{{Lagos} {et~al.}(2020){Lagos}, {da Cunha}, {Robotham}, {Obreschkow},
  {Valentino}, {Fujimoto}, {Magdis}, \& {Tobar}}]{Lagos2020}
{Lagos}, C. d.~P., {da Cunha}, E., {Robotham}, A. S.~G., {et~al.} 2020, \mnras,
  499, 1948

\bibitem[{{Laigle} {et~al.}(2016){Laigle}, {McCracken}, {Ilbert}, {Hsieh},
  {Davidzon}, {Capak}, {Hasinger}, {Silverman}, {Pichon}, {Coupon}, {Aussel},
  {Le Borgne}, {Caputi}, {Cassata}, {Chang}, {Civano}, {Dunlop}, {Fynbo},
  {Kartaltepe}, {Koekemoer}, {Le F{\`e}vre}, {Le Floc'h}, {Leauthaud}, {Lilly},
  {Lin}, {Marchesi}, {Milvang-Jensen}, {Salvato}, {Sanders}, {Scoville},
  {Smolcic}, {Stockmann}, {Taniguchi}, {Tasca}, {Toft}, {Vaccari}, \&
  {Zabl}}]{Laigle2015}
{Laigle}, C., {McCracken}, H.~J., {Ilbert}, O., {et~al.} 2016, \apjs, 224, 24

\bibitem[{{Le F{\`e}vre} {et~al.}(2020){Le F{\`e}vre}, {B{\'e}thermin},
  {Faisst}, {Jones}, {Capak}, {Cassata}, {Silverman}, {Schaerer}, {Yan},
  {Amorin}, {Bardelli}, {Boquien}, {Cimatti}, {Dessauges-Zavadsky},
  {Giavalisco}, {Hathi}, {Fudamoto}, {Fujimoto}, {Ginolfi}, {Gruppioni},
  {Hemmati}, {Ibar}, {Koekemoer}, {Khusanova}, {Lagache}, {Lemaux}, {Loiacono},
  {Maiolino}, {Mancini}, {Narayanan}, {Morselli}, {M{\'e}ndez-Hern{\`a}ndez},
  {Oesch}, {Pozzi}, {Romano}, {Riechers}, {Scoville}, {Talia}, {Tasca},
  {Thomas}, {Toft}, {Vallini}, {Vergani}, {Walter}, {Zamorani}, \&
  {Zucca}}]{Le_Fevre2020}
{Le F{\`e}vre}, O., {B{\'e}thermin}, M., {Faisst}, A., {et~al.} 2020, \aap,
  643, A1

\bibitem[{{Lee} {et~al.}(2009){Lee}, {Giavalisco}, {Conroy}, {Wechsler},
  {Ferguson}, {Somerville}, {Dickinson}, \& {Urry}}]{Lee2009}
{Lee}, K.-S., {Giavalisco}, M., {Conroy}, C., {et~al.} 2009, \apj, 695, 368

\bibitem[{{Legrand} {et~al.}(2019){Legrand}, {McCracken}, {Davidzon}, {Ilbert},
  {Coupon}, {Aghanim}, {Douspis}, {Capak}, {Le F{\`e}vre}, \&
  {Milvang-Jensen}}]{Legrand2019}
{Legrand}, L., {McCracken}, H.~J., {Davidzon}, I., {et~al.} 2019, \mnras, 486,
  5468

\bibitem[{{Lidz} \& {Taylor}(2016)}]{Lidz2016}
{Lidz}, A. \& {Taylor}, J. 2016, \apj, 825, 143

\bibitem[{{Lin} {et~al.}(2012){Lin}, {Dickinson}, {Jian}, {Merson}, {Baugh},
  {Scott}, {Foucaud}, {Wang}, {Yan}, {Yan}, {Cheng}, {Guo}, {Helly}, {Kirsten},
  {Koo}, {Lagos}, {Meger}, {Messias}, {Pope}, {Simard}, {Grogin}, \&
  {Wang}}]{Lin2012}
{Lin}, L., {Dickinson}, M., {Jian}, H.-Y., {et~al.} 2012, \apj, 756, 71

\bibitem[{{Liu} {et~al.}(2015){Liu}, {Gao}, {Isaak}, {Daddi}, {Yang}, {Lu}, \&
  {van der Werf}}]{Liu2015}
{Liu}, D., {Gao}, Y., {Isaak}, K., {et~al.} 2015, \apjl, 810, L14

\bibitem[{{Loiacono} {et~al.}(2021){Loiacono}, {Decarli}, {Gruppioni}, {Talia},
  {Cimatti}, {Zamorani}, {Pozzi}, {Yan}, {Lemaux}, {Riechers}, {Le F{\`e}vre},
  {B{\`e}thermin}, {Capak}, {Cassata}, {Faisst}, {Schaerer}, {Silverman},
  {Bardelli}, {Boquien}, {Burkutean}, {Dessauges-Zavadsky}, {Fudamoto},
  {Fujimoto}, {Ginolfi}, {Hathi}, {Jones}, {Khusanova}, {Koekemoer}, {Lagache},
  {Lubin}, {Massardi}, {Oesch}, {Romano}, {Vallini}, {Vergani}, \&
  {Zucca}}]{Loiacono2021}
{Loiacono}, F., {Decarli}, R., {Gruppioni}, C., {et~al.} 2021, \aap, 646, A76

\bibitem[{{Lupi} \& {Bovino}(2020)}]{Lupi2020}
{Lupi}, A. \& {Bovino}, S. 2020, \mnras, 492, 2818

\bibitem[{{Madau} \& {Dickinson}(2014)}]{Madau2014}
{Madau}, P. \& {Dickinson}, M. 2014, \araa, 52, 415

\bibitem[{{Magdis} {et~al.}(2012){Magdis}, {Daddi}, {B{\'e}thermin}, {Sargent},
  {Elbaz}, {Pannella}, {Dickinson}, {Dannerbauer}, {da Cunha}, {Walter},
  {Rigopoulou}, {Charmandaris}, {Hwang}, \& {Kartaltepe}}]{Magdis2012b}
{Magdis}, G.~E., {Daddi}, E., {B{\'e}thermin}, M., {et~al.} 2012, \apj, 760, 6

\bibitem[{{Magliocchetti} {et~al.}(2011){Magliocchetti}, {Santini},
  {Rodighiero}, {Grazian}, {Aussel}, {Altieri}, {Andreani}, {Berta}, {Cepa},
  {Casta{\~n}eda}, {Cimatti}, {Daddi}, {Elbaz}, {Genzel}, {Gruppioni}, {Lutz},
  {Magnelli}, {Maiolino}, {Popesso}, {Poglitsch}, {Pozzi}, {Sanchez-Portal},
  {F{\"o}rster Schreiber}, {Sturm}, {Tacconi}, \&
  {Valtchanov}}]{Magliocchetti2011}
{Magliocchetti}, M., {Santini}, P., {Rodighiero}, G., {et~al.} 2011, \mnras,
  416, 1105

\bibitem[{{Maiolino} {et~al.}(2015){Maiolino}, {Carniani}, {Fontana},
  {Vallini}, {Pentericci}, {Ferrara}, {Vanzella}, {Grazian}, {Gallerani},
  {Castellano}, {Cristiani}, {Brammer}, {Santini}, {Wagg}, \&
  {Williams}}]{Maiolino2015}
{Maiolino}, R., {Carniani}, S., {Fontana}, A., {et~al.} 2015, \mnras, 452, 54

\bibitem[{{Maniyar} {et~al.}(2021){Maniyar}, {B{\'e}thermin}, \&
  {Lagache}}]{Maniyar2021}
{Maniyar}, A., {B{\'e}thermin}, M., \& {Lagache}, G. 2021, \aap, 645, A40

\bibitem[{{Maniyar} {et~al.}(2018){Maniyar}, {B{\'e}thermin}, \&
  {Lagache}}]{Maniyar2018}
{Maniyar}, A.~S., {B{\'e}thermin}, M., \& {Lagache}, G. 2018, Astronomy and
  Astrophysics, 614, A39

\bibitem[{{McCracken} {et~al.}(2015){McCracken}, {Wolk}, {Colombi},
  {Kilbinger}, {Ilbert}, {Peirani}, {Coupon}, {Dunlop}, {Milvang-Jensen},
  {Caputi}, {Aussel}, {B{\'e}thermin}, \& {Le F{\`e}vre}}]{McCracken2015}
{McCracken}, H.~J., {Wolk}, M., {Colombi}, S., {et~al.} 2015, \mnras, 449, 901

\bibitem[{{Moster} {et~al.}(2013){Moster}, {Naab}, \& {White}}]{Moster2013}
{Moster}, B.~P., {Naab}, T., \& {White}, S.~D.~M. 2013, \mnras, 428, 3121

\bibitem[{{Moster} {et~al.}(2018){Moster}, {Naab}, \& {White}}]{Moster2018}
{Moster}, B.~P., {Naab}, T., \& {White}, S. D.~M. 2018, \mnras, 477, 1822

\bibitem[{{Murmu} {et~al.}(2021){Murmu}, {Majumdar}, \& {Datta}}]{Murmu2021}
{Murmu}, C.~S., {Majumdar}, S., \& {Datta}, K.~K. 2021, \mnras, 507, 2500

\bibitem[{{Neben} {et~al.}(2017){Neben}, {Stalder}, {Hewitt}, \&
  {Tonry}}]{Neben2017}
{Neben}, A.~R., {Stalder}, B., {Hewitt}, J.~N., \& {Tonry}, J.~L. 2017, \apj,
  849, 50

\bibitem[{{Novak} {et~al.}(2017){Novak}, {Smol{\v{c}}i{\'c}}, {Delhaize},
  {Delvecchio}, {Zamorani}, {Baran}, {Bondi}, {Capak}, {Carilli}, {Ciliegi},
  {Civano}, {Ilbert}, {Karim}, {Laigle}, {Le F{\`e}vre}, {Marchesi},
  {McCracken}, {Miettinen}, {Salvato}, {Sargent}, {Schinnerer}, \&
  {Tasca}}]{Novak2017}
{Novak}, M., {Smol{\v{c}}i{\'c}}, V., {Delhaize}, J., {et~al.} 2017, \aap, 602,
  A5

\bibitem[{{Olsen} {et~al.}(2017){Olsen}, {Greve}, {Narayanan}, {Thompson},
  {Dav{\'e}}, {Niebla Rios}, \& {Stawinski}}]{Olsen2017}
{Olsen}, K., {Greve}, T.~R., {Narayanan}, D., {et~al.} 2017, \apj, 846, 105

\bibitem[{{Ota} {et~al.}(2014){Ota}, {Walter}, {Ohta}, {Hatsukade}, {Carilli},
  {da Cunha}, {Gonz{\'a}lez-L{\'o}pez}, {Decarli}, {Hodge}, {Nagai}, {Egami},
  {Jiang}, {Iye}, {Kashikawa}, {Riechers}, {Bertoldi}, {Cox}, {Neri}, \&
  {Weiss}}]{Ota2014}
{Ota}, K., {Walter}, F., {Ohta}, K., {et~al.} 2014, \apj, 792, 34

\bibitem[{{Ouchi} {et~al.}(2013){Ouchi}, {Ellis}, {Ono}, {Nakanishi}, {Kohno},
  {Momose}, {Kurono}, {Ashby}, {Shimasaku}, {Willner}, {Fazio}, {Tamura}, \&
  {Iono}}]{Ouchi2013}
{Ouchi}, M., {Ellis}, R., {Ono}, Y., {et~al.} 2013, The Astrophysical Journal,
  778, 102

\bibitem[{{Pallottini} {et~al.}(2019){Pallottini}, {Ferrara}, {Decataldo},
  {Gallerani}, {Vallini}, {Carniani}, {Behrens}, {Kohandel}, \&
  {Salvadori}}]{Pallottini2019}
{Pallottini}, A., {Ferrara}, A., {Decataldo}, D., {et~al.} 2019, \mnras, 487,
  1689

\bibitem[{{Pallottini} {et~al.}(2022){Pallottini}, {Ferrara}, {Gallerani},
  {Behrens}, {Kohandel}, {Carniani}, {Vallini}, {Salvadori}, {Gelli},
  {Sommovigo}, {D'Odorico}, {Di Mascia}, \& {Pizzati}}]{Pallottini2022}
{Pallottini}, A., {Ferrara}, A., {Gallerani}, S., {et~al.} 2022, arXiv
  e-prints, arXiv:2201.02636

\bibitem[{{Pallottini} {et~al.}(2015){Pallottini}, {Gallerani}, {Ferrara},
  {Yue}, {Vallini}, {Maiolino}, \& {Feruglio}}]{Pallottini2015}
{Pallottini}, A., {Gallerani}, S., {Ferrara}, A., {et~al.} 2015, \mnras, 453,
  1898

\bibitem[{pandas~development team(2020)}]{pandas}
pandas~development team, T. 2020, pandas-dev/pandas: Pandas

\bibitem[{{Papadopoulos} \& {Greve}(2004)}]{Papadopoulos2004}
{Papadopoulos}, P.~P. \& {Greve}, T.~R. 2004, \apjl, 615, L29

\bibitem[{{P{\'e}nin} {et~al.}(2012){P{\'e}nin}, {Dor{\'e}}, {Lagache}, \&
  {B{\'e}thermin}}]{Penin2012a}
{P{\'e}nin}, A., {Dor{\'e}}, O., {Lagache}, G., \& {B{\'e}thermin}, M. 2012,
  \aap, 537, A137

\bibitem[{{Pillepich} {et~al.}(2018){Pillepich}, {Springel}, {Nelson}, {Genel},
  {Naiman}, {Pakmor}, {Hernquist}, {Torrey}, {Vogelsberger}, {Weinberger}, \&
  {Marinacci}}]{Pillepich2018}
{Pillepich}, A., {Springel}, V., {Nelson}, D., {et~al.} 2018, \mnras, 473, 4077

\bibitem[{{Planck Collaboration} {et~al.}(2014){Planck Collaboration}, {Ade},
  {Aghanim}, {Armitage-Caplan}, {Arnaud}, {Ashdown}, {Atrio-Barandela},
  {Aumont}, {Baccigalupi}, {Banday}, \& et~al.}]{Planck_CIB2013}
{Planck Collaboration}, {Ade}, P.~A.~R., {Aghanim}, N., {et~al.} 2014, \aap,
  571, A30

\bibitem[{{Planck Collaboration} {et~al.}(2011){Planck Collaboration}, {Ade},
  {Aghanim}, {Arnaud}, {Ashdown}, {Aumont}, {Baccigalupi}, {Balbi}, {Banday},
  {Barreiro}, \& et~al.}]{Planck_CIB2011}
{Planck Collaboration}, {Ade}, P.~A.~R., {Aghanim}, N., {et~al.} 2011, \aap,
  536, A18

\bibitem[{{Planck Collaboration} {et~al.}(2016){Planck Collaboration}, {Ade},
  {Aghanim}, {Arnaud}, {Ashdown}, {Aumont}, {Baccigalupi}, {Banday},
  {Barreiro}, {Bartlett}, \& et~al.}]{Planck2015_cosmo}
{Planck Collaboration}, {Ade}, P.~A.~R., {Aghanim}, N., {et~al.} 2016, \aap,
  594, A13

\bibitem[{{Popping} {et~al.}(2019){Popping}, {Narayanan}, {Somerville},
  {Faisst}, \& {Krumholz}}]{Popping2019}
{Popping}, G., {Narayanan}, D., {Somerville}, R.~S., {Faisst}, A.~L., \&
  {Krumholz}, M.~R. 2019, \mnras, 482, 4906

\bibitem[{{Popping} {et~al.}(2014){Popping}, {Somerville}, \&
  {Trager}}]{Popping2014}
{Popping}, G., {Somerville}, R.~S., \& {Trager}, S.~C. 2014, \mnras, 442, 2398

\bibitem[{{Price-Whelan} {et~al.}(2018){Price-Whelan}, {Sip{\H{o}}cz},
  {G{\"u}nther}, {Lim}, {Crawford}, {Conseil}, {Shupe}, {Craig}, {Dencheva},
  {Ginsburg}, {VanderPlas}, {Bradley}, {P{\'e}rez-Su{\'a}rez}, {de Val-Borro},
  {Paper Contributors}, {Aldcroft}, {Cruz}, {Robitaille}, {Tollerud},
  {Coordination Committee}, {Ardelean}, {Babej}, {Bach}, {Bachetti}, {Bakanov},
  {Bamford}, {Barentsen}, {Barmby}, {Baumbach}, {Berry}, {Biscani}, {Boquien},
  {Bostroem}, {Bouma}, {Brammer}, {Bray}, {Breytenbach}, {Buddelmeijer},
  {Burke}, {Calderone}, {Cano Rodr{\'\i}guez}, {Cara}, {Cardoso}, {Cheedella},
  {Copin}, {Corrales}, {Crichton}, {D{\textquoteright}Avella}, {Deil},
  {Depagne}, {Dietrich}, {Donath}, {Droettboom}, {Earl}, {Erben}, {Fabbro},
  {Ferreira}, {Finethy}, {Fox}, {Garrison}, {Gibbons}, {Goldstein}, {Gommers},
  {Greco}, {Greenfield}, {Groener}, {Grollier}, {Hagen}, {Hirst}, {Homeier},
  {Horton}, {Hosseinzadeh}, {Hu}, {Hunkeler}, {Ivezi{\'c}}, {Jain}, {Jenness},
  {Kanarek}, {Kendrew}, {Kern}, {Kerzendorf}, {Khvalko}, {King}, {Kirkby},
  {Kulkarni}, {Kumar}, {Lee}, {Lenz}, {Littlefair}, {Ma}, {Macleod},
  {Mastropietro}, {McCully}, {Montagnac}, {Morris}, {Mueller}, {Mumford},
  {Muna}, {Murphy}, {Nelson}, {Nguyen}, {Ninan}, {N{\"o}the}, {Ogaz}, {Oh},
  {Parejko}, {Parley}, {Pascual}, {Patil}, {Patil}, {Plunkett}, {Prochaska},
  {Rastogi}, {Reddy Janga}, {Sabater}, {Sakurikar}, {Seifert}, {Sherbert},
  {Sherwood-Taylor}, {Shih}, {Sick}, {Silbiger}, {Singanamalla}, {Singer},
  {Sladen}, {Sooley}, {Sornarajah}, {Streicher}, {Teuben}, {Thomas},
  {Tremblay}, {Turner}, {Terr{\'o}n}, {van Kerkwijk}, {de la Vega}, {Watkins},
  {Weaver}, {Whitmore}, {Woillez}, {Zabalza}, \& {Contributors}}]{astropy:2018}
{Price-Whelan}, A.~M., {Sip{\H{o}}cz}, B.~M., {G{\"u}nther}, H.~M., {et~al.}
  2018, \aj, 156, 123

\bibitem[{{Riechers} {et~al.}(2019){Riechers}, {Pavesi}, {Sharon}, {Hodge},
  {Decarli}, {Walter}, {Carilli}, {Aravena}, {da Cunha}, {Daddi}, {Dickinson},
  {Smail}, {Capak}, {Ivison}, {Sargent}, {Scoville}, \& {Wagg}}]{Riechers2019}
{Riechers}, D.~A., {Pavesi}, R., {Sharon}, C.~E., {et~al.} 2019, \apj, 872, 7

\bibitem[{{Rodr{\'{\i}}guez-Puebla} {et~al.}(2016){Rodr{\'{\i}}guez-Puebla},
  {Behroozi}, {Primack}, {Klypin}, {Lee}, \&
  {Hellinger}}]{Rodriguez-Puebla2016}
{Rodr{\'{\i}}guez-Puebla}, A., {Behroozi}, P., {Primack}, J., {et~al.} 2016,
  \mnras, 462, 893

\bibitem[{{Rosenberg} {et~al.}(2015){Rosenberg}, {van der Werf}, {Aalto},
  {Armus}, {Charmandaris}, {D{\'{\i}}az-Santos}, {Evans}, {Fischer}, {Gao},
  {Gonz{\'a}lez-Alfonso}, {Greve}, {Harris}, {Henkel}, {Israel}, {Isaak},
  {Kramer}, {Meijerink}, {Naylor}, {Sanders}, {Smith}, {Spaans}, {Spinoglio},
  {Stacey}, {Veenendaal}, {Veilleux}, {Walter}, {Wei{\ss}}, {Wiedner}, {van der
  Wiel}, \& {Xilouris}}]{Rosenberg2015}
{Rosenberg}, M.~J.~F., {van der Werf}, P.~P., {Aalto}, S., {et~al.} 2015, \apj,
  801, 72

\bibitem[{{Rowan-Robinson} {et~al.}(2016){Rowan-Robinson}, {Oliver}, {Wang},
  {Farrah}, {Clements}, {Gruppioni}, {Marchetti}, {Rigopoulou}, \&
  {Vaccari}}]{Rowan-Robinson2016}
{Rowan-Robinson}, M., {Oliver}, S., {Wang}, L., {et~al.} 2016, \mnras, 461,
  1100

\bibitem[{{Saintonge} {et~al.}(2013){Saintonge}, {Lutz}, {Genzel}, {Magnelli},
  {Nordon}, {Tacconi}, {Baker}, {Bandara}, {Berta}, {F{\"o}rster Schreiber},
  {Poglitsch}, {Sturm}, {Wuyts}, \& {Wuyts}}]{Saintonge2013}
{Saintonge}, A., {Lutz}, D., {Genzel}, R., {et~al.} 2013, \apj, 778, 2

\bibitem[{{Sargent} {et~al.}(2014){Sargent}, {Daddi}, {B{\'e}thermin},
  {Aussel}, {Magdis}, {Hwang}, {Juneau}, {Elbaz}, \& {da Cunha}}]{Sargent2014}
{Sargent}, M.~T., {Daddi}, E., {B{\'e}thermin}, M., {et~al.} 2014, \apj, 793,
  19

\bibitem[{{Schaerer} {et~al.}(2020){Schaerer}, {Ginolfi}, {Bethermin},
  {Fudamoto}, {Oesch}, {Le Fevre}, {Faisst}, {Capak}, {Cassata}, {Silverman},
  {Yan}, {Jones}, {Amorin}, {Bardelli}, {Boquien}, {Cimatti},
  {Dessauges-Zavadsky}, {Giavalisco}, {Hathi}, {Fujimoto}, {Ibar}, {Koekemoer},
  {Lagache}, {Lemaux}, {Loiacono}, {Maiolino}, {Narayanan}, {Morselli},
  {Mendez-Hernandez}, {Pozzi}, {Riechers}, {Talia}, {Toft}, {Vallini},
  {Vergani}, {Zamorani}, \& {Zucca}}]{Schaerer2020}
{Schaerer}, D., {Ginolfi}, M., {Bethermin}, M., {et~al.} 2020, arXiv e-prints,
  arXiv:2002.00979

\bibitem[{{Schenker} {et~al.}(2013){Schenker}, {Robertson}, {Ellis}, {Ono},
  {McLure}, {Dunlop}, {Koekemoer}, {Bowler}, {Ouchi}, {Curtis-Lake}, {Rogers},
  {Schneider}, {Charlot}, {Stark}, {Furlanetto}, \& {Cirasuolo}}]{Schenker2013}
{Schenker}, M.~A., {Robertson}, B.~E., {Ellis}, R.~S., {et~al.} 2013, \apj,
  768, 196

\bibitem[{{Schreiber} {et~al.}(2015){Schreiber}, {Pannella}, {Elbaz},
  {B{\'e}thermin}, {Inami}, {Dickinson}, {Magnelli}, {Wang}, {Aussel}, {Daddi},
  {Juneau}, {Shu}, {Sargent}, {Buat}, {Faber}, {Ferguson}, {Giavalisco},
  {Koekemoer}, {Magdis}, {Morrison}, {Papovich}, {Santini}, \&
  {Scott}}]{Schreiber2015}
{Schreiber}, C., {Pannella}, M., {Elbaz}, D., {et~al.} 2015, \aap, 575, A74

\bibitem[{{Shang} {et~al.}(2012){Shang}, {Haiman}, {Knox}, \& {Oh}}]{Shang2012}
{Shang}, C., {Haiman}, Z., {Knox}, L., \& {Oh}, S.~P. 2012, \mnras, 421, 2832

\bibitem[{{Silva} {et~al.}(2015){Silva}, {Santos}, {Cooray}, \&
  {Gong}}]{Silva2015}
{Silva}, M., {Santos}, M.~G., {Cooray}, A., \& {Gong}, Y. 2015, \apj, 806, 209

\bibitem[{{Solomon} {et~al.}(1997){Solomon}, {Downes}, {Radford}, \&
  {Barrett}}]{Solomon1997}
{Solomon}, P.~M., {Downes}, D., {Radford}, S.~J.~E., \& {Barrett}, J.~W. 1997,
  \apj, 478, 144

\bibitem[{{Solomon} \& {Vanden Bout}(2005)}]{Solomon2005}
{Solomon}, P.~M. \& {Vanden Bout}, P.~A. 2005, \araa, 43, 677

\bibitem[{{Stacey} {et~al.}(2018){Stacey}, {Aravena}, {Basu}, {Battaglia},
  {Beringue}, {Bertoldi}, {Bond}, {Breysse}, {Bustos}, {Chapman}, {Chung},
  {Cothard}, {Erler}, {Fich}, {Foreman}, {Gallardo}, {Giovanelli}, {Graf},
  {Haynes}, {Herrera-Camus}, {Herter}, {Hlo{\v{z}}ek}, {Johnstone}, {Keating},
  {Magnelli}, {Meerburg}, {Meyers}, {Murray}, {Niemack}, {Nikola}, {Nolta},
  {Parshley}, {Riechers}, {Schilke}, {Scott}, {Stein}, {Stevens}, {Stutzki},
  {Vavagiakis}, \& {Viero}}]{Stacey2018}
{Stacey}, G.~J., {Aravena}, M., {Basu}, K., {et~al.} 2018, in Society of
  Photo-Optical Instrumentation Engineers (SPIE) Conference Series, Vol. 10700,
  Ground-based and Airborne Telescopes VII, ed. H.~K. {Marshall} \&
  J.~{Spyromilio}, 107001M

\bibitem[{{Sun} {et~al.}(2018){Sun}, {Moncelsi}, {Viero}, {Silva}, {Bock},
  {Bradford}, {Chang}, {Cheng}, {Cooray}, {Crites}, {Hailey-Dunsheath},
  {Uzgil}, {Hunacek}, \& {Zemcov}}]{Sun2018}
{Sun}, G., {Moncelsi}, L., {Viero}, M.~P., {et~al.} 2018, \apj, 856, 107

\bibitem[{{Tacconi} {et~al.}(2020){Tacconi}, {Genzel}, \&
  {Sternberg}}]{Tacconi2020}
{Tacconi}, L.~J., {Genzel}, R., \& {Sternberg}, A. 2020, \araa, 58, 157

\bibitem[{{Tacconi} {et~al.}(2013){Tacconi}, {Neri}, {Genzel}, {Combes},
  {Bolatto}, {Cooper}, {Wuyts}, {Bournaud}, {Burkert}, {Comerford}, {Cox},
  {Davis}, {F{\"o}rster Schreiber}, {Garc{\'{\i}}a-Burillo}, {Gracia-Carpio},
  {Lutz}, {Naab}, {Newman}, {Omont}, {Saintonge}, {Shapiro Griffin}, {Shapley},
  {Sternberg}, \& {Weiner}}]{Tacconi2013}
{Tacconi}, L.~J., {Neri}, R., {Genzel}, R., {et~al.} 2013, \apj, 768, 74

\bibitem[{{Talia} {et~al.}(2021){Talia}, {Cimatti}, {Giulietti}, {Zamorani},
  {Bethermin}, {Faisst}, {Le F{\`e}vre}, \& {Smol{\c{c}}i{\'c}}}]{Talia2021}
{Talia}, M., {Cimatti}, A., {Giulietti}, M., {et~al.} 2021, \apj, 909, 23

\bibitem[{{Tielens} \& {Hollenbach}(1985)}]{Tielens1985}
{Tielens}, A.~G.~G.~M. \& {Hollenbach}, D. 1985, \apj, 291, 722

\bibitem[{{Tinker} {et~al.}(2010){Tinker}, {Robertson}, {Kravtsov}, {Klypin},
  {Warren}, {Yepes}, \& {Gottl{\"o}ber}}]{Tinker2010}
{Tinker}, J.~L., {Robertson}, B.~E., {Kravtsov}, A.~V., {et~al.} 2010, \apj,
  724, 878

\bibitem[{{Uzgil} {et~al.}(2021){Uzgil}, {Oesch}, {Walter}, {Aravena},
  {Boogaard}, {Carilli}, {Decarli}, {D{\'\i}az-Santos}, {Fudamoto}, {Inami},
  {Bouwens}, {Cortes}, {Cox}, {Daddi}, {Gonz{\'a}lez-L{\'o}pez}, {Labbe},
  {Popping}, {Riechers}, {Stefanon}, {Van der Werf}, \& {Weiss}}]{Uzgil2021}
{Uzgil}, B.~D., {Oesch}, P.~A., {Walter}, F., {et~al.} 2021, \apj, 912, 67

\bibitem[{{Valentino} {et~al.}(2020{\natexlab{a}}){Valentino}, {Daddi},
  {Puglisi}, {Magdis}, {Liu}, {Kokorev}, {Cortzen}, {Madden}, {Aravena},
  {G{\'o}mez-Guijarro}, {Lee}, {Le Floc'h}, {Gao}, {Gobat}, {Bournaud},
  {Dannerbauer}, {Jin}, {Dickinson}, {Kartaltepe}, \&
  {Sanders}}]{Valentino2020b}
{Valentino}, F., {Daddi}, E., {Puglisi}, A., {et~al.} 2020{\natexlab{a}}, \aap,
  641, A155

\bibitem[{{Valentino} {et~al.}(2020{\natexlab{b}}){Valentino}, {Magdis},
  {Daddi}, {Liu}, {Aravena}, {Bournaud}, {Cortzen}, {Gao}, {Jin}, {Juneau},
  {Kartaltepe}, {Kokorev}, {Lee}, {Madden}, {Narayanan}, {Popping}, \&
  {Puglisi}}]{Valentino2020}
{Valentino}, F., {Magdis}, G.~E., {Daddi}, E., {et~al.} 2020{\natexlab{b}},
  \apj, 890, 24

\bibitem[{{Vallini} {et~al.}(2015){Vallini}, {Gallerani}, {Ferrara},
  {Pallottini}, \& {Yue}}]{Vallini2015}
{Vallini}, L., {Gallerani}, S., {Ferrara}, A., {Pallottini}, A., \& {Yue}, B.
  2015, The Astrophysical Journal, 813, 36

\bibitem[{{Vallini} {et~al.}(2016){Vallini}, {Gruppioni}, {Pozzi}, {Vignali},
  \& {Zamorani}}]{Vallini2016}
{Vallini}, L., {Gruppioni}, C., {Pozzi}, F., {Vignali}, C., \& {Zamorani}, G.
  2016, \mnras, 456, L40

\bibitem[{{Vallini} {et~al.}(2018){Vallini}, {Pallottini}, {Ferrara},
  {Gallerani}, {Sobacchi}, \& {Behrens}}]{Vallini2018}
{Vallini}, L., {Pallottini}, A., {Ferrara}, A., {et~al.} 2018, \mnras, 473, 271

\bibitem[{{Viero} {et~al.}(2009){Viero}, {Ade}, {Bock}, {Chapin}, {Devlin},
  {Griffin}, {Gundersen}, {Halpern}, {Hargrave}, {Hughes}, {Klein},
  {MacTavish}, {Marsden}, {Martin}, {Mauskopf}, {Moncelsi}, {Negrello},
  {Netterfield}, {Olmi}, {Pascale}, {Patanchon}, {Rex}, {Scott}, {Semisch},
  {Thomas}, {Truch}, {Tucker}, {Tucker}, \& {Wiebe}}]{Viero2009}
{Viero}, M.~P., {Ade}, P.~A.~R., {Bock}, J.~J., {et~al.} 2009, \apj, 707, 1766

\bibitem[{{Viero} {et~al.}(2013){Viero}, {Wang}, {Zemcov}, {Addison},
  {Amblard}, {Arumugam}, {Aussel}, {B{\'e}thermin}, {Bock}, {Boselli}, {Buat},
  {Burgarella}, {Casey}, {Clements}, {Conley}, {Conversi}, {Cooray}, {De
  Zotti}, {Dowell}, {Farrah}, {Franceschini}, {Glenn}, {Griffin},
  {Hatziminaoglou}, {Heinis}, {Ibar}, {Ivison}, {Lagache}, {Levenson},
  {Marchetti}, {Marsden}, {Nguyen}, {O'Halloran}, {Oliver}, {Omont}, {Page},
  {Papageorgiou}, {Pearson}, {P{\'e}rez-Fournon}, {Pohlen}, {Rigopoulou},
  {Roseboom}, {Rowan-Robinson}, {Schulz}, {Scott}, {Seymour}, {Shupe}, {Smith},
  {Symeonidis}, {Vaccari}, {Valtchanov}, {Vieira}, {Wardlow}, \&
  {Xu}}]{Viero2013}
{Viero}, M.~P., {Wang}, L., {Zemcov}, M., {et~al.} 2013, \apj, 772, 77

\bibitem[{{Walter} {et~al.}(2016){Walter}, {Decarli}, {Aravena}, {Carilli},
  {Bouwens}, {da Cunha}, {Daddi}, {Ivison}, {Riechers}, {Smail}, {Swinbank},
  {Weiss}, {Anguita}, {Assef}, {Bacon}, {Bauer}, {Bell}, {Bertoldi}, {Chapman},
  {Colina}, {Cortes}, {Cox}, {Dickinson}, {Elbaz}, {G{\'o}nzalez-L{\'o}pez},
  {Ibar}, {Inami}, {Infante}, {Hodge}, {Karim}, {Le Fevre}, {Magnelli}, {Neri},
  {Oesch}, {Ota}, {Popping}, {Rix}, {Sargent}, {Sheth}, {van der Wel}, {van der
  Werf}, \& {Wagg}}]{Walter2016}
{Walter}, F., {Decarli}, R., {Aravena}, M., {et~al.} 2016, \apj, 833, 67

\bibitem[{{Walter} {et~al.}(2014){Walter}, {Decarli}, {Sargent}, {Carilli},
  {Dickinson}, {Riechers}, {Ellis}, {Stark}, {Weiner}, {Aravena}, {Bell},
  {Bertoldi}, {Cox}, {Da Cunha}, {Daddi}, {Downes}, {Lentati}, {Maiolino},
  {Menten}, {Neri}, {Rix}, \& {Weiss}}]{Walter2014}
{Walter}, F., {Decarli}, R., {Sargent}, M., {et~al.} 2014, \apj, 782, 79

\bibitem[{{Wang} {et~al.}(2013){Wang}, {Farrah}, {Oliver}, {Amblard},
  {B{\'e}thermin}, {Bock}, {Conley}, {Cooray}, {Halpern}, {Heinis}, {Ibar},
  {Ilbert}, {Ivison}, {Marsden}, {Roseboom}, {Rowan-Robinson}, {Schulz},
  {Smith}, {Viero}, \& {Zemcov}}]{Wang2013}
{Wang}, L., {Farrah}, D., {Oliver}, S.~J., {et~al.} 2013, \mnras, 431, 648

\bibitem[{{Wang} {et~al.}(2021){Wang}, {Gao}, {Best}, {Duncan}, {Hardcastle},
  {Kondapally}, {Ma{\l}ek}, {McCheyne}, {Sabater}, {Shimwell}, {Tasse},
  {Bonato}, {Bondi}, {Cochrane}, {Farrah}, {G{\"u}rkan}, {Haskell}, {Pearson},
  {Prandoni}, {R{\"o}ttgering}, {Smith}, {Vaccari}, \& {Williams}}]{Wang2021}
{Wang}, L., {Gao}, F., {Best}, P.~N., {et~al.} 2021, \aap, 648, A8

\bibitem[{{Wang} {et~al.}(2019{\natexlab{a}}){Wang}, {Pearson}, {Cowley},
  {Trayford}, {B{\'e}thermin}, {Gruppioni}, {Hurley}, \&
  {Micha{\l}owski}}]{WangL2019}
{Wang}, L., {Pearson}, W.~J., {Cowley}, W., {et~al.} 2019{\natexlab{a}}, \aap,
  624, A98

\bibitem[{{Wang} {et~al.}(2019{\natexlab{b}}){Wang}, {Schreiber}, {Elbaz},
  {Yoshimura}, {Kohno}, {Shu}, {Yamaguchi}, {Pannella}, {Franco}, {Huang},
  {Lim}, \& {Wang}}]{WangT2019}
{Wang}, T., {Schreiber}, C., {Elbaz}, D., {et~al.} 2019{\natexlab{b}}, \nat,
  572, 211

\bibitem[{{Wang} \& {Hu}(2006)}]{Wang2006}
{Wang}, X. \& {Hu}, W. 2006, \apj, 643, 585

\bibitem[{{Wilkinson} {et~al.}(2017){Wilkinson}, {Almaini}, {Chen}, {Smail},
  {Arumugam}, {Blain}, {Chapin}, {Chapman}, {Conselice}, {Cowley}, {Dunlop},
  {Farrah}, {Geach}, {Hartley}, {Ivison}, {Maltby}, {Micha{\l}owski},
  {Mortlock}, {Scott}, {Simpson}, {Simpson}, {van der Werf}, \&
  {Wild}}]{Wilkinson2017}
{Wilkinson}, A., {Almaini}, O., {Chen}, C.-C., {et~al.} 2017, \mnras, 464, 1380

\bibitem[{{Willott} {et~al.}(2015){Willott}, {Carilli}, {Wagg}, \&
  {Wang}}]{Wilott2015}
{Willott}, C.~J., {Carilli}, C.~L., {Wagg}, J., \& {Wang}, R. 2015, \apj, 807,
  180

\bibitem[{{Wolfire} {et~al.}(2022){Wolfire}, {Vallini}, \&
  {Chevance}}]{Wolfire2022}
{Wolfire}, M.~G., {Vallini}, L., \& {Chevance}, M. 2022, arXiv e-prints,
  arXiv:2202.05867

\bibitem[{{Yan} {et~al.}(2020){Yan}, {Sajina}, {Loiacono}, {Lagache},
  {B{\'e}thermin}, {Faisst}, {Ginolfi}, {F{\`e}vre}, {Gruppioni}, {Capak},
  {Cassata}, {Schaerer}, {Silverman}, {Bardelli}, {Dessauges-Zavadsky},
  {Cimatti}, {Hathi}, {Lemaux}, {Ibar}, {Jones}, {Koekemoer}, {Oesch}, {Talia},
  {Pozzi}, {Riechers}, {Tasca}, {Toft}, {Vallini}, {Vergani}, {Zamorani}, \&
  {Zucca}}]{Yan2020}
{Yan}, L., {Sajina}, A., {Loiacono}, F., {et~al.} 2020, \apj, 905, 147

\bibitem[{{Yang} {et~al.}(2017){Yang}, {Omont}, {Beelen}, {Gao}, {van der
  Werf}, {Gavazzi}, {Zhang}, {Ivison}, {Lehnert}, {Liu}, {Oteo},
  {Gonz{\'a}lez-Alfonso}, {Dannerbauer}, {Cox}, {Krips}, {Neri}, {Riechers},
  {Baker}, {Micha{\l}owski}, {Cooray}, \& {Smail}}]{Yang2017}
{Yang}, C., {Omont}, A., {Beelen}, A., {et~al.} 2017, \aap, 608, A144

\bibitem[{{Yang} {et~al.}(2021{\natexlab{a}}){Yang}, {Popping}, {Somerville},
  {Pullen}, {Breysse}, \& {Maniyar}}]{Yang2021}
{Yang}, S., {Popping}, G., {Somerville}, R.~S., {et~al.} 2021{\natexlab{a}},
  arXiv e-prints, arXiv:2108.07716

\bibitem[{{Yang} {et~al.}(2021{\natexlab{b}}){Yang}, {Somerville}, {Pullen},
  {Popping}, {Breysse}, \& {Maniyar}}]{Yang2021sam}
{Yang}, S., {Somerville}, R.~S., {Pullen}, A.~R., {et~al.} 2021{\natexlab{b}},
  \apj, 911, 132

\bibitem[{{Yue} \& {Ferrara}(2019)}]{Yue2019}
{Yue}, B. \& {Ferrara}, A. 2019, \mnras, 490, 1928

\bibitem[{{Yue} {et~al.}(2015){Yue}, {Ferrara}, {Pallottini}, {Gallerani}, \&
  {Vallini}}]{Yue2015}
{Yue}, B., {Ferrara}, A., {Pallottini}, A., {Gallerani}, S., \& {Vallini}, L.
  2015, \mnras, 450, 3829

\bibitem[{{Zavala} {et~al.}(2021){Zavala}, {Casey}, {Manning}, {Aravena},
  {Bethermin}, {Caputi}, {Clements}, {Cunha}, {Drew}, {Finkelstein},
  {Fujimoto}, {Hayward}, {Hodge}, {Kartaltepe}, {Knudsen}, {Koekemoer}, {Long},
  {Magdis}, {Man}, {Popping}, {Sanders}, {Scoville}, {Sheth}, {Staguhn},
  {Toft}, {Treister}, {Vieira}, \& {Yun}}]{Zavala2021}
{Zavala}, J.~A., {Casey}, C.~M., {Manning}, S.~M., {et~al.} 2021, \apj, 909,
  165

\end{thebibliography}


\begin{appendix}

\section{Description of the release}

\label{app:release}

The code and a set of products are released together with this paper. The Python code is released publicly at \url{https://gitlab.lam.fr/mbethermin/sides-public-release}. It contains the routines used to generate the catalogs, the cubes, and the maps used in this paper. The jupyter notebooks used to produce the plots are also released. We also release the simulated catalog used in this paper (\url{https://cesamsi.lam.fr/instance/sides/home}). The columns are described in Table\,\ref{tab:release}. Finally, the cubes are available at the same address.

\begin{table*}
\caption{\label{tab:release} Description of the columns of the simulated catalog released together with this paper.}
\centering
\begin{tabular}{lll}
\hline
\hline
Name & Unit & Description\\
\hline
\verb redshift & & observed redshift\\
\verb ra & deg & Right Ascension \\
\verb dec & deg & Declination \\
\verb Mhalo & M$_\odot$ & Host halo mass \\
\verb Mstar & M$_\odot$ & Stellar mass \\
\verb qflag & & quenching flag (\textit{True} if quenched) \\
\verb SFR & M$_\odot$/yr & star formation rate \\
\verb issb & & starburst flag (\textit{True} if starburst)\\
\verb mu & & lensing magnification \\
\verb Dlum & Mpc & luminosity distance \\
\verb Umean & & intensity of the radiation field \\
\verb LIR & L$_\odot$ & infrared luminosity (8--1000\,$\mu$m)\\
\verb S[W] & Jy & Monochromatic flux density at [W] microns ([W]: wavelength) \\
\verb S[F] & Jy & Flux density in the [F] filter ([F]: photometric filter name) \\
\verb LprimCO10 & K\,km/s\,pc$^2$ & Pseudo luminosity of CO(1-0)\\
\verb ICO[J][J-1] & Jy\,km/s & Flux of the CO[J][J-1] line ([J]: upper rotational level of the transition) \\ 
\verb LCII_Lagache & L$_\odot$ & Luminosity of [CII] line using the L18 relation\\ 
\verb ICII_Lagache & Jy\,km/s & Flux of the [CII] line using the L18 relation\\
\verb LCII_de_Looze & L$_\odot$ & Luminosity of [CII] line using the DL14 relation\\ 
\verb ICII_de_Looze & Jy\,km/s & Flux of the [CII] line using the DL14 relation\\
\verb ICI10 & Jy\,km/s & Flux of the [CI](1-0) line\\
\verb ICI21 & Jy\,km/s & Flux of the [CI](2-1) line\\
\hline
\end{tabular}
\end{table*}

\section{Relation between the CO and infrared luminosity at low redshift}

\label{sect:lowz_scaling_CO}

\begin{figure*}
\centering
\includegraphics[width=15cm]{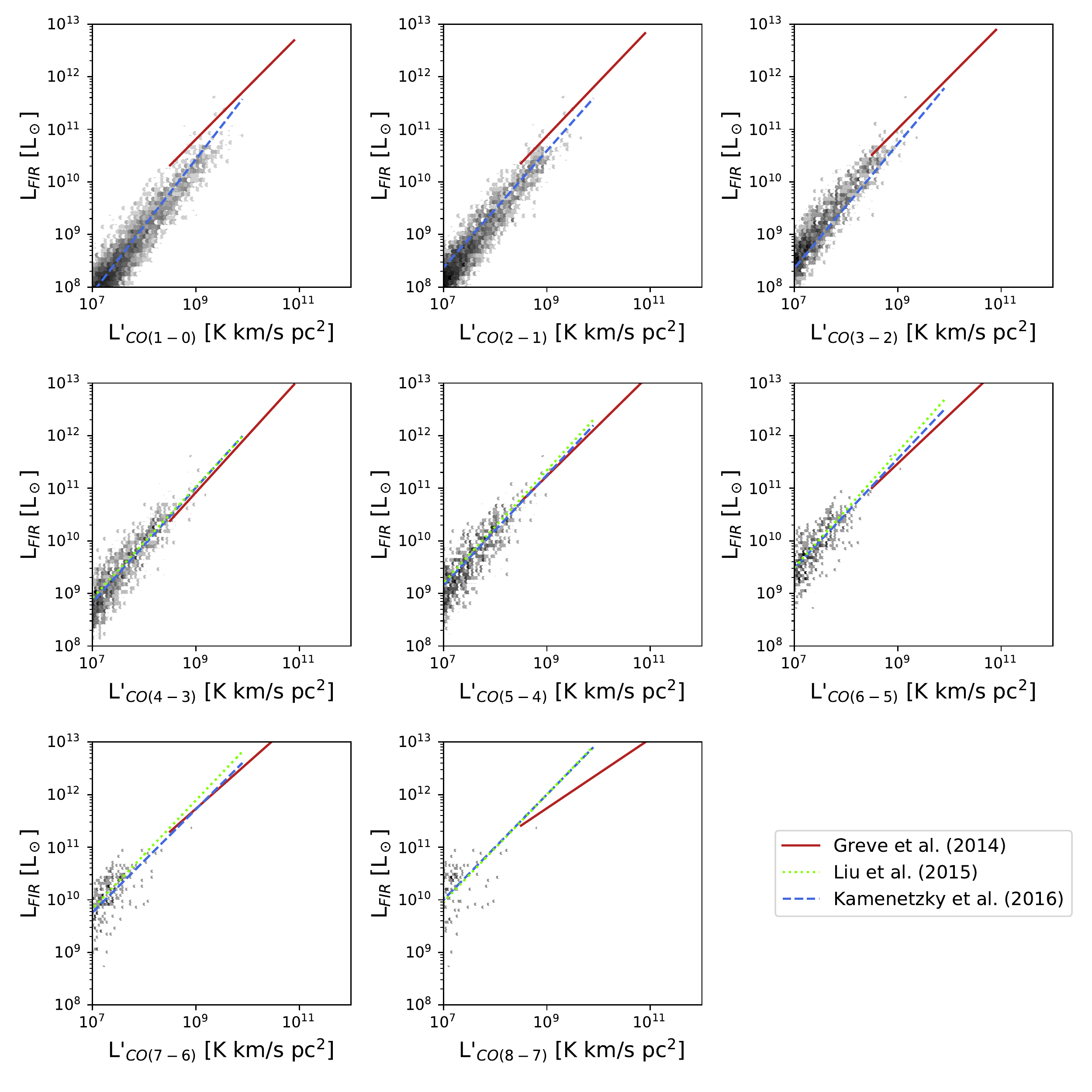}
\caption{\label{fig:lowz_scaling_CO} Same plot as Fig.\,\ref{fig:CO_scale}, but considering on the z$<0.2$ objects in SIDES.}
\end{figure*}

As we discussed in Sect.\,\ref{sect:CO_scale}, there is a small offset between the observed L$_{\rm IR}$-L'$_{\rm CO}$ relations and SIDES. However, these relations are measured in the local Universe. As shown by Fig.\,\ref{fig:lowz_scaling_CO}, there is no offset if we consider only z$<$0.2 objects in SIDES. The 0.2 redshift limit was chosen as a compromise between keeping the lowest possible redshift and having a sufficiently large volume to have statistics.

\section{Analytical computation of the [CII] background from the luminosity function}

\label{app:CII_bkg}

The [CII] background at a frequency $\nu_{\rm obs}$ is directly connected to the [CII] luminosity function at a redshift $z = (\nu_{\rm [CII], rest} / \nu_{\rm obs})-1$ through the equation:
\begin{equation}
\label{eq:CII_background}
B_\nu^{\rm [CII]} (\nu_{\rm obs}) = \int S_\nu^{\rm [CII]} \frac{d^2 N}{dL_{\rm [CII]} dV_C} \frac{d V_C}{d \Omega} \, dL_{\rm [CII]} ,
\end{equation}
where $B_\nu^{\rm [CII]}$ is the surface brightness density of the [CII] background, $S_\nu^{\rm [CII]}$ is flux density corresponding a line with a luminosity $L_{\rm [CII]}$ at the redshift z, and $\frac{d^2 N}{dL_{\rm [CII]} dV}$ is the [CII] luminosity function, and $\frac{dV}{d \Omega}$ is the differential comoving volume associated to a solid angle $d \Omega$ and a small frequency interval d$\nu_{\rm obs}$. 

The [CII] flux density depends on the frequency interval $d \nu$ and can be obtained using the \citet{Carilli2013} formula to convert luminosities into fluxes:
\begin{equation}
\label{eq:flux_from_lum}
S_\nu^{\rm [CII]} = \frac{L_{\rm [CII]}}{C d \upsilon \, \nu_{\rm obs} \, D_L^2} = \frac{L_{\rm [CII]}}{C \, c \, d \nu_{\rm obs} \, D_L^2},
\end{equation}
where $S_\nu^{\rm [CII]}$, $L_{\rm [CII]}$, and $D_L$ are expressed in Jy, L$_\odot$, and Mpc, respectively. Velocities as c and $d \upsilon$ are expressed in km/s and frequencies are expressed in GHz. $C$ is constant and equal to $1.04 \times 10^{-3}$\,L$_\odot$\,(Jy\,km\,s$^{-1}$\,GHz\,Mpc$^2$)$^{-1}$.\footnote{We do not simply this unit to make it easier to compute from observational units.}

The differential comoving volume elements $\frac{dV_C}{d \Omega}$ is (see, e.g., \citealt{Hogg1999}):
\begin{equation}
\label{eq:diff_volume}
\frac{dV_C}{d \Omega} = \frac{c \, D_c^2 \, dz}{H(z)} = \frac{c \, D_c^2 \, (1+z) \, d \nu_{\rm obs}}{H(z)\,\nu_{\rm obs} },  
\end{equation}
where $H(z)$ is the Hubble parameter at a redshift $z$ ($H_0 \, \sqrt{\Omega_m (1+z)^3 + \Omega_\Lambda}$ for a flat cosmology).

By combining Eq.\,\ref{eq:CII_background}, \ref{eq:flux_from_lum}, and \ref{eq:diff_volume}, we obtain:
\begin{equation}
B_\nu^{\rm [CII]} (\nu_{\rm obs}) = \frac{1}{C \, (1+z) \, \nu_{\rm obs}  \, H(z)} \int L_{\rm [CII]}  \, \frac{d^2 N}{dL_{\rm [CII]} dV_C} \, dL_{\rm [CII]}.
\end{equation}
The result is in Jy/sr if the units mentioned previously are used. We can remark that the two terms in $d \nu$ from the previous equation cancel each other. This is normal, since the the background does not depend on the resolution.

The exact same computation can be performed for the various CO and [CI] transitions.

\section{Analytical computation of the [CII] shot-noise from the luminosity function}

\label{app:CII_poisson}

The Poisson part of the 2-dimensional power spectrum of the [CII] ($P_{\rm Poi}^{\rm [CII]}$) from a thin frequency slice with a center $\nu_{\rm obs}$ and a width $\Delta \nu$ can be computed using a similar approach as in Sect.\,\ref{app:CII_bkg}.
\begin{equation}
\label{eq:CII_SN_2D}
P_{\rm Poi}^{\rm [CII]} (\nu_{\rm obs}) = \int \left ( S_\nu^{\rm [CII]} \right )^2 \frac{d^2 N}{dL_{\rm [CII]} dV_C} \frac{d V_C}{d \Omega} \, dL_{\rm [CII]}.
\end{equation}
The flux density squared term is similar to the one in the computation of the CIB shot noise \citep[e.g.,][]{Lagache2003}.

We then combine the Eq.\,\ref{eq:CII_SN_2D}, \ref{eq:flux_from_lum}, and \ref{eq:diff_volume} and obtain:
\begin{equation}
P_{\rm Poi}^{\rm [CII]} (\nu_{\rm obs}) = \frac{1}{C^2 \, c \, D_L^2 \, (1+z) \, H(z) \, \nu_{\rm obs} \Delta \nu_{\rm obs}} \int L_{\rm [CII]}^2 \, \frac{d^2 N}{dL_{\rm [CII]} dV_C} \, dL_{\rm [CII]}.
\end{equation}
We remark that the terms in $d \nu$ does not simplify and the Poisson term is inversely proportional to $\Delta \nu$. This behavior is expected, since doubling the frequency width dilutes the line fluxes by a factor of 2 and the power spectrum by a factor of 4, while the power spectra of two independent redshift slices sum linearly (factor of 2). However we can define a quantity independent of the spectral resolution:
\begin{equation}
\label{eq:norm_poisson}
\widetilde{P}_{\rm Poi}^{\rm [CII]} = {P}_{\rm Poi}^{\rm [CII]} \times \Delta \nu_{\rm obs}.
\end{equation}
This quantity is used for instance by \citet{Keating2020} to report the CO shot-noise measured around 100\,GHz. This normalization applies only to the shot-noise component and not to the large-scale clustering term, since the frequency slices are no longer independent in presence of clustering.

The 3-dimensional power spectrum can be derived using Eq.\,\ref{eq:2d_to_3d}:

\begin{align}
P_{3D} &= D_C^2 \, dD_C \, P_{2D} = D_C^2 \, \frac{c}{H(z)} \, dz \, P_{2D} = \frac{D_C^2 \, c \, (1+z)}{H(z) \, \nu_{\rm obs}} \, d \nu_{\rm obs} \, P_{2D}\\
 &= \frac{D_C^2 \, c \, (1+z)}{H(z) \, \nu_{\rm obs}} \, \widetilde{P}_{2D}. 
\end{align}
We note that the 3D power spectrum is independent on the spectral resolution.
 
\section{Conversion of the shot-noise from $\mu$K$^2$\,Hz\,sr to Jy$^2$\,sr$^{-1}$\,GHz}

\label{sect:mmIME_conversion}

The shot noise measured by the mmIME experiment \citet{Keating2020} is expressed in $\mu$K$^2$\,Hz\,sr, while we use Jy$^2$\,sr$^{-1}$\,GHz in our simulation. We detail here the conversion from one unit to the other.\\

The conversion from sky temperature $T$ at a given observed frequency $\nu_{\rm obs}$ and surface brightness density $B_\nu$ expressed in SI units:
\begin{align}
B_\nu [{\rm W m^{-2} sr^{-1}}] &= 2 k_B \left ( \frac{\nu_{\rm obs} [{\rm Hz]}}{c} \right )^2 T[K] = 2 \times 1.38 \times 10^{-23} \left ( \frac{\nu_{\rm obs} [{\rm Hz}]}{3 \times 10^{8}} \right )^2\\
& = 3.07 \times 10^{-40} \nu_{\rm obs}^2 [{\rm Hz^2}] T[{\rm K}],
\end{align}
where $k_B$ is the Boltzmann constant and $c$ is the speed of light. The conversion from $\mu$K to Jy$^2$/sr involves two extra factors: $10^{-6}$ (from $\mu$K to K) and $10^{26}$ (from SI to Jy). An additional factor of $(10^9)^2$ must be applied if we want to use GHz instead of Hz. We thus get:
\begin{equation}
B_\nu [{\rm Jy^2\,sr^{-1}}] = 0.0307 \, \nu_{\rm obs}^2 [{\rm GHz^2}] \, T[ \mu {\rm K}].
\end{equation}
This conversion factor must be squared for the power spectrum units. We have also to convert the bandwidth normalization from Hz to GHz. We finally obtain:

\begin{align}
\widetilde{P}_{\rm Poi} [ {\rm Jy^2 \, sr^{-1} \, GHz}] &= 0.0307^2 \times 10^{-9}\, \nu^4 [{\rm GHz^2}] \, \widetilde{P}_{\rm Poi} [ {\rm \mu K^2 \,Hz\,sr} ]\\
&=  9.44 \times 10^{-13} \, \nu_{\rm obs}^4 [{\rm GHz}] \, \widetilde{P}_{\rm Poi} \, [ {\rm \mu K^2 \,Hz\,sr} ],
\end{align}
where $\widetilde{P}_{\rm Poi}$ is the shot-noise normalized to be bandwidth independent (see appendix\,\ref{app:CII_poisson}).

\end{appendix}

\end{document}